\documentclass[pdflatex,sn-mathphys-num]{sn-jnl}

\usepackage{graphicx}%
\usepackage{multirow}%
\usepackage{amsmath,amssymb,amsfonts}%
\usepackage{amsthm}%
\usepackage{mathrsfs}%
\usepackage[title]{appendix}%
\usepackage{xcolor}%
\usepackage{textcomp}%
\usepackage{manyfoot}%
\usepackage{booktabs}%
\usepackage[linesnumbered,ruled,vlined]{algorithm2e}
\usepackage{algpseudocode}%
\usepackage{listings}%

\usepackage{caption}
\usepackage{enumitem}
\usepackage{subfigure}

\usepackage{xspace}
\def\tool{{SPES}\xspace}
\def\newtool{{SPES-Ext}\xspace}

\usepackage[dvipsnames]{xcolor}
\definecolor{background}{HTML}{FFFFF5}
\definecolor{edge}{HTML}{87C2B1}
\definecolor{bisque}{HTML}{B86500}
\definecolor{myblue}{HTML}{2A4597}

\usepackage{cleveref}

\usepackage[most]{tcolorbox}

\definecolor{beauty}{HTML}{58A0A4}
\definecolor{myblue}{HTML}{2A4597}
\definecolor{luoshen}{HTML}{C5563B}
\definecolor{background}{HTML}{D0EEF3}
\definecolor{edge}{HTML}{6DAECD}

\newtcolorbox{mybox}{
colback=background!20, colframe=edge,
width=\columnwidth,
arc=1.5mm,
auto
outer
arc
}

\theoremstyle{thmstyleone}%
\theoremstyle{thmstyletwo}%

\theoremstyle{thmstylethree}%

\begin{document}

\title[Article Title]{Priority-Aware Scheduler for Serverless Functions Provision and Load Balancing}


\author[1]{\fnm{Cheryl} \sur{Lee}}\email{cheryllee@link.cuhk.edu.hk}

\author[1]{\fnm{Zishan} \sur{Su}}\email{zshsu@link.cuhk.edu.hk}

\author[2]{\fnm{Zhouruixing} \sur{Zhu}}\email{zhouruixingzhu@link.cuhk.edu.cn}

\author[1]{\fnm{Tianyi} \sur{Yang}}\email{tyyang@cse.cuhk.edu.hk}

\author*[3]{\fnm{Yuxin} \sur{Su}}\email{suyx35@mail.sysu.edu.cn}

\author[2]{\fnm{Pinjia} \sur{He}}\email{hepinjia@cuhk.edu.cn}

\author[1]{\fnm{Michael R.} \sur{Lyu}}\email{lyu@cse.cuhk.edu.hk}

\affil[1]{\orgname{The Chinese University of Hong Kong}, \orgaddress{\postcode{999077}, \state{Hong Kong}, \country{China}}}

\affil[2]{\orgname{The Chinese University of Hong Kong}, \orgaddress{\postcode{518172}, \state{Shenzhen}, \country{China}}}

\affil*[3]{\orgname{Sun Yat-sen University}, \orgaddress{\postcode{519000}, \state{Zhuhai}, \country{China}}}


\abstract{
Serverless computing has emerged as a pivotal cloud deployment paradigm due to its efficient utilization of on-demand resources. However, the persistent \textit{cold start} problem
continues to hinder system performance. 
Current strategies often employ oversimplified preloading/unloading mechanisms that inadequately exploit invocation patterns, failing to optimize the critical trade-off between startup latency and resource efficiency. 
Furthermore, existing approaches disregard practical constraints of memory-limited multi-worker environments, exacerbating function eviction and load imbalance issues.
To address these limitations, we present \textbf{\newtool}, a rule-based priority-aware scheduler designed for memory-constrained distributed worker nodes.
Our methodology stems from two key observations: (1) serverless systems exhibit concentrated invocation patterns enabling behavioral predictability, and (2) these patterns permit function categorization for granular instance management. 
This insight posits that characteristic-driven prioritization enables optimal scheduling rules, leading to our system design, which integrates three coordinated components: predictive function provisioning, eviction optimization, and dynamic load balancing.
Experimental validation demonstrates the superiority of \newtool over state-of-the-art solutions: 80\% reduction in $75^{th}$-percentile cold starts, 83.92\% decrease in memory waste, and 52.49\% improvement in worker load variance. 
}




\maketitle
\section{Introduction}
\begin{figure}[htb]
    \centering
    \vspace{-0.1in}
        {\includegraphics[width=0.8\linewidth]{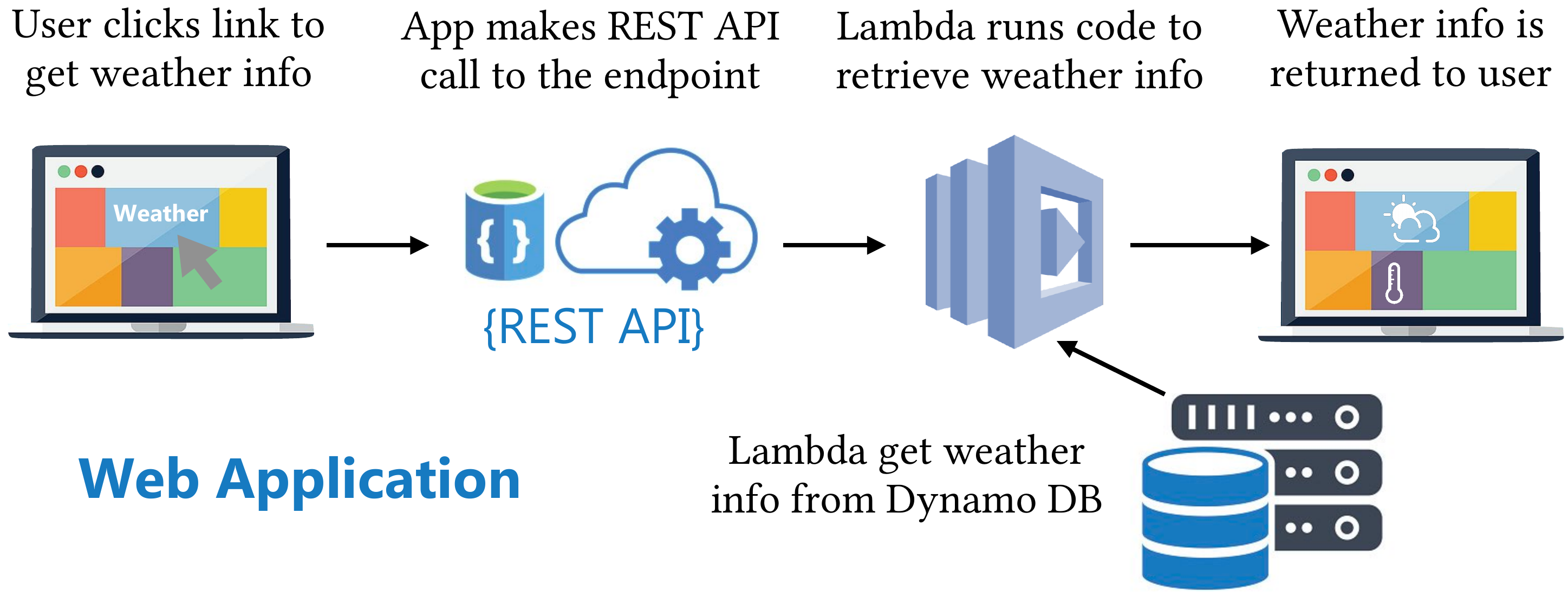}}
    \caption{A weather inquiry website based on a serverless web application.}
    \vspace{-0.1in}
    \label{fig:intro_example}
\end{figure}

Function-as-a-Service (FaaS) constitutes the core implementation model of serverless computing, facilitating streamlined cloud resource utilization while abstracting infrastructure complexities.
Prominent cloud platforms, including AWS Lambda, Google Cloud Functions, and Azure Functions, have implemented FaaS solutions supporting diverse applications: web services~\cite{WebServerless, Firestore}, machine learning algorithms~\cite{INFless, Cirrus}, and other distributed computing frameworks~\cite{Freyr, ServerlessBFT}.
This paradigm shifts infrastructure management responsibilities from developers to providers~\cite{Valve}, enabling focused application logic development.
FaaS architectures exhibit three principal advantages: (1) native support for stateless event-driven operations enabling agile deployment cycles, (2) elastic resource allocation synchronized with runtime demand~\cite{FaaSnap}, and (3) granular cost optimization to customers through pay-per-execution billing.
A representative application architecture is demonstrated in Figure~\ref{fig:intro_example}.

Despite its advantages, serverless computing introduces significant cold start challenges that create tension between latency optimization and resource efficiency.
Functions are instantiated ephemerally with execution durations typically spanning milliseconds to seconds~\cite{Formalfoundations}, yet initializing fresh instances (cold starts) requires non-trivial environment preparation latency~\cite{Firecracker} compared to reusing preloaded \textit{warm} instances.
Empirical studies indicate cold starts contribute over 80\% of total response latency~\cite{StaticAnalysis}, directly impacting user experience and service economics~\cite{Economics}.
Meanwhile, maintaining warm instances for all functions proves impractical for two fundamental reasons:
(1) Most functions experience low invocation frequencies~\cite{ServerlessWild}, making persistent memory retention economically unjustifiable due to unbilled idle allocation.
(2) Providers achieve server consolidation through colocating thousands of short-lived functions, but preloading all instances would require hundreds of GBs of memory~\cite{vHive}, exceeding typical server capacities.
This dual constraint establishes the core optimization challenge: achieving cold start mitigation while preventing memory bloat~\cite{FaasCache, INFless, SCache}.

Existing mitigation approaches can be categorized as: (1) system-layer optimizations accelerating environment initialization~\cite{Catalyzer, FaaSFlow, FaaSLight} and (2) occurrence mitigation strategies through instance management~\cite{Defuse, ServerlessWild, Ensure}. 
This work focuses on the latter category, specifically investigating non-intrusive cold start reduction within existing infrastructure constraints. Prior instance management techniques~\cite{Defuse, ServerlessWild, Ensure} employ either fixed-duration instance retention or basic interval-based heuristics, exhibiting three critical limitations:
(1) \textbf{Pattern Oversight}: Failure to account for event-driven invocation triggers and their temporal correlations.
(2) \textbf{Sparse Invocation Neglect}: Ineffective handling of infrequently called functions due to static retention policies.
(3) \textbf{Resource Agnosticism}: Ignorance of memory-constrained multi-worker environments requiring coordinated load distribution.

In this context, we aim to design a predictive, non-intrusive scheduling framework that synergistically addresses these gaps. We identify four challenges in developing such a method:

\begin{itemize}
\item \textit{Low-Latency}: Operational constraints demand subsecond provisioning decisions given FaaS platforms process thousands of concurrent invocations across millions of functions per minute~\cite{ServerlessData}, with typical execution windows under one second.

\item \textit{Burst Tolerance with Multi-Worker Scalability}: Scheduling mechanisms must exhibit (1) sublinear complexity growth to accommodate request burst~\cite{Firestore}, and (2) elastic coordination across dynamically scaled worker pools without architectural modification.

\item \textit{Skewed Invocation Distribution}: Invocation data can be sparse, where 95\% of functions experience less than one invocation per day~\cite{ServerlessWild}, limiting training data availability. Plus, the inherent invocation heterogeneity creates significant memory utilization variance across workers, making load balancing nontrivial.

\item \textit{Temporal Concept Drift}: The lightweight nature of FaaS makes applications evolve faster than traditional software, so scheduling strategies must withstand both short-term concept drift (e.g., hourly user activity fluctuations affecting) and long-term distribution shifts (e.g., quarterly software updates), altering function call patterns and complicating predictive modeling. 

\end{itemize}

\textbf{Insight:} \textit{The event-driven paradigm of serverless architectures enables systematic identification of recurrent invocation patterns, permitting optimized function categorization and strategic provisioning.}
Three empirical observations grounded in both domain expertise and data analysis substantiate this premise:
\textbf{Trigger-Pattern Correlation}: 82\% of functions exhibit predictable execution sequences tied to specific event triggers (e.g., scheduled data processing workflows), as verified through analysis of production traces (in Section~\ref{sec:motivation}).
\textbf{Behavioral Convergence}: functions within an application often display behavior convergence. For example, functions within a multi-stage processing workload are typically invoked in turn. 
\textbf{Transient Temporal Locality}: Even irregular functions can exhibit temporal locality, maintaining a short execution window during peak activity periods, permitting cost-effective transient retention.
These observations demonstrate that functions can present limited predictable patterns, facilitating a dual-phase scheduling mechanism: 1) anticipate invocations through trigger-pattern mapping and historical sequence analysis to minimize cold starts with minimal memory influence; 2) implement priority-based placement across distributed workers, considering both temporal urgency and memory constraints to reduce loading variances.

We therefore present \textbf{\newtool}, a \underline{Ext}ension versions of the \underline{S}cheduler for \underline{P}rovisioning runtim\underline{E} \underline{S}erverless functions.
This is a pattern-aware adaptive scheduler that optimizes the latency-resource tradeoff while enforcing load balancing across memory-constrained distributed workers. The design synthesizes three key innovations:
(1) We summarize five dominant invocation patterns driven by the empirical analysis of production traces, supplemented by three adaptive retention policies for function patterns that are less predictive.
(2) We propose a computationally lightweight yet effective metric--co-occurrence rate--that quantifies inter-function dependencies, enabling predictive chaining of related invocations, \textit{i.e.,} highly correlated functions can be predictive indicators of each other.
(3) We design dynamic policy adjustments for continuous pattern recalibration to address concept drift as FaaS evolves.
In this way, \newtool operates through two coordinated phases. It anticipates invocations using pattern matching and correlation analysis to pre-load functions with fine-grained strategies. When the invocations come, \newtool orchestrates the load and distribution of functions toward workers.
As a result, our rule-based \newtool addresses all the above challenges. Its tailored strategies can handle diverse functions with remarkable scalability and adaptivity, introducing a small computation overhead. 

We evaluate \newtool on the most widely used industrial dataset released by Microsoft Azure~\cite{ServerlessWild}.
Experiments demonstrate the superiority of \newtool, which achieves both-side improvement:
reducing the $75^{th}$-percentile cold start rate by 80\%--90\% and the wasted memory time by 83.92\%--88.82\% compared to baselines, and also reducing the proportion of ``always-cold'' functions significantly. Meanwhile, \newtool lowers the mean variances of workers space by 52.49\%--96.14\%.
Extensive ablation studies further confirm the effectiveness of our designs of inter-function connection and adaptive strategies in our provision policy, as well as the validity of our load-balancing scheduling.

In summary, the main contributions of this paper are:
\begin{itemize}
\item We present \newtool, the first rule-based priority-aware scheduler for memory-constrained distributed worker clusters, integrating three modules, i.e., function provision, eviction decision, and load-balancing.

\item We summarize typical invocation patterns and categorize functions accordingly as the basis of our novel differentiated provision strategy.
We also create an effective ``co-occurrence'' metric to gauge inter-function correlations. 
Contributions of designs for adaptivity and unseen functions have also been validated.

\item We conduct extensive experiments under both a single-worker scenario without memory constraint and a multi-worker scenario with memory limitation to demonstrate the significant improvement of \newtool in the cold start and wasted memory reduction, as well as the improvement in load balancing effects.

\end{itemize}

Part of the results in this paper has been reported in \tool~\cite{lee2024spes}, published at ICDE. For clarity, we should point out that \tool represents the provision policy we propose in~\cite{lee2024spes}, and \newtool refers to the general scheduler we propose in this paper that involves provision, eviction, and load-balancing. For readers who have read the conference version, this extension mainly includes a general scheduler that couples both provisioning and load-balancing (See Section \ref{sec:method}), and new experiments under the memory-limited, multi-worker scenario (See Section \ref{sec:exp:eval:spes-ext}). We also include a more comprehensive background, preliminary analysis, discussion, and related work version.

\section{Background}\label{sec:background}
\subsection{Serverless Computing}
Serverless computing constitutes an evolving cloud execution model where cloud providers assume full responsibility for infrastructure management and dynamic resource allocation, enabling developers to focus exclusively on application logic implementation.
As a core subset of serverless architectures, Function-as-a-Service (FaaS) specifically operationalizes this paradigm by abstracting runtime environments into ephemeral, \textit{event-triggered} computational units.
In FaaS, services are structured as stateless computational units (functions) that execute in response to specific event triggers defined through declarative rules~\cite{Netherite}.
This operational model offers inherent scalability and cost-efficiency through granular pay-per-use billing for actual resource consumption~\cite{ServerlessDataScience}.
Practically, cloud-native \textit{applications} are typically decomposed into discrete, loosely-coupled functions that may adopt explicit orchestration patterns~\cite{Kraken}.
The technology has gained significant traction across domains, including web services and distributed machine learning training~\cite{TowardsML}.
Despite its advantages, key technical challenges persist, particularly in guaranteeing strict Quality-of-Service (QoS) compliance for metrics such as response latency and tail latency optimization~\cite{AQUATOPE}.

\subsection{Cold Start Challenge}\label{sec:background:challenge}
\begin{figure}[htb]
    \centering
     \vspace{-0.1in}
        {\includegraphics[width=0.75\linewidth]{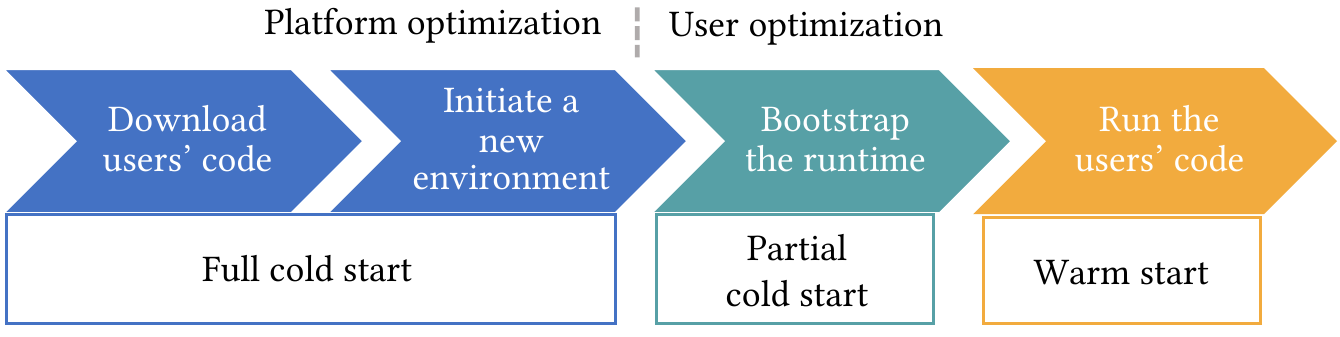}}
    \caption{A serverless function's lifecycle.}
     \vspace{-0.1in}
    \label{fig:function_process}
\end{figure}

Figure~\ref{fig:function_process} displays the lifecycle of a serverless function under FaaS.
The FaaS provider initially provisions the function by retrieving user-submitted code and instantiating an isolated runtime sandbox (e.g., container or VM) on cluster nodes—a process termed \textit{cold start}. 
This phase sequentially involves code retrieval from persistent storage, sandbox initialization, memory loading, and handler execution.
A \textit{warm start}, conversely, bypasses provisioning phases by reusing pre-initialized environments, thereby minimizing latency.
After execution for a request, the environment persists in an idle state until evicted by the orchestration layer after a provider-defined timeout~\cite{Atoll}.
Naturally, environment reuse inversely correlates with cold start frequency.
A critical trade-off emerges: while retaining active environments reduces provisioning overhead and accelerates subsequent invocations, it introduces more \textit{idle resource retention} quantified via \textit{wasted memory time (WMT)}.
Formally, WMT represents the cumulative duration where a function's runtime image occupies memory without processing requests, serving as a key empirical metric for infrastructure efficiency analysis~\cite{ServerlessWild}.

The cold start problem poses a critical challenge in serverless computing, as it introduces substantial runtime latency while providers must strictly adhere to QoS constraints and sustain a perceived ``always-ready'' service illusion.
Empirical studies indicate two compounding factors.
First, cold starts dominate the overall execution time~\cite{FaaSLight, FaasCache, Peeking}, accounting for over 70\% of total invocation time in mainstream platforms like gVisor, as reported in~\cite{Catalyzer}.
On the other hand, transient function lifetimes trigger frequent cold starts across diverse workloads~\cite{AJAX, SOCK, FaaSNet}.
These render cold-start-induced latency prohibitively detrimental to QoS adherence.
Given memory's cost constraints and finite availability, mitigation strategies necessitate either predictive instance optimization to balance memory utilization and readiness or accelerated provisioning (e.g., parallelized sandbox initialization).

\section{Preliminary Empirical Analysis}\label{sec:motivation}
\subsection{Challenges of Serverless Function Provision}\label{sec:motivation:challenge}

\subsubsection{Low-Latency}
Serverless provisioning systems face dual operational constraints. First, they must execute sub-second scheduling decisions for bursty invocation volumes (e.g., tens of thousands per minute~\cite{LaptopLambda}), precluding computationally intensive optimization models.
Second, short-lived function lifetimes (typically $<$ 60s) demand dynamic strategy updates within tight scheduling windows~\cite{Hermod}, as delayed provisioning directly impacts billing efficiency.
Consequently, lightweight schedulers outperform complex algorithms in practice, prioritizing decision velocity over theoretical optimality.

\subsubsection{Scalability under invocation burst}
Serverless provisioning systems must inherently support autonomous elastic scaling as FaaS invocations are extremely bursty and dynamic~\cite{FaaSNet}, and resource allocation in a FaaS platform must synchronize with demand spikes within milliseconds to prevent QoS violations.
Since invocation rates exhibit order-of-magnitude fluctuations within minute-level intervals (e.g., 10x traffic surges during e-commerce peak events), a highly scalable solution is required.

\subsubsection{Skewed Invocation Distribution}
\begin{figure}[htb]
    \centering
     \vspace{-0.1in}
        {\includegraphics[width=0.8\linewidth]{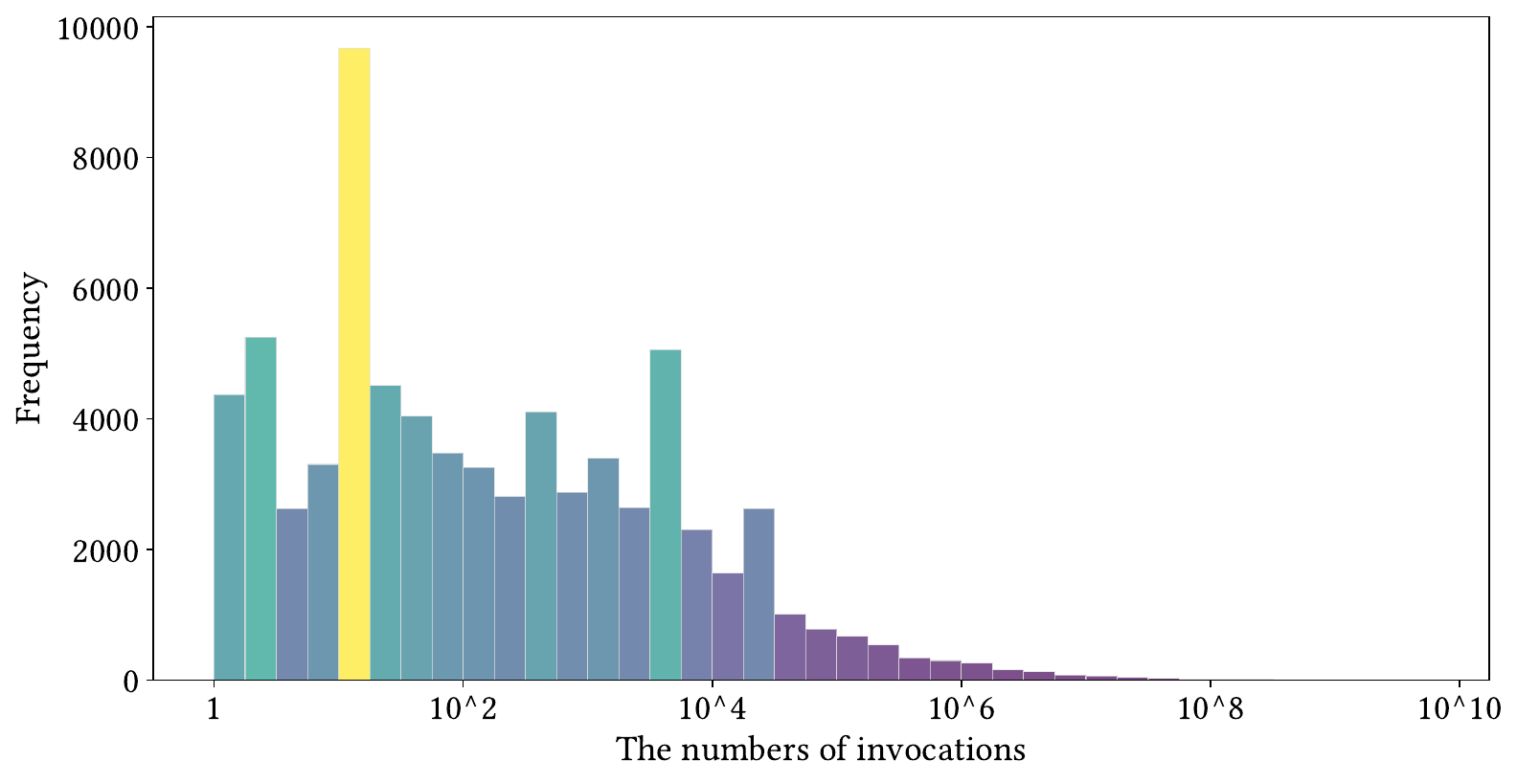}}
    \caption{The distribution of function invocations. The x-axis represents the range of function invocation counts, and the y-axis indicates the number of functions falling into the corresponding ranges.}
     \vspace{-0.1in}
    \label{fig:lognum_dist}
\end{figure}

Serverless functions demonstrate multidimensional heterogeneity in functionality.
They are implemented with varied runtime environments and dependency libraries, creating distinct execution profiles~\cite{AQUATOPE}. 
Moreover, trigger mechanisms span HTTP endpoints (API gateways), database transactions (e.g., Cosmos DB), and message queues, each imposing unique invocation patterns~\cite{RiseServerless}. 
This heterogeneity induces extreme invocation disparity, evidenced by Azure's 14-day production traces collected by~\cite{ServerlessWild}, as shown in Figure~\ref{fig:lognum_dist}.
We can observe that the invocations are highly non-uniform, with most functions being rarely invoked.
Such a power-law distribution creates inherent challenges for learning-based provisioning that require lots of records of each function for training.

\subsubsection{Evolution in invocation behavior}\label{sec:motivation:challenges:evolution}
\begin{figure}[htb]
    \centering
     \vspace{-0.1in}
        {\includegraphics[width=0.75\linewidth]{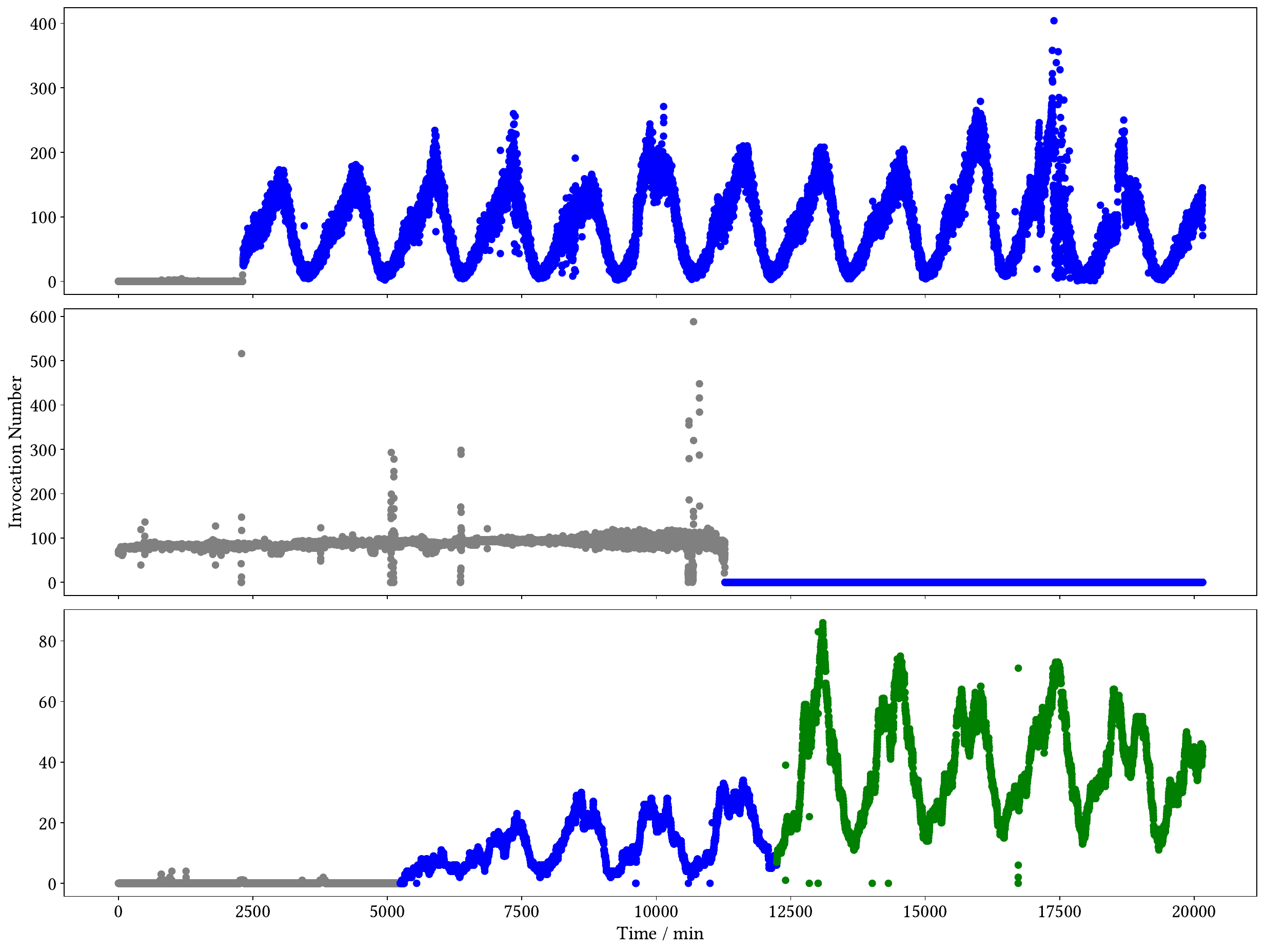}}
    \caption{The function invocations can experience distinct concept shifts that may degrade the provision performance. Different colors distinguish changes over time in the function invocation patterns.}
     \vspace{-0.1in}
    \label{fig:show_shift}
\end{figure}

Invocation dynamics evolution introduces critical adaptability challenges for serverless provisioning systems, particularly those employing predictive pre-loading to minimize cold starts. 
This issue is exacerbated in learning-based approaches due to their need for frequent, resource-intensive retraining.
Many factors can cause such shifts, including temporal fluctuations (e.g., holiday-induced traffic spikes in e-commerce platforms), business logic updates (e.g., geographic expansion in business propagating to function dependencies and invocation patterns), trigger reconfigurations (17\% of functions annually rebind to new triggers or call chains~\cite{ServerlessWild}), etc.
The Azure dataset (Fig.~\ref{fig:show_shift}) demonstrates short-term volatility, with representative functions showing abrupt invocation shifts. Such dynamics necessitate provisioning methods with inherent adaptability to concept shifts without requiring constant retraining.

\subsection{Observations and Our Insight}\label{sec:motivation:insight}
\subsubsection{Invocation pattern and triggers}\label{sec:motivation:insight:trigger}
Serverless systems employ an event-triggered invocation paradigm where predefined trigger types (HTTP, message queues, scheduled timers) establish deterministic invocation modalities~\cite{RiseServerless}. This architectural constraint enables predictable behavioral patterns for most functions, as evidenced by two empirical findings from the Azure dataset.

\begin{figure}[htb]
    \centering
        {\includegraphics[width=0.75\linewidth]{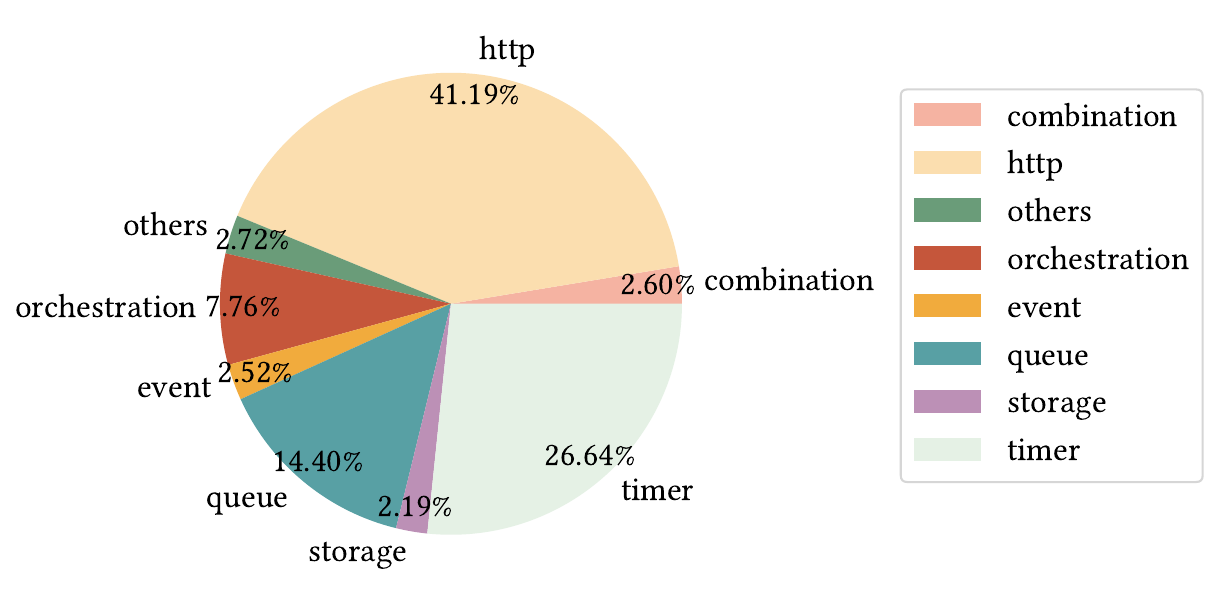}}
    \caption{The proportion of trigger types among functions.}
     \vspace{-0.1in}
    \label{fig:trigger_prop} 
\end{figure}

First, most functions correspond to only one trigger type.
As shown in Figure~\ref{fig:trigger_prop}, most of functions bind to a single trigger type, and only 2.6\% exhibit multi-trigger bindings (maximum 3 types, categorized as ``combination'').

Second, we find that trigger-specific invocation regularity emerges as a predictive signal, validated through distributional analysis, where we applied the Kolmogorov-Smirnov test on functions with $\geq$10 invocations ($pValue \geq 0.05$=0.05~\cite{KStest}). 
Results show that 68.12\% timer-triggered functions are invoked periodically or quasi-periodically, excluding 6.65\% do not have enough counts.
Also, 45.02\% HTTP-triggered invocations align with Poisson arrivals, conforming to the expectation of human activities, excluding 36.20\% without enough samples.
Though the Poisson process is memoryless and unsuitable for exact prediction, this statistical alignment demonstrates that meaningful invocation behaviors are associated with triggers. 

\subsubsection{Application workflow}\label{sec:motivation:insight:app}
Serverless applications often orchestrate functions into logical workflows through explicit chaining or callback mechanisms~\cite{Kraken}, creating inherent temporal and semantic dependencies between invocations. Empirical analysis reveals two key properties of these dependencies.
We create a metric \textit{co-occurrence rate (COR)} to quantify functional cohesion, inspired by linguistic co-occurrence analysis. For any function pair sharing the same application or user context, COR quantifies their invocation simultaneity as:
\begin{equation}
    COR(f_i,f_j) = \frac{\|\{t| f_i, f_j \text{ invoked at }t\}\|}{\|\{t|f_i \text{ invoked at }t\}\|}
\end{equation}
Analysis of 50 randomly sampled negative pairs (non-associated functions) demonstrates strong workflow cohesion: application/user-linked functions (\textit{candidate function}) exhibit 4.6× higher COR (0.231 vs. 0.050) than unrelated pairs.
Trigger-mediated convergence further amplifies this cohesion. When functions share trigger types (e.g., HTTP endpoints), their COR increases to 0.271—2.1× higher than cross-trigger pairs (0.131). This aligns with our earlier observation in Section~\ref{sec:motivation:insight:trigger} that triggers dictate invocation semantics. 

These findings confirm that workflow structures imprint measurable invocation patterns, particularly under shared operational triggers. Such regularity provides critical signals for dependency-aware provisioning strategies.

\subsubsection{Temporal locality in invocations}
\begin{figure}[htb]
    \centering
    \vspace{-0.12in}
        {\includegraphics[width=0.95\linewidth]{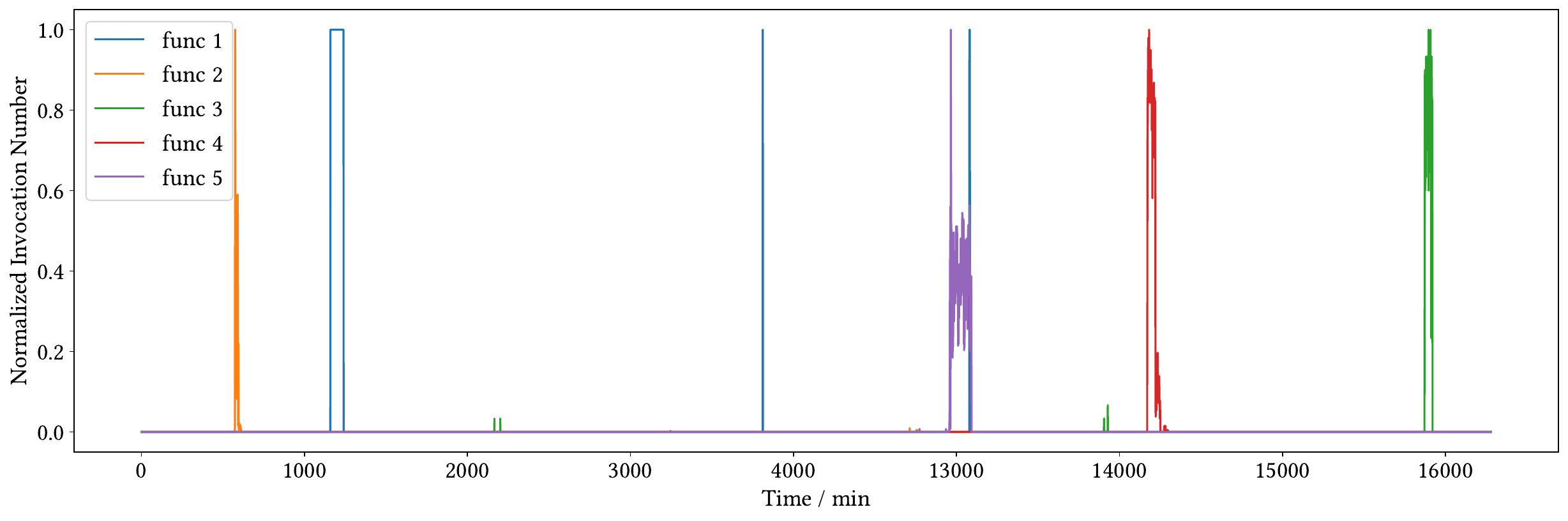}}
        \vspace{-0.1in}
    \caption{Invocations of these infrequently invoked five functions are centralized over several periods. We omit the no-invocation period (4000--13000).}
    \vspace{-0.1in}
    \label{fig:temporal_locality}
\end{figure}

Temporal invocation locality—the propensity for serverless functions to exhibit clustered invocation bursts within short timespans—emerges as a critical operational pattern. This phenomenon arises from two complementary factors.
First, real-world triggers like flash sales in e-commerce platforms generate rapid request successions, directly propagating to function invocation spikes.
Second, common architectural patterns systematically induce bursty behaviors.
For example, in a fan-out application, a primary function spawns parallel invocations of multiple downstream functions simultaneously (e.g., processing 100 image thumbnails in one API call).
Callback functions that handle asynchronous operations also trigger recursive invocations until task completion (e.g., retry loops in payment systems).
Empirical validation via the Azure dataset (Fig.~\ref{fig:temporal_locality}) reveals even infrequently invoked functions sustain temporal locality.
For infrequently invoked functions presenting temporal locality, invocation prediction can be difficult, yet simply keeping a loaded function for some time can considerably reduce cold starts without wasting much memory.

\begin{mybox}
  \small
  \textbf{Insight}: The microservice and event-driven architecture of serverless systems results in the concentration of recurring and systematic invocation behaviors.
  By analyzing the invocation patterns and categorizing functions, we can accurately predict and pre-load function instances to reduce cold starts.
\end{mybox}

Motivated by these insights, we present \newtool, a lightweight scheduling framework for FaaS systems that synergistically integrates three core mechanisms: rule-based provisioning, adaptive eviction, and distributed load balancing
It leverages the application, user, and trigger information to design tailored and evolution-adaptive strategies to pre-load, evict, and dispatch categorized functions without requiring large amounts of training data.
In general, \newtool provides a comprehensive, developer-free, language-agnostic function scheduler that can be seamlessly adopted on systems to optimize cold starts and memory usage.

\section{Methodology}
\label{sec:method}
\begin{figure*}[htb]
    \centering
    \vspace{-0.1in}
        {\includegraphics[width=0.9\linewidth]{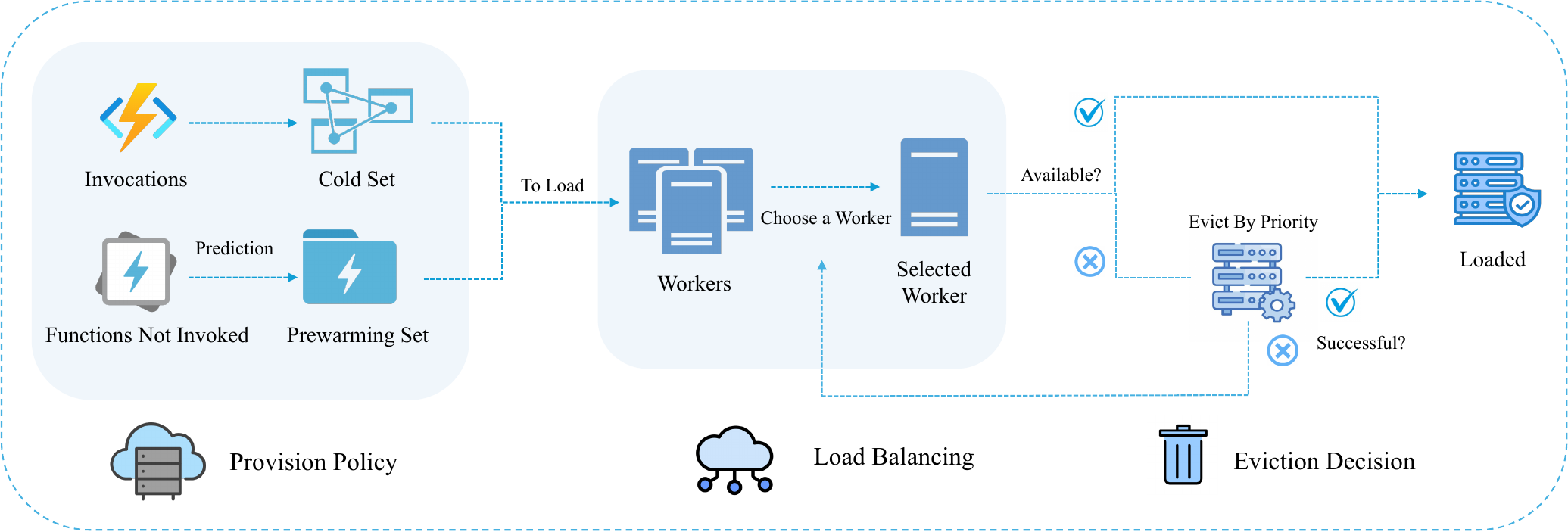}}
    \caption{Overview of \newtool. In the provisioning stage, our \tool scheduler predicts and decides which functions to prewarm and load into the workers. Among multiple workers, we apply a priority-based load-balancing policy to select a worker for the coming function. If the worker is out of memory, it will conduct the eviction decision and evict functions with low priority until it gets enough space. However, if it still does not have enough space for the function after evicting all the possible functions, we will turn to the next appropriate worker.}
    \vspace{0.1in}
    \label{fig:flow_spes_ext}
\end{figure*}

To address the aforementioned challenges, we propose \newtool, a general-purpose, rule-based scheduler with priority awareness.
As illustrated in Figure~\ref{fig:flow_spes_ext}, \newtool adopts a three-component architecture: Provision Policy, Eviction Decision, and Workers Load Balancing.
The design of \newtool is driven by the following key considerations:
\begin{itemize}
    \item To alleviate cold start, how do we predict function invocations and decide function provision?
    \item When there are functions to load, how to dispatch each function to the workers?
    \item When the workers are out of memory, which functions should be evicted?
\end{itemize}

\subsection{Provision Policy}

\begin{figure*}[htb]
    \centering
    \vspace{-0.1in}
        {\includegraphics[width=0.9\linewidth]{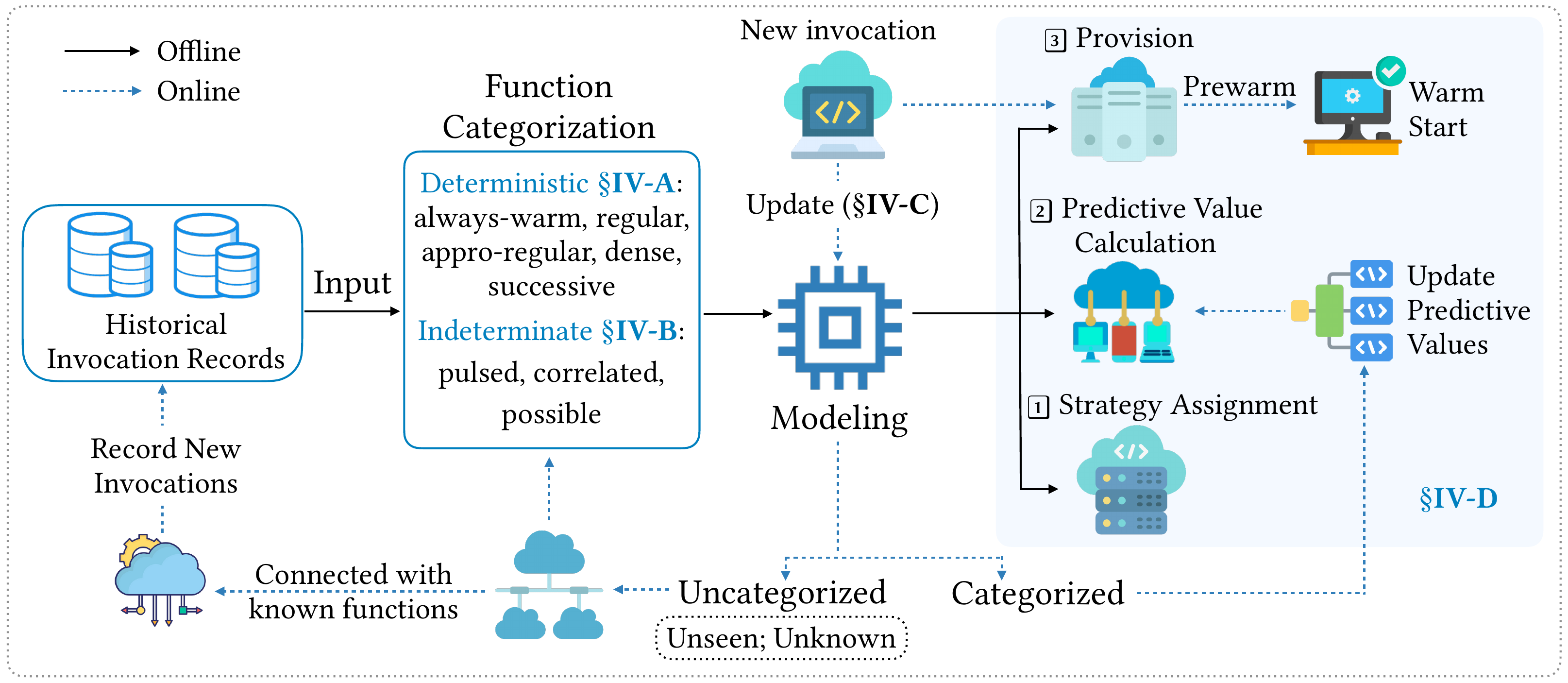}}
    \caption{Overview of \tool, the provision policy for \newtool. In the offline phase, \tool models the historical invocation records for each function and categorizes functions according to their mined invocation patterns. During the online phase, if the new invocation comes from a categorized function, its function's provision directly follows its provision rule, and uncategorized functions are tried to be connected with known ones. The invocation prediction and predictive values/indicators are then updated upon new invocations.}
    \vspace{-0.1in}
    \label{fig:overview}
\end{figure*}

To address the first challenge, we employ a provision policy that predicts function invocations and dynamically determines which functions to load. 
Our design rationale is that an optimal scheduler should decide to load a function immediately before its invocation and evict it from memory right away after the execution if no more invocations, so as to minimize idle memory retention.
Achieving this requires fine-grained invocation prediction, which forms the core of \tool's design, whose overview is presented in Figure~\ref{fig:overview}.
\tool consists of four key components: deterministic function categorization, indeterminate function assignment, adaptive strategy application, and function provision based on invocation prediction.
Specifically, we first summarize typical invocation patterns and formulate their corresponding definitions.
These patterns result from pre-defined triggers~\cite{AzureTriggers} and the combinatorial methods of function calling (in common application scenarios~\cite{AzureScenario}).
If a function satisfies the definitions, it is categorized; Otherwise, \tool assigns uncategorized functions to three supplementary types.
We also introduce two dynamic strategies to accommodate evolving FaaS workloads.
Finally, \tool provisions functions according to rule-based invocation prediction, where each type follows its own predicting rule.

It is important to note that distinct triggers may demonstrate identical temporal patterns in function invocations. For instance, both service bus triggers and event grid triggers can process streaming data packets, despite specializing in different message formats. In addressing cold start optimization, maintaining a one-to-one mapping between trigger types and function categories is neither practical nor advantageous.

Let us start with three definitions: 
\begin{itemize}
    \item \textit{Waiting time (WT)}: the length of successive idle time. 
    Take an invocation sequence $(28, 0, 12, 1, 0, 0, 0, 7)$ as an example, where each value denotes the invocation count at each sampling slot, derived to a sequence of WTs $\{WT\}=(1, 3)$, as no invocation exists at the 2nd slot or in slots 5--8.
    If we foreknow the next WT, for example, at the 4th slot, we know no invocations will arrive in slots 5-8, then we can easily make a perfect decision: evict the function now and re-load it at the end of slot 9.
    Hence, we can transform invocation prediction to WT prediction.
    Different from a previously proposed inter-arrival time (IAT)~\cite{ServerlessWild}, WT depicts the slot-grained interval between two successive invocation sequences, from the last's end to the next's start.
    WT is also different from idle time (IT). IT is the time without invocations or subsequent function executions, so IT exists after any single execution. 
    WT, instead, only appears when a successive series of invocations ends, and the end of a single execution does not necessarily incur a WT.

    \item  \textit{Active time (AT)}: the length of successive time with invocations. 
    The above-mentioned invocation sequence delivers an AT sequence $\{AT\}=(1, 2, 1)$ because the function is invoked in slots 1, 3--4, and 9. 
    
    \item \textit{Active number (AN)}: a descendable definition, the number of successive invocations during AT. Again, the above sequence delivers $\{AN\}=(28, 13, 7)$.

\end{itemize}

\subsubsection{Deterministic Function Categorization}\label{sec:method:rule}

This section defines five invocation types based on typical invocation patterns and categorizes functions accordingly. Table~\ref{tab:define} shows an overview of the five types.
The guiding principle for definition is from easy to difficult, also serving as the categorization priority: if a function fits a former type, it will not fit any latter type.
Our introduction follows this order.

\begin{table*}[htb]
    \small\centering
    \vspace{-0.1in}
    \captionsetup{justification=centering}
    \caption{A brief overview of the well-defined function types.}
    \resizebox{0.9\textwidth}{!}{%
        \begin{tabular}{cccc}
        \toprule
        \textbf{Type name} & \textbf{Characteristics} & \textbf{Definition} & \textbf{Predictive values} \\
        \midrule
        \multirow{2}*{Always warm} & \multirow{2}*{Almost invoked all the time} & {Invoked at every time} & \multirow{2}*{--} \\
        && Inter-invocation time $\leq$ observing time\textperthousand & \\
    
        \midrule
        \multirow{2}*{Regular} & \multirow{2}*{Almost invoked periodically} & {(Processed) $P_{95}(\{WT\})-P_5(\{WT\}) \leq 1$} & \multirow{2}*{Median of WTs} \\
        && CV of $\{WT\} \leq 0.01$ & \\
    
        \midrule
        \multirow{2}*{Appro-regular} & \multirow{2}*{Invoked quasi-periodically} & (Processed) the count of Mode$_n(\{WT\}) \geq$  & \multirow{2}*{Mode$_n(\{WT\})$}\\
        && $0.9 \times$ sequence length & \\
    
        \midrule
        Dense & Frequently invoked &   $P_{90}(\{WT\}) \leq$ a small constant & $[\min, \max]$ of $\text{Mode}_k(\{WT\})$ \\
    
        \midrule
        \multirow{2}*{Successive} & \multirow{2}*{Successively invoked} & $\min(\{AT\}) \geq \gamma_1$; $\min(\{AN\}) \geq \gamma_2$ & \multirow{2}*{--} \\
        && $\gamma_1 < \gamma_2$ & \\
        
        \bottomrule
        \end{tabular}%
    }
    \label{tab:define}
     \vspace{-0.1in}
\end{table*}


\textit{i) Always warm:} 
This type describes consistently active functions, such as long-running operations and hyperfrequent calls, which usually involve durable or timer-triggered functions (whose interval is small enough).
For example, functions in continuous integration/deployment (CI/CD) pipelines are consistently invoked to automate code building, testing, and deployment.
A function is defined as ``always warm'' if 1) it is invoked at every sampling time or 2) the sum of inter-invocation time is $\leq$ one-thousandth the observing time.
Such functions are undoubtedly always loaded.

\textit{ii) Regular:}
Regular task processing, usually using timer-triggered functions, is a basic functionality (e.g., polling).
Yet, their actual invocations may not be strictly periodic (with a constant WT) since there are undesired fluctuations:
1) The exact first/last WTs during an observing period are almost impossible to record.
2) Periodically generated events can be blocked or delayed by contingencies such as concurrency limits, network connectivity issues, etc. 
3) Other events can invoke a mostly regularly invoked function occasionally.

Here comes the slacked definition:
A ``regular'' function satisfies either 1) the difference between the 5th and 95th percentile of its WT sequence is $\leq 1$ or 2) the coefficient of variation of WTs is close to zero ($\leq 0.01$ in practice). 
Otherwise, we remove the first and last WTs and re-check if the function follows the definition.
If the answer is still no, we apply another slacking rule by merging adjacent small WTs.
In particular, for each WT closely valued to the WT mode, its adjacent small WTs are gradually merged until reaching \textcircled{{\scriptsize 1}} the sequence's end or \textcircled{{\scriptsize 2}} another WT close to the mode or \textcircled{{\scriptsize 3}} an already merged WT. 
In this way, a WT sequence valued by $(1439, 1438, 1, 1439, 1438, 1)$ is processed to be $(1439, 1439, 1439, 1439)$, and then it seems ``regular''.
With a processed WT sequence satisfying the above definition, the function belongs to the ``regular'' type.
We partially eliminate accidental factors by processing WTs with slacking rules and modeling essential invocation behaviors.
Finally, we record the median of WTs as the \textit{predictive value} for invocation prediction, leveraged in Section~\ref{sec:method:provision}.

\textit{iii) Approximatively regular:} 
This is the derived type from ``regular''.
A regular function may experience invocation variance and long-term disturbance.
For example, a data-processing function is expected to receive updates from a data station every three minutes (IoT Hub trigger functions). Yet, limited by the data transmission capability, the function is actually invoked every 3--5 minutes.
We define such functions as ``approximatively (appro-)regular''. 
Particularly, a function is ``appro-regular'' if the occurring count of the first $n$ frequently appearing values (modes) of WTs is $\geq$ a large percent (90\% practically) of the WT sequence's length. $n$ is a pre-defined integer.
The predictive values for these ``appro-regular'' functions are the first $n$ modes.

\textit{iv) Dense:}
This characterizes irregular but frequent invocations with intermittent idle durations.
``Dense'' functions can involve several triggers: queues (or service bus using the queue flavor) for frequent messages in asynchronous processing, influenced by variations in message arrival times; Cosmon DB triggers for frequent real-time data processing; HTTP triggers for processing frequent HTTP requests.
Functions with a 90$^{th}$ percentile of the WTs $\leq$ a small constant are defined as ``dense''.
The range of the first $k$ modes derives predictive values, \textit{i.e.} $[\min(\text{Mode}_k(\text{WTs}))$, $\max(\text{Mode})_k(\text{WTs})]$, where $k$ is an empirical integer.
Such a function should be unloaded only if its idle time is larger than the constant.

\textit{v) Successive:}
This describes inactive functions experiencing consecutive invocations in a short timeframe (temporal locality) before returning to an inactive state.
It can involve HTTP trigger functions (e.g., bursty social media trends), storage trigger functions (e.g., uploading files to Blob Storage), etc.
In practice, feedback loops engaged functions, caching mechanism implementations, and load balancing-involved applications can also exhibit such behavior.
We define this as $\min(\{AT\}) \geq \gamma_1$ or $\{AN\} \geq \gamma_2$, where $\gamma_1 < \gamma_2$ and they are pre-defined lower bounds.
Predicting the start of such an invocation wave is highly difficult.
Nevertheless, we can tolerate the first invocation and keep the function alive until the wave ends. With limited cold starts, we save lots of memory.

\subsubsection{Indeterminate Function Assignment}\label{sec:method:difficult}

This section handles functions that do not satisfy the above five definitions using two strategies: \textcircled{{\scriptsize 1}} ``forgetting'' previous records to re-check whether the not-too-long history invocations satisfy the five definitions, and \textcircled{{\scriptsize 2}} assigning the remaining functions with three extra types.

\textit{i) Forgetting:}
As discussed in $\S$\ref{sec:motivation:challenges:evolution}, function invocations experience concept shifts. For example, a regular invoked function can be categorized as “always warm” if the interval is adjusted to be smaller than the recording unit.
Therefore, the near-the-present records should be given priority.
We slice out the sequential invocation observations in days, from the beginning to the half.
Then, we gradually remove the older observations from the original $d$-day observations and check whether invocations in 2nd--$d$th days conform to existing definitions. 
If not, we check the invocations in 3rd--$d$th days, and so on, until the $\lfloor d/2 \rfloor$th day.

\textit{ii) Assigning:}
The remaining indeterminate functions are assigned to three types according to the validation results:
\begin{itemize}

    \item[D1.] \textit{Pulsed} describes functions showing less obvious temporal locality in invocations than ``successive" ones.
    Similarly, we allow a cold start at the first invocation time and keep the function warm until its idle time reaches a pre-defined threshold.
    
    \item[D2.] \textit{Correlated} invocations are predicted based on other functions.
    Functions usually work in a logic workflow topology, interacting with other functions. This is present in several common application patterns, including function chaining (where a sequence of functions executes in a specific order), fan-out (where a function calls multiple functions in parallel), and fan-in (where a function waits for multiple functions to finish)~\cite{AzureScenario}.
    We define such functions using $T$-lagged co-occurrence rate ($T$-COR), derived from COR in $\S$\ref{sec:motivation:insight}. $T$-COR is the COR between the target invocation sequence and an $T$-time-lagged invocation sequence of its candidate functions (those sharing an application/user).
    The lagged COR can better reflect how the invocation of candidate functions indicates near-future invocations of the target function.
    If the $T$-COR reaches a threshold (practically 0.5, $T \leq 10$), the candidate invocation is an important predictive indicator, and these two functions are linked.
    A function can have multiple linkages with different functions, including categorized and uncategorized ones.

    \item[D3.] \textit{Possible} functions are infrequently invoked but have at least one mode of WTs (appears more than once). By taking the modes as predictive values, it is possible to predict invocations, though such prediction is deemed to be insufficiently satisfactory.

\end{itemize}

If any of the three strategies yields both the least cold starts and the minimum wasted memory during validation, the indeterminate function is directly assigned accordingly.
Otherwise, the rate of rise is decided. 
Denote the cold starts of the above three strategies as $(cs_1, cs_2, cs_3)$ and the corresponding wasted memory are ${wm_1, wm_2, wm_3}$.
Assume definition D1 delivers the minimum cold starts while strategy D2 delivers the minimum wasted memory.
We compute the rise rates: $\Delta cs=(cs_2-cs_1)/cs_1$ and $\Delta wm=(wm_1-wm_2)/wm_2$. 
If $\Delta cs \times \alpha \leq \Delta wm $, then the function is assigned to D1, \textit{i.e.} ``plused''; otherwise, D2 prevails, where $\alpha \in (0,1)$ is a scaling factor, and the smaller $\alpha \in (0,1)$, the more importance is put on cold starts than wasted memory.

Most functions have been categorized so far, except those that have never been invoked in the validation. 
We simply leave them as ``unknown'' since they are likely very infrequently invoked and hardly have meaningful predictive indicators. 
Some unknown functions may be categorized during the online provision based on our adaptive strategies illustrated in the next section ($\S$\ref{sec:method:adaptive}).

\subsubsection{Adaptive Strategy Application}\label{sec:method:adaptive}
This section designs two adaptive strategies to handle concept shifts during the online provision: one adjusts the predictive values; the other is based on inter-function correlation. 
These strategies can further address some unknown or unseen functions (those that never appear in the training data) with meaningful patterns during the online provision.
The application of these strategies is also presented in Figure~\ref{fig:overview}, where the invocation prediction and the prediction values are adaptively updated upon online invocations.

\textit{i) Adjusting:}
This adaptive strategy contains three steps:
\begin{itemize}

    \item[S1.] Record the WTs during the online provision. If there are enough WTs, then initiate the adaptive updating.
    
    \item[S2.] Update the predictive values if the corresponding values computed from newly collected WTs change significantly with the mean of the old and new ones.
    
    \item[S3.] If the new WTs for an unknown or unseen function conform to previous definitions with enough samples, categorize the function as the corresponding type.
    
\end{itemize}
S2 requires more explanations.
We record the predictive values for four types: ``regular'', ``appro-regular'', ``dense'', and ``possible''.
Their predictive values are the median, the first $n$ modes, the range of the first $k$ modes, and the WT values occurring more than twice, respectively.
Take ``regular'' as an example. 
Suppose the absolute difference between the median of offline WTs (\textit{i.e.} the predictive value) and that of online WTs is larger than the standard of offline WTs. The new predictive value is updated by the mean of the old median and the new median.
Other types adopt a similar adjusting strategy. 

\textit{ii) Online correlation:}
This strategy correlates unseen functions with known functions or appeared unseen functions, as the ``correlated'' strategy does.
To speed up the computation and obtain informative indicators, we only consider candidate functions sharing the same trigger with the target function (\textit{aka.} unseen function).
Initially, if one of the candidates is invoked, we also pre-load the target function.
Afterward, we gradually remove less correlated candidates.
Again, we adopt COR to measure the degree of correlation.
We count the pair-wise target-candidate COR at each slot and record the maximum.
If the difference between a COR and the maximum is large enough, the corresponding candidate is kicked out from consideration unless its COR returns close to the maximum.

\subsubsection{Next Invocation Prediction and Provision}\label{sec:method:provision}
This section predicts invocations and pre-loads functions based on the predictive values or indicators.
``(Appro-)regular'' functions have discrete predictive values and ``dense'' functions use continuous ones.
In terms of ``possible'' functions, if the range of predictive values is larger than a threshold, the values are regarded as discrete; otherwise, we consider continuous integers inside the range of predictive values.
The predicted invocation times are naturally the last invoked time added by each of the predictive values.

For the online provision, if one of the predicted invocation times falls in $[t-\theta_{prewarm}, t+\theta_{prewarm}]$ with a pre-defined parameter $\theta_{prewarm}$ at the time of $t$, the function will be pre-loaded. 
If a loaded function's current WT is $\geq \theta_{givenup}$, the function will be evicted from the memory, where the parameter $\theta_{givenup}$ differs among function types.
Algorithm~\ref{algo:scheduler} shows our provision optimization, which is rule-based with good scalability and implements our defined strategies faithfully.
Every time the algorithm returns the updated MemSet, based on which we load function instances.

\begin{algorithm}[htbp]
    \footnotesize
    \caption{The provision algorithm of \newtool.} \label{algo:scheduler}
    \LinesNumbered
    \KwIn{
        FList: hash ids for all functions; 
    
        FState: recording the necessary information about all functions, such as the current WT, history WTs, predictive values, the time of the last invocation, etc; 
        
        Invo$^{(t)}$: invocation numbers of each function at the time of $t$; 
        
        InvoSet: set of functions that are invoked at current minute;
        
        ColdSet: set of functions that are invoked but not loaded;
        
        PrewarmSet: set of functions that need to be preloaded but have not yet been loaded;
        
        PendSet: set of functions that are loaded but not invoked currently;
        
        UnloadSet: set of functions that need to be unloaded;
        
        UCorr: the correlations of unseen functions.

        WPool: set of functions that are currently loaded in the worker pools.
    }
    \KwOut{The updated InvoSet, ColdSet, PrewarmSet, PendSet, UnloadSet, FState and UCorr}

    \SetKwFunction{FDeg}{Provision}
    \SetKwProg{Fn}{Function}{:}{End}
    \Fn{\FDeg{FList, FState, Invo$^{(t)}$, InvoSet, ColdSet, PrewarmSet, PendSet, UnloadSet, UCorr, WPool}}{
    \For{$f \in$ FList}{
    
        \eIf{Invo$^{(t)}$[$f$] $>$ 0}
        {
            InvoSet.add(f)\;
                FState[$f$].last\_invoked $\leftarrow t$\;
                FState[$f$].update\_WTs(FState[$f$].current\_WT)\;
                FState[$f$].current\_WT $\leftarrow 0$\; 
                
                \tcp{Adaptively adjusting predictive values.}
                FState[$f$].update\_predictive\_values($t$)\;
            \uIf {$f$ is not loaded}{
                FState[$f$].cold\_start\_record($t$)\;
                ColdSet.add($f$)\;
            }

        }
        { 
                FState[$f$].current\_WT += 1\;
                pre\_load\_flag $\leftarrow$ FState[$f$].pre\_load(FState[$f$].$\theta_{prewarm}$)\;
                \uIf{pre\_load\_flag
                is \textbf{False} \textbf{AND} FState[$f$].current\_WT $\geq$ FState[$f$].$\theta_{givenup}$}{ 
                    UnloadSet.add($f$)\;
                    }
                \uElseIf {pre\_load\_flag is \textbf{True}}{
                    PrewarmSet.add($f$)\;
                }
        }
        \tcp{Adaptively processing unseen functions.}
        UCorr.update()\;
    }
    
    Unload all the functions in UnloadSet\;
    
    PendSet $\leftarrow$ WPool.difference(InvoSet)\;

    PrewarmSet $\leftarrow$ PrewarmSet.difference(InvoSet).difference(PendSet)\;

    Sort PrewarmSet by memory in decreasing order\;
    
    Sort InvoSet by memory in decreasing order\;

    \Return{InvoSet, ColdSet, PrewarmSet, PendSet, UnloadSet, FState, UCorr}

}

\end{algorithm}

\subsection{Workers Load Balancing}

We apply a greedy algorithm to select the server to load in order to optimize load balancing.
Upon a new function to load, our policy selects the server with the most idle memory. If the remaining space is enough to accommodate the function, this function can be easily loaded.
However, if the remaining space is not enough (smaller than the required memory size of the target function), we will evict the functions in this server according to the policy introduced in section \ref{sec:method:evict}.
Under an extreme circumstance where all the functions of this server that are predicted not to be invoked in the next moment are evicted, this function still cannot be accommodated, this server will be given up, and we will choose the next server--which has the most remaining space among other servers. The process will be repeated until reaching the last server.
Note that it is impossible for a function to be loaded according to our primary assumption built on practical expertise.

\subsection{Eviction Decision} \label{sec:method:evict}
This section considers the extended memory-limited scenario and answers the second question at the beginning of this chapter -- which functions should be evicted when workers are out of memory.

\subsubsection{Functions to be Loaded and Evicted}

At the start of each minute, two sets of function candidates are considered for loading.
The first set, denoted by the \textit{Cold Set}, includes functions requested by downstream services that have not yet been loaded and must be immediately available for invocation.
The second set, referred to as the \textit{Prewarming Set}, consists of functions predicted by \tool to be beneficial for prewarming, even though they will not be invoked within the current minute.
The loading process begins by prioritizing the \textit{Cold Set} to ensure that there is enough space for all necessary functions. 
After that, the remaining space is prepared to arrange the functions in the \textit{Prewarming Set} for potential future invocations.
Loaded functions without invocations at this minute are grouped in the \textit{Pending Set}. Only the functions in this set are considered to be evicted at the beginning.

We assume that the space allows all invoked functions to be loaded within one minute. In practice, when the theoretical capacity for function invocations is exceeded—such as during periods of high user demand—serverless providers typically respond by dynamically scaling their infrastructure. This scalability enables providers to increase supply and, in turn, generate more profit, thereby maintaining a balance between supply (memory and infrastructure) and demand (user requests), resulting in more available resources for concurrent function executions.

\subsubsection{Eviction Policy}

A fundamental principle is that when there is insufficient memory to load all the functions in the \textit{Cold Set} or the \textit{Prewarming Set}, it is required to evict the functions in the \textit{Pending Set}. 
Consequently, we assign priorities to the functions in the \textit{Pending Set}. The priorities are formulated on the basis of the previously calculated prediction values (which are directly related to the timing of the next invocation) and the memory occupied by the functions.
Apparently, those functions with larger prediction values (which means they will be invoked later) and larger memory occupation (which means they may waste more resources) are supposed to have a higher priority in being evicted.

Specifically, we first evict functions predicted to have a long idle time. In each minute, the prediction value of each function is updated to predict the next invocation time.
According to the previous function categorization, ``warm'', ``successive'', and ``pulsed'' functions have a high probability to be invoked at ${(t_i+1)}^{th}$ min after invocations at $t_i^{th}$ min.
Prediction values of functions categorized in other types may fall into an interval, and the lower bound of this interval is taken as the prediction value when calculating the priority.
For functions with an equal prediction value, those with larger memory will be evicted first.

Once there is not enough space and we must trade-off functions in the \textit{Pending Set} and the \textit{Prewarming Set}, we first keep the functions in the \textit{Pending Set} that are predicted to be invoked in the next minute.
Afterward, we union these two sets and sort them by their confidence scores. The scores are defined by their function types and standard deviations of prediction values.

\begin{figure}[htb]
    \centering
    \vspace{-0.12in}
        {\includegraphics[width=0.3\linewidth]{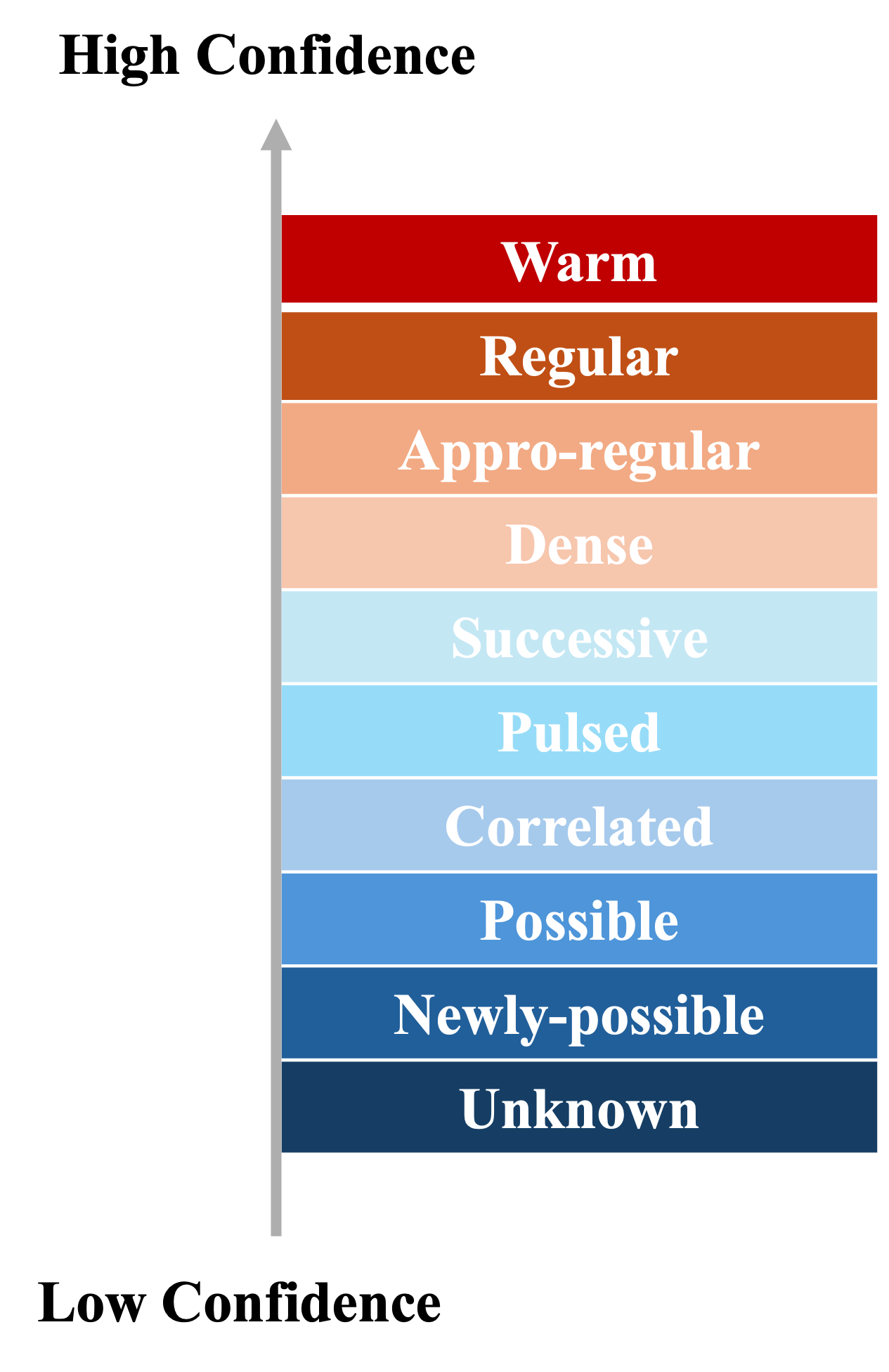}}
    \caption{Different function types are ranked by confidence. This presents that functions categorized as Warm have the highest confidence, and functions categorized as Unknown have the lowest confidence.}
    \vspace{-0.1in}
    \label{fig:confidence}
\end{figure}

The confidence scores of the functions are defined as shown in Figure~\ref{fig:confidence}, following the order of the function categorization.
As previously discussed in Section~\ref{sec:method}, the reliability of predicting the next invocation decreases as we move from the first category to the last, with high-frequency calls being the most predictable, and finally ``unknown'' calls, which exhibit the least predictable behavior.
Thus, functions with a higher degree of predictability should be given a higher confidence score in their likelihood of future invocation, thus making them less likely to be evicted. This approach ensures that more reliably invoked functions remain warm, thereby reducing cold start latency and optimizing system performance in a resource-constrained environment.

In the eviction process, we first evict the function types with lower confidence. For the functions that fall into the same type, we sort them by standard deviations and evict the ones with larger deviations. This is reasonable since larger deviations mean they fluctuate more dramatically which may lead to inaccurate predictions with higher probabilities.

\begin{figure}[htb]
    \centering
    \vspace{-0.12in}
        {\includegraphics[width=0.8\linewidth]{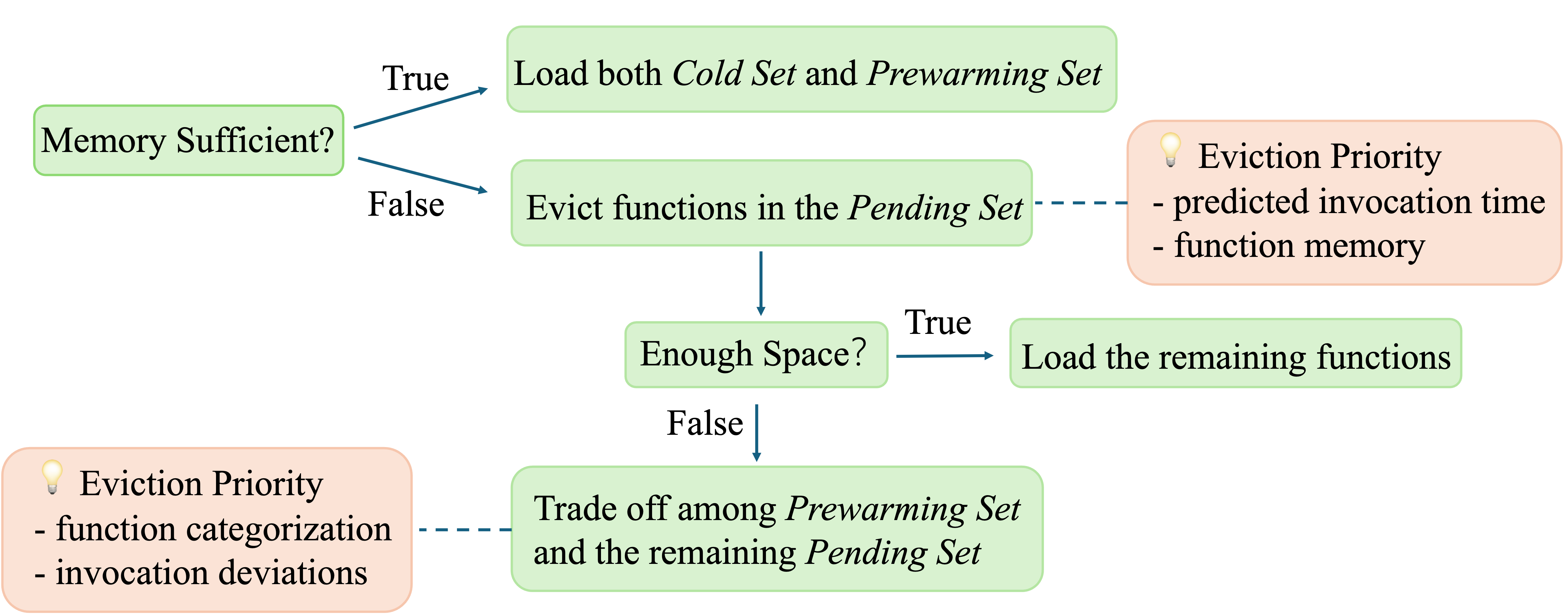}}
    \caption{The workflow of the eviction process. When there's not sufficient space for the demanded functions, we need to trade-off among functions in different sets. Here we apply our rule-based and priority-aware eviction policy.}
    \vspace{-0.1in}
    \label{fig:evict_flow}
\end{figure}

Figure~\ref{fig:evict_flow} depicts the workflow of our priority-aware eviction policy.
Algorithm~\ref{algo:scheduler2} shows our eviction and load-balancing algorithm for \newtool. 

\begin{algorithm}[htbp] 
    \footnotesize
    \caption{The eviction and load-balancing algorithm of \newtool.} \label{algo:scheduler2}
    \LinesNumbered
    \SetKwFunction{FDeg}{Eviction and Load-balancing}
    \SetKwProg{Fn}{Function}{:}{End}
    \Fn{\FDeg{FList, FState, InvoSet, ColdSet, PrewarmSet, PendSet, UnloadSet, WorkerList}}{
    ColdSet $\leftarrow$ sorted (ColdSet, key = lambda $f$: FState[$f$].memory, reverse = True);
    
    PrewarmSet $\leftarrow$ sorted (PrewarmSet, key = lambda $f$: FState[$f$].memory, reverse = True);
    
    cold\_flag $\leftarrow$ \textbf{True};
    
    prewarm\_flag $\leftarrow$ \textbf{True};
    
    \For{$f$ in ColdSet}
    {
        WorkerList $\leftarrow$ sorted (WorkerList, key = lambda $w$: $w$.space, reverse = True);
        
            \If{ CanLoad(WorkerList[0], $f$) }{
                Load $f$ into worker;
            }
            \Else{
                cold\_flag $\leftarrow$ \textbf{False};
                
                RMColdSet  $\leftarrow$ Remaining ColdSet;

                break;
            }
    }

    \uIf{cold\_flag is \textbf{True}}{
        \For{$f$ in PrewarmSet}
        {
            WorkerList $\leftarrow$ sorted (WorkerList, key = lambda $w$: $w$.space, reverse = True);
            
                \If{ CanLoad(WorkerList[0], $f$) }{
                    Load $f$ into worker;
                }
                \Else{
                    prewarm\_flag $\leftarrow$ \textbf{False};

                    Unload previous PreWarmSet functions;

                    break;
                }
        }
    }

    

    \If{cold\_flag is \textbf{False} \textbf{OR} prewarm\_flag is \textbf{False}}
    {    
        \For{$func$ $\in$ RMColdSet }{
            \For{each worker by memory decreasing order}
            {
                \uIf{$func.memory$ $\leq$  remaining space of the worker \textbf{OR} CanEvictForCold (PendSet, $func$) }
                    {   Load $func$ into worker;
                    
                        break;
                    }
            }
        

        }
        
        \For{$func$ $\in$ PrewarmSet}{
            \For{each worker by memory decreasing order}
            {
                \uIf{$func.memory$ $\leq$ remaining space of the worker}
                    {   Load $func$ into worker;
                    
                        break;
                    }
                \uElse{
                    \uIf{$get\_min\_t(PendSet)$ $>$ $t_i+1$ \textbf{AND} $func.memory$ $>$ remaining space of the worker}
                    {
                        \uIf{CanEvictForPrewarm (PendSet, $func$) }
                        {   Load $func$ into worker;
                        
                            break;
                        }
                    }
                    
                }
            }
            \uIf{for all workers $get\_min\_t(PendSet)$ $\leq$ $t_i+1$ \textbf{AND} there's not enough space for $func$ }
                {
                    Unload previous PreWarmSet functions;

                    Tradeoff (PendSet, PrewarmSet);
                }

        }

    }

                
                
                
                

    }
\end{algorithm}

\section{Evaluation}
\noindent\textbf{Workload.}
We evaluate \tool and \newtool on the most widely used industrial dataset~\cite{AzureDataset} released by Microsoft Azure Function~\cite{ServerlessWild} with real-world invocation traces. 
The dataset contains the invocation counts per minute for 14 days. The first 12 days are used for pattern modeling (training), and the last two days are used for the simulation, conducted in a container with 32-core Intel Xeon(R) Platinum 8360H CPU and 256 GB memory.
The records involve 15,097 users, 24,964 applications, and 83,137 functions. 71,616 functions appear in the training data, 39,388 functions appear in the simulation data, and 743 never appeared during training.

\noindent\textbf{Assumptions.} 
We adopt the simulation principles following~\cite{ServerlessWild}.
First, assume all executions finish within one minute since 1) most (96\%) functions have very short execution time (less than 60 seconds), and 2) we can thereby calculate the worst-case wasted resource time.
Second, we assume that cold-start latency for each function is the same, so we only need to care about the number of cold starts and the wasted memory time.

Since the dataset only contains the memory of the app, to evaluate under a memory-constrained scenario, we evenly split the memory allocation among all functions in an application following~\cite{FaasCache}. For the functions that only appear in the simulation, we assign them the average memory of all applications, split among all functions.

\subsection{Evaluating \tool Without Memory Limitation Under Single-Worker Scenario}

We evaluate \tool by answering four research questions:
\begin{itemize}[leftmargin=12pt, topsep=0pt]
\item \textbf{RQ1}: How effectively does \tool decrease cold starts?
\item \textbf{RQ2}: How much memory waste and computation overhead does \tool incur?
\item \textbf{RQ3}: How does \tool trade off memory waste with latency reduction?
\item \textbf{RQ4}: How do complementary designs influence \tool?
\end{itemize}

\subsubsection{Baselines}\label{sec:exp:baselines}
We compare \tool with five state-of-the-art baselines applied to the application layer. 
Approaches involving system renovations are out of this paper's scope and will be discussed in Section~\ref{sec:review}.
Our baselines include FaaSCache~\cite{FaasCache}, Defuse~\cite{Defuse}, Hybrid~\cite{ServerlessWild}, and a fixed keep-alive policy.
We reproduce these methods as they do not provide open-source code by strictly following the original papers and referring to a reproduction attempt~\cite{Reproduce}.
Note that the original Hybrid method works at the application unit (Hybrid-Application, HA), so we derive a Hybrid-Function method by employing its design on the function granularity (Hybrid-Function, HF), following~\cite{Defuse}.
All the parameters are set according to their original papers. The fixed keep-alive policy adopts a length of 10 minutes. FaaSCache requires a pre-defined memory limit, so we adopt the maximum memory size of \tool during the whole simulation.

\subsubsection{Metrics and parameters}\label{sec:exp:metrics}
To quantify the cold-start optimization, we measure the function-wise (application-wise for HA) \textit{cold-start rate (CSR)}, \textit{i.e.} the number of cold starts divided by the number of invocations. 
We apply wasted memory time (\textit{WMT}, $\S$\ref{sec:background:challenge}) to gauge the idle resource waste.
Naturally, the lower the CSR or WMT, the fewer cold starts or wasted resources, the better.

We also monitor the \textit{effective memory consumption ratio (EMCR)}, which measures the fraction of invoked function instances relative to the total loaded instances on each host machine, serving to assess resource efficiency. A higher EMCR signifies wiser memory allocation, as it indicates a greater proportion of memory is used by active instances rather than remaining idle.
As for the parameters, we set $\theta_{prewarm}$ as two. 
The $\theta_{givenup}$ for ``dense'' and ``pulsed'' is five, whereas one for the other types. 
Section~\ref{sec:exp:trade-off} further discusses the impact of these pre-defined parameters.

\subsubsection{RQ1: Effectiveness in Cold-Start Reduction}\label{sec:exp:cold-start}

\begin{figure}[htb]
    \centering
    \vspace{-0.13in}
        {\includegraphics[width=0.8\linewidth]{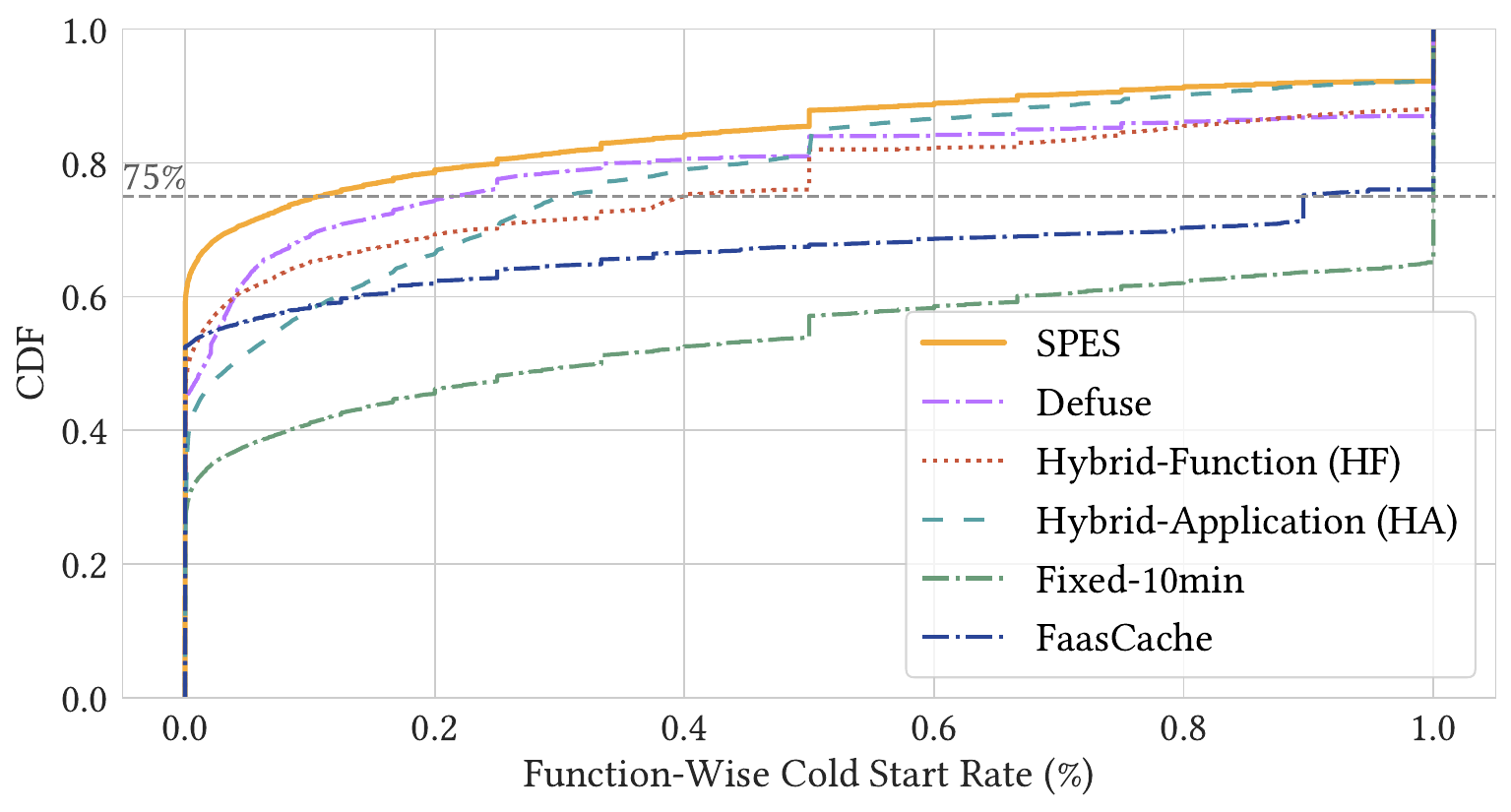}}
    \caption{Cold start behaviors of \tool and its competitors.}
    \vspace{-0.1in}
    \label{fig:CDF_cold_start}
\end{figure}

Figrue~\ref{fig:CDF_cold_start} displays the cumulative distribution (CDF) of CSR under the provision decisions of \tool and baselines. 
With a fixed y-axis value, the line representing \tool is positioned to the left of other baseline lines, which indicates that \tool consistently leads to fewer cold starts across a variety of functions, each corresponding to different invocation frequencies.
In particular, \tool reduces the 75th percentile cold-start rate (Q3-CSR, for simplicity) from 0.215 to 0.108 compared to Defuse, the best-performing baseline, achieving 49.77\% improvement, and reduces 75-CSR by 64.06\%--89.20\% compared to other baselines. 
We care more about Q3-CSR because infrequently invoked functions benefit most from optimization~\cite{ServerlessWild}.
Moreover, 57.99\% functions experience no cold starts with \tool, indicating that \tool can allow most functions to be warmly invoked. In contrast, 25.61\%--52.59\% functions completely experience warm startup using baselines where FaaSCache is the best one.
On the other hand, regarding infrequently invoked functions, \tool reduces the 90th percentile CSR by 19.87\% compared to the best-performing baseline, HA. 
Such improvements imply that \tool exhibits significant optimization performance for both frequently and infrequently invoked functions, while no baseline can achieve second-best performance with different percentile CSRs.

\begin{figure}[htbp]
\centering
\subfigure[Normalized memory usage]{
\begin{minipage}{0.4\textwidth}
\centering 
\includegraphics[width=0.95\linewidth]{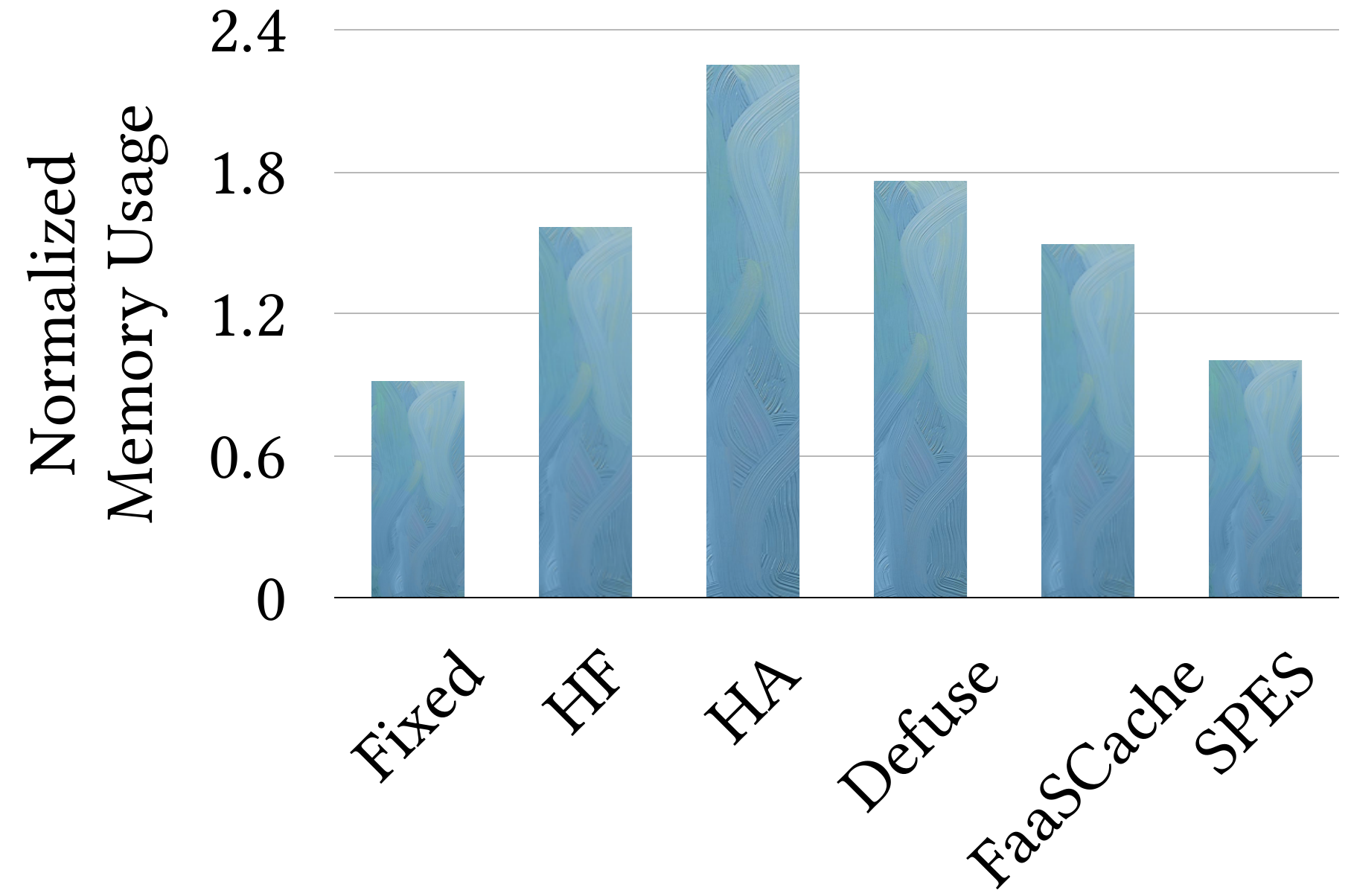}
\label{fig:normalized_mem_usage}
\end{minipage}
}
\subfigure[Percentage of always-cold functions]{
\begin{minipage}{0.4\textwidth}
\centering 
\includegraphics[width=0.95\linewidth]{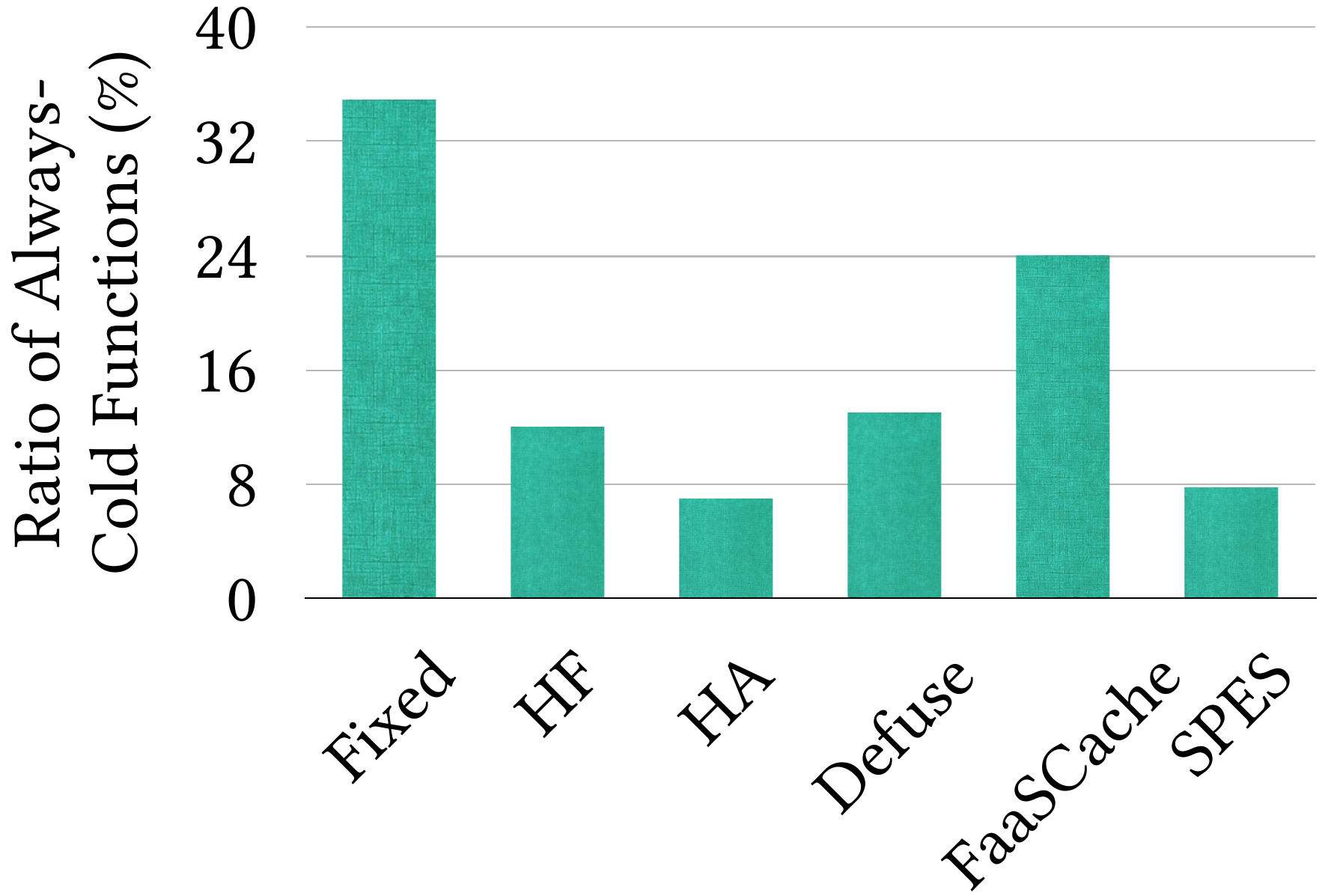}
\label{fig:total_cold_perc}
\end{minipage}
}
\caption{\tool is memory-efficient compared to most baselines with relatively few always-cold functions.}
\vspace{-0.1in}
\label{fig:high_csr_normalized_memory}
\end{figure}

Furthermore, \tool significantly reduces ``always-cold'' functions, which always experience cold starts upon invocations (with CSR equal to 1.0).
Figure~\ref{fig:total_cold_perc} further presents the always-cold percentage, and that of \tool is just less than 8\%.
Among the baselines, HA has the fewest always-cold functions, which is closest to that of \tool, whereas Defuse and HF, both function-grained methods, increase always-cold functions dramatically.
This can be attributed to infrequently invoked functions.
About 3.82\% functions are invoked less than twice during training, and 6.14\% are only once during the simulation, so always-cold functions seem inevitable with limited records.
HA mitigates this issue by grouping and loading functions together, yet resulting in more memory consumption and waste.
\tool instead connects unseen and unpredictable functions with known functions using trigger, application, and user information and gradually updates the associated functions based on actual invocations. 
Hence, without much extra memory, \tool performs close to HA.

In addition to cold start reduction, Figure~\ref{fig:normalized_mem_usage} presents the memory usage normalized by the averaged one of \tool.
\tool's memory usage is only 8.08\% more than the most resource-efficient method, the fixed keep-alive policy, and saves 36.07\%--55.55\% memory compared to the other baselines on average. 
It only consumes about half of the memory of Defuse, the best baseline for cold start reduction.

\begin{figure}[htb]
    \centering
    \vspace{-0.1in}
        {\includegraphics[width=0.8\linewidth]{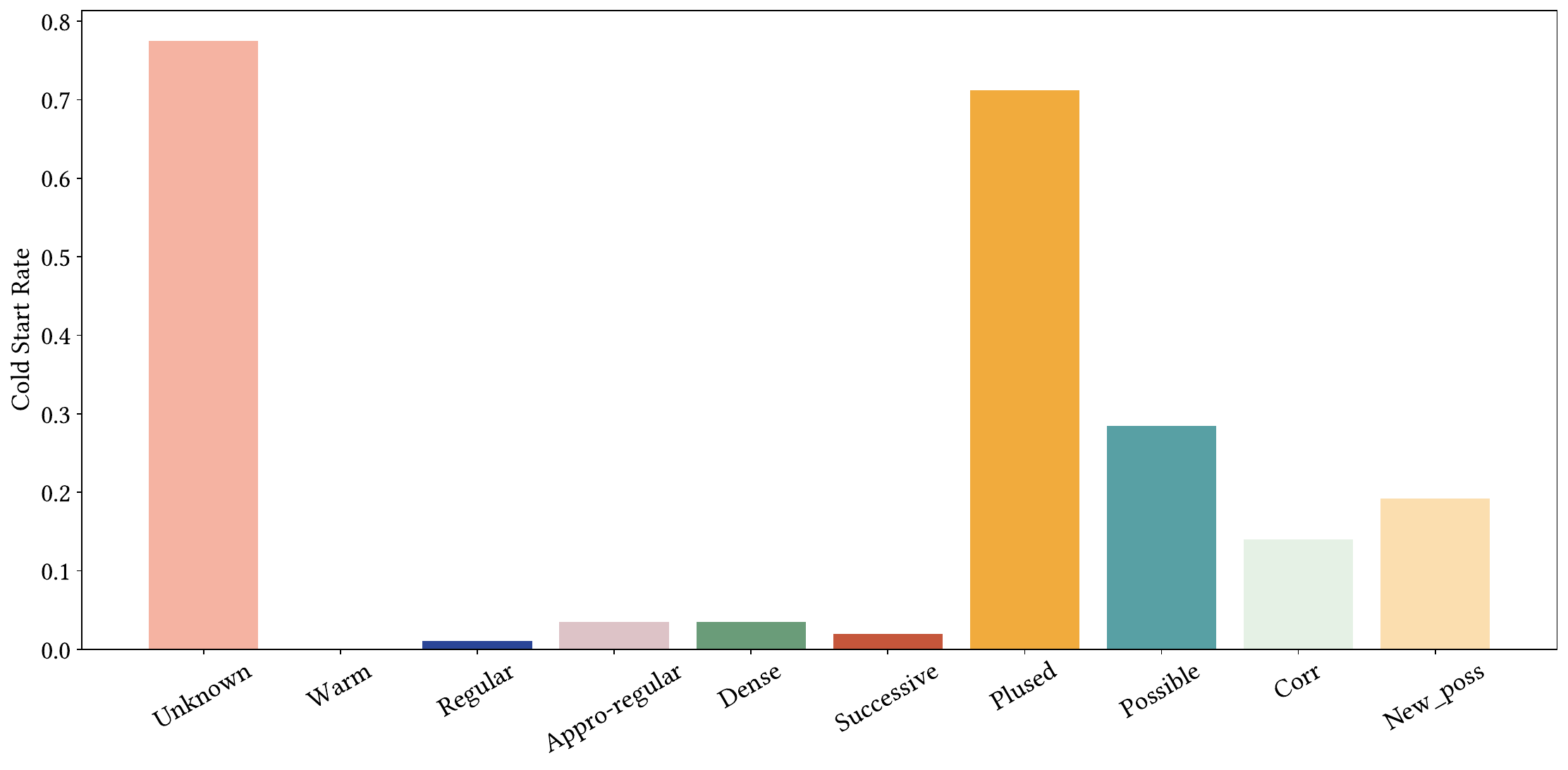}}
        \vspace{-0.1in}
    \caption{Averaged cold start rate of each type.}
    \vspace{-0.1in}
    \label{fig:cold_start_per_type}
\end{figure}

Figure~\ref{fig:cold_start_per_type} shows the averaged CSR among different types. As the simulation period is not long enough, we only categorize unknown functions into the type of ``possible'', denoted by ``newly-possible''. 
We can see that ``unknown'' functions contribute the most to cold starts, and ``pulsed'' functions also incur high CSRs. This is attributed to insufficient historical invocation records.
Actually, \tool intentionally connives a cold start when an ``unknown'' or ``pulsed'' invocation arrives after a long idle time.
Though we can leverage less predictive information, such as the averaged WT, to predict the next invocation or even keep such functions always warm, this can lead to considerable unbillable and wasted memory, undesired for FaaS providers.
This outcome is deliberating on the trade-off between performance and resource allocation.
Such an issue can be mitigated with more invocation histories revealing predictive indicators for better provision.

Meanwhile, though ``correlated'' and ``possible/newly-possible'' functions also lack predictive information with infrequent invocations, their invocations are more likely to be predicted. Thus, we connect some of them with deterministic functions or adaptively extract meaningful behavior indicators from incoming invocations. 
In this way, their cold starts are suppressed effectively. We will further discuss the usefulness of such strategy designs in Section~\ref{sec:exp:ablation}.

\subsubsection{RQ2: Wasted Memory Time and Overhead}\label{sec:exp:wasted-memory-time}

\begin{figure}[htbp]
\centering
\subfigure[Normalized wasted memory time (WMT, lower is better)]{
\begin{minipage}{0.4\textwidth}
\centering 
\includegraphics[width=0.95\linewidth]{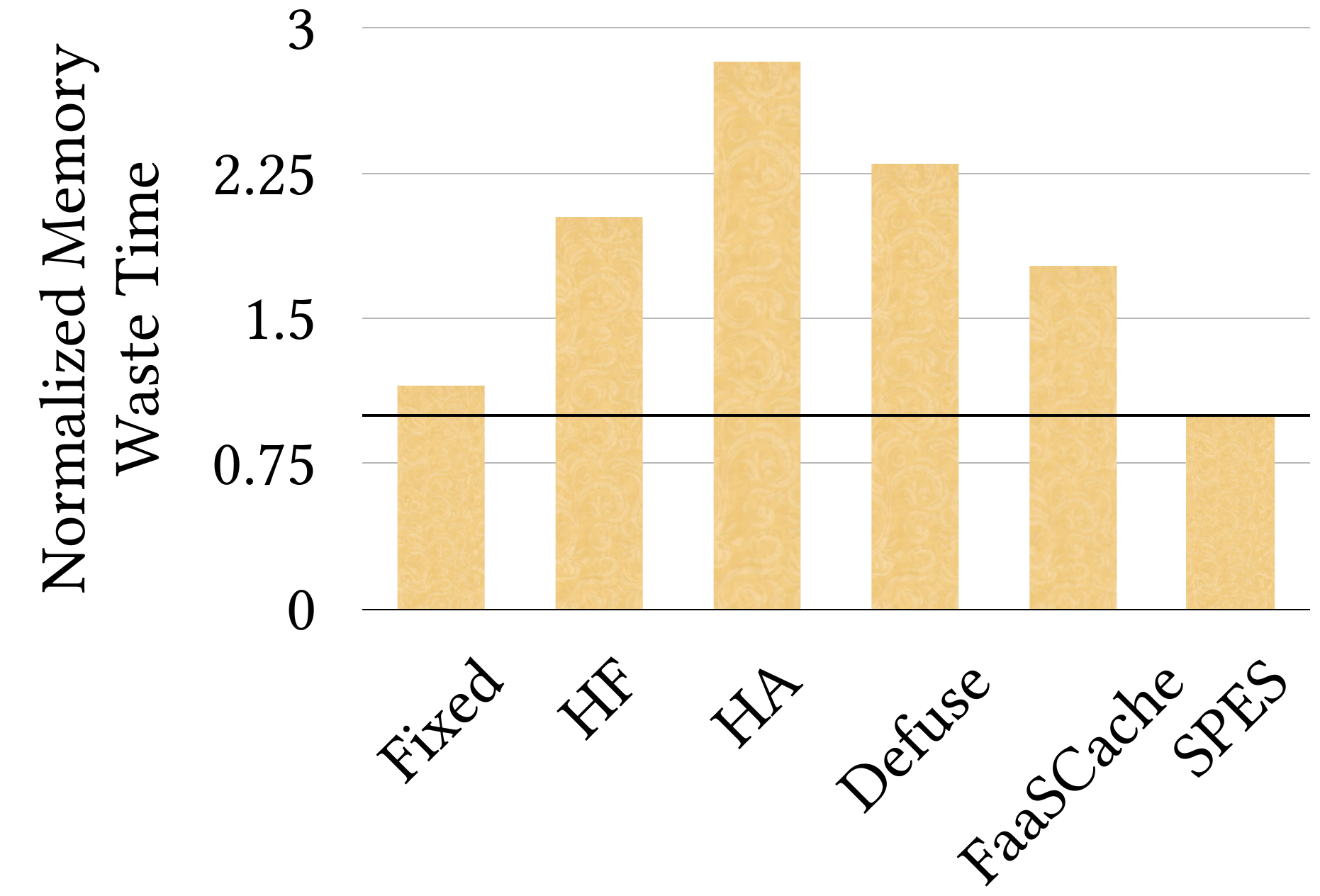}
\label{fig:waste_memory_time}
\end{minipage}
}
\subfigure[Effective memory consumption ratio (EMCR, higher is better)]{
\begin{minipage}{0.4\textwidth}
\centering 
\includegraphics[width=0.95\linewidth]{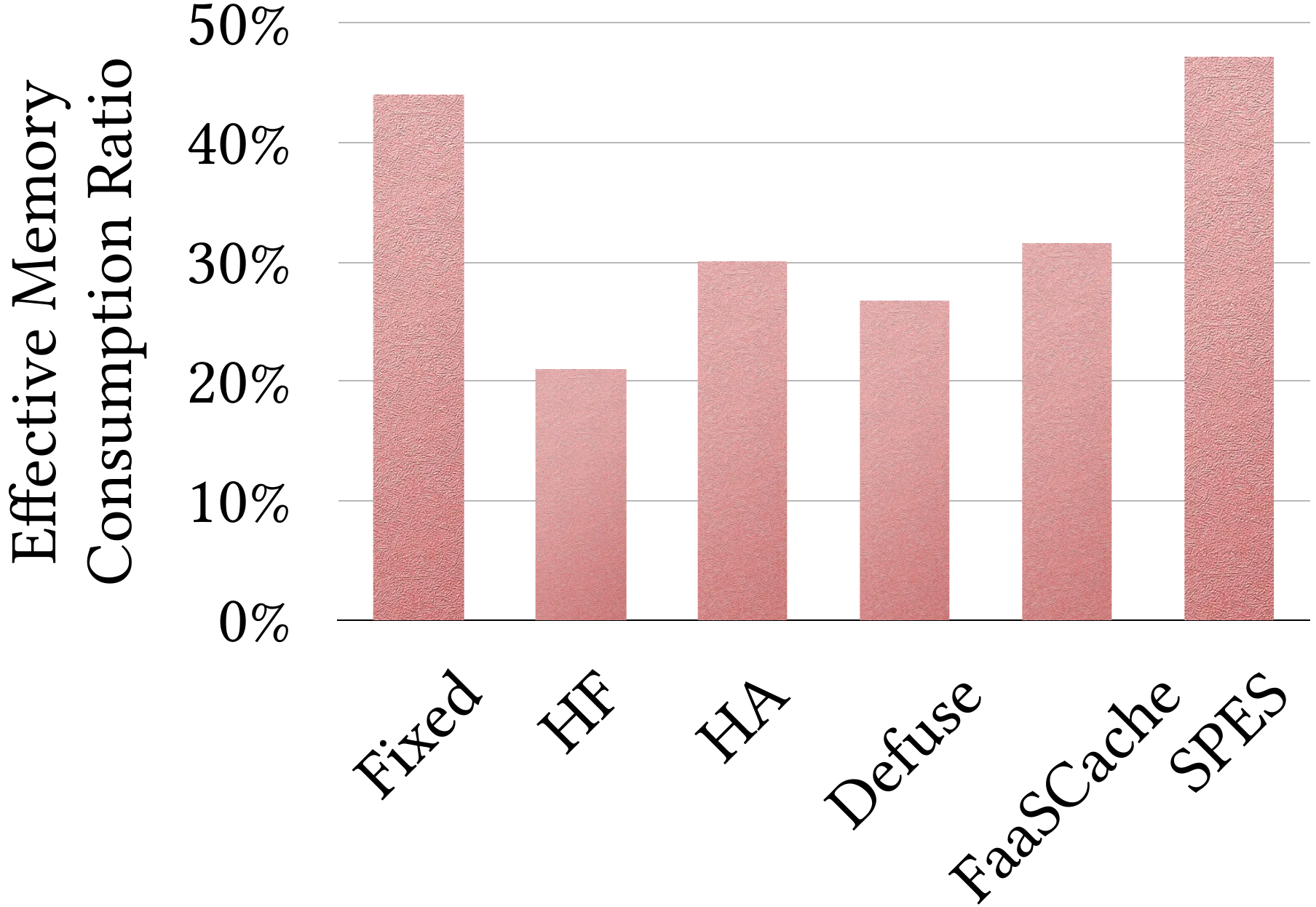}
\label{fig:memory_usage_ratio}
\end{minipage}
}
\caption{\tool significantly reduces memory waste and achieves more effective memory consumption.}
\vspace{-0.1in}
\label{fig:wasted_memory_and_usage}
\end{figure}

\textbf{Waste Memory Time.}
Figure~\ref{fig:waste_memory_time} presents that \tool significantly decreases the wasted memory time by 10.89\%--63.50\% compared to all baselines. Particularly, compared to Defuse, the most effective baseline on cold-start reduction, we reduce 57.06\% of the WMT.
Figure~\ref{fig:memory_usage_ratio} also demonstrates that \tool efficiently uses memory resources, whose EMCR is 46.32\%, 5.20\%--120.89\% higher than compared approaches.
The success of \tool can be attributed to careful pattern modeling and differentiated strategies.

\begin{figure}[htb] 
    \centering
    \vspace{-0.1in}
        {\includegraphics[width=0.8\linewidth]{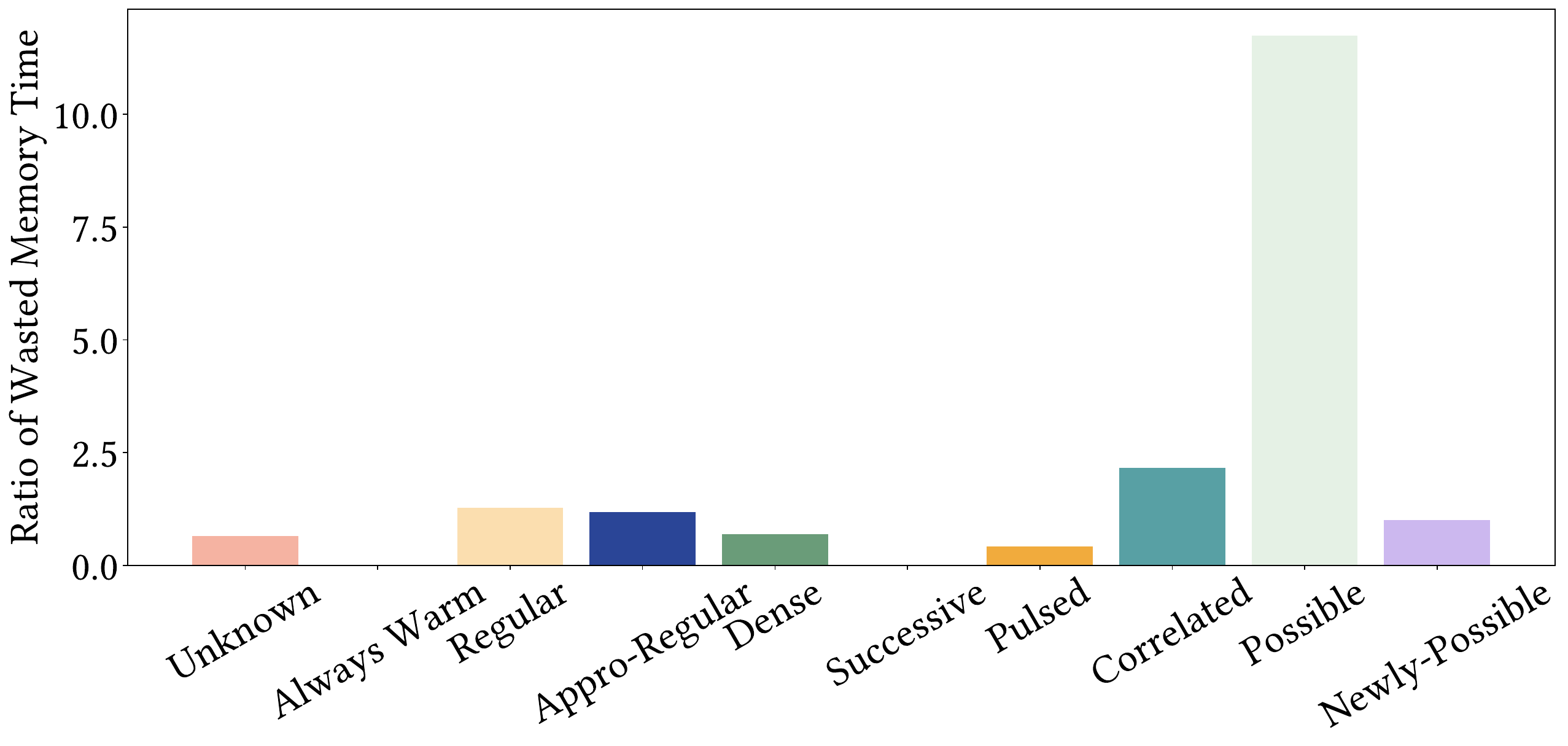}}
        \vspace{-0.1in}
    \caption{The ratio of WMT of each function type.}
    \vspace{-0.1in}
    \label{fig:ratio_WMT_per_type}
\end{figure}

To further demonstrate how each type contributes to WMT, we derive a metric called the \textit{ratio of WMT}, which is the WMT divided by the number of invoked times for each serverless function.
If a type contains more functions or function invocations, the bespoke scheduling strategy tends to make more proactive provisions, likely resulting in more WMT.
Thus, the ratio of WMT can better portray the accuracy of invocation prediction and the usefulness of pre-loading.
Figure~\ref{fig:ratio_WMT_per_type} shows the distribution of the ratio of WMT among different function types, from which we can see the ``possible'' functions have the highest probability of generating WMT. 
The predominant reason for this is the infrequent invocation of possible functions, which leads to a scarcity of patterns in the historical database and complicates establishing associations with other functions. Despite these challenges, we persist in our efforts to forecast and pre-warm the invocation of these functions, aiming to reduce cold starts as much as possible. 
Different from ``pulsed'' or ``unknown'' functions that are allowed to generate cold starts, we encourage aggressive prediction attempts for ``possible'' functions since the latter have at least a duplicated WT, enabling potential predictive value obtaining.
However, this strategy unavoidably results in augmented resource wastage.

\textbf{\tool Overhead.}
Herein, we measure the additional latency induced by our implementation. The time complexity within each time window of our approach and the fixed keep-alive policy is $\mathcal{O}(n)$, where $n$ represents the number of functions, as it only requires accessing the corresponding function's category and its predictive value(s). For each invocation, the time complexity of pre-load/unload operations is merely $\mathcal{O}(1)$. Within each time window, the time complexity for the fixed keep-alive method is $\mathcal{O}(m)$, where $m$ is the number of loaded functions. In contrast, FaaSCache has a time complexity of $\mathcal{O}(\log m)$, which involves identifying the container with the lowest priority. Other methods, such as HA, HF, and Defuse, have higher complexities, mainly due to the computational bottleneck in updating histograms. 
The simulation results align with these expectations. 
Compared to the fastest fixed keep-alive baseline, which has an average overhead of 0.024 sec per minute, our method adds 0.44 sec's overhead per minute, mainly due to our more complex strategies for each provision action.
Contrasted with the second fastest, FaaSCache, our overhead is reduced by 6.8\%. 
In summary, our overhead is inconsequential compared to the typical latency found in most existing serverless platforms.

\subsubsection{RQ3: Trading-off resources and latency}\label{sec:exp:trade-off}

We control the trade-off through two parameters: $\theta_{prewarm}$ and $\theta_{givenup}$. 
As introduced in Section~\ref{sec:method:provision}, the former decides how long to pre-load a function with a predicted nearby invocation, and the latter decides how long to keep an idle function warm.
Intuitively, the larger these two parameters, the more likely a function is to pre-load or keep loaded with more memory usage and potential wasted memory, and the less cold starts.
In RQ1 ($\S$\ref{sec:exp:cold-start}), we set $\theta_{prewarm}$ as two, and the $\theta_{givenup}$ for ``dense'' and ``plused'' is five, whereas one for the other types. Our original simulation setting is denoted by the red star (\textcolor{luoshen}{$\star$}) in the following figures.
 
Figure~\ref{fig:prewarm} shows the trade-off under different $\theta_{prewarm}$, where a point $(x, y)$ represents using $x$-unit memory and obtaining the 75-CSR of $y$, under a certain $\theta_{prewarm}$.
The memory is normalized to that under the original setting.
Normalized memory usage and 75-CSR are nearly linearly correlated.
We can conveniently choose a proper setting by controlling $\theta_{prewarm}$. 
Since the red star is below and the most distant from the fitting line, $\theta_{prewarm}=2$ is the optimal value.

As $\theta_{givenup}$ should be integers and different from each type, we simply multiply the original $\theta_{givenup}$ setting by 2, 3, 4, and 5, respectively.
The results are shown in Figure~\ref{fig:givenup}, where the linear relationship remains approximately the same, but larger $\theta_{givenup}$s have less impact on the reduction of cold start. 
This indicates that keeping invoked functions too long is sub-optimal and idle functions should be evicted promptly.

\begin{figure}[htbp]
\centering
\subfigure[Parameter $\theta_{prewarm}$]{
\begin{minipage}{0.4\textwidth}
\centering 
\includegraphics[width=0.98\linewidth]{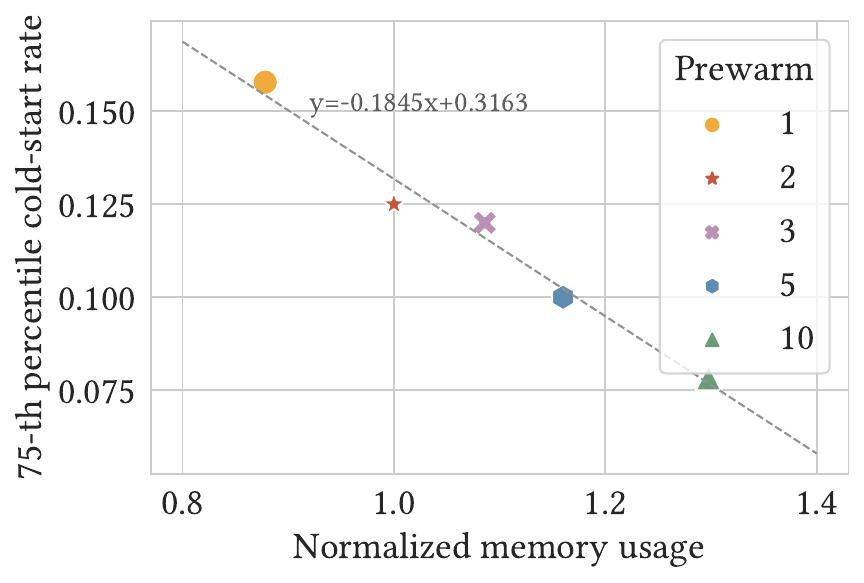}
\label{fig:prewarm}
\end{minipage}
}
\subfigure[Parameter $\theta_{given}$]{
\begin{minipage}{0.4\textwidth}
\centering 
\includegraphics[width=0.98\linewidth]{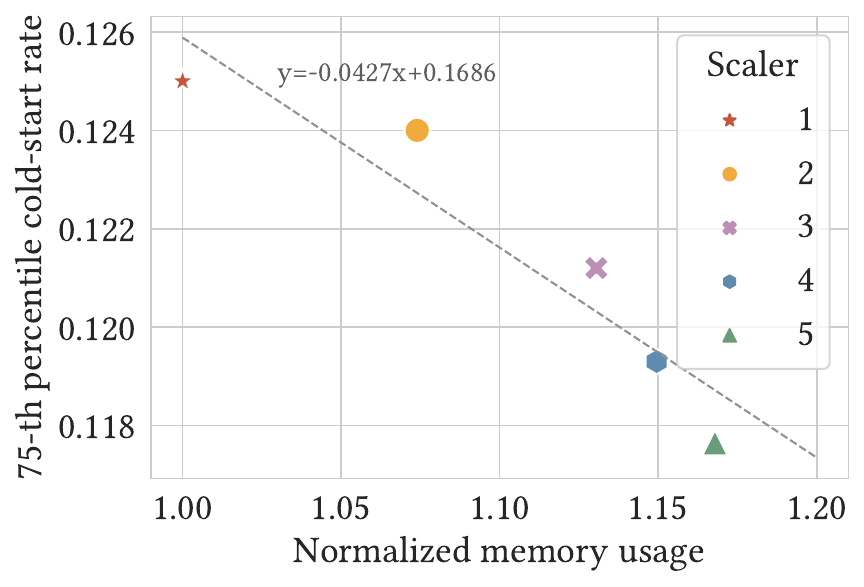}
\label{fig:givenup}
\end{minipage}
}
\vspace{-0.1in}
\caption{Under different $\theta_{prewarm}$, the memory usage and Q3-CSR exhibit an approximately linear relationship. Similarly, under different $\theta_{given}$, the linear relationship still approximately holds, yet dramatically increasing the memory does not guarantee cold start mitigation.}
\vspace{-0.1in}
\label{fig:prewarm-givenup}
\end{figure}

\subsubsection{RQ4: Impact of Strategy Designs}\label{sec:exp:ablation}

\begin{figure}[htb]
    \centering
        \vspace{-0.08in}
        {\includegraphics[width=0.6\linewidth]{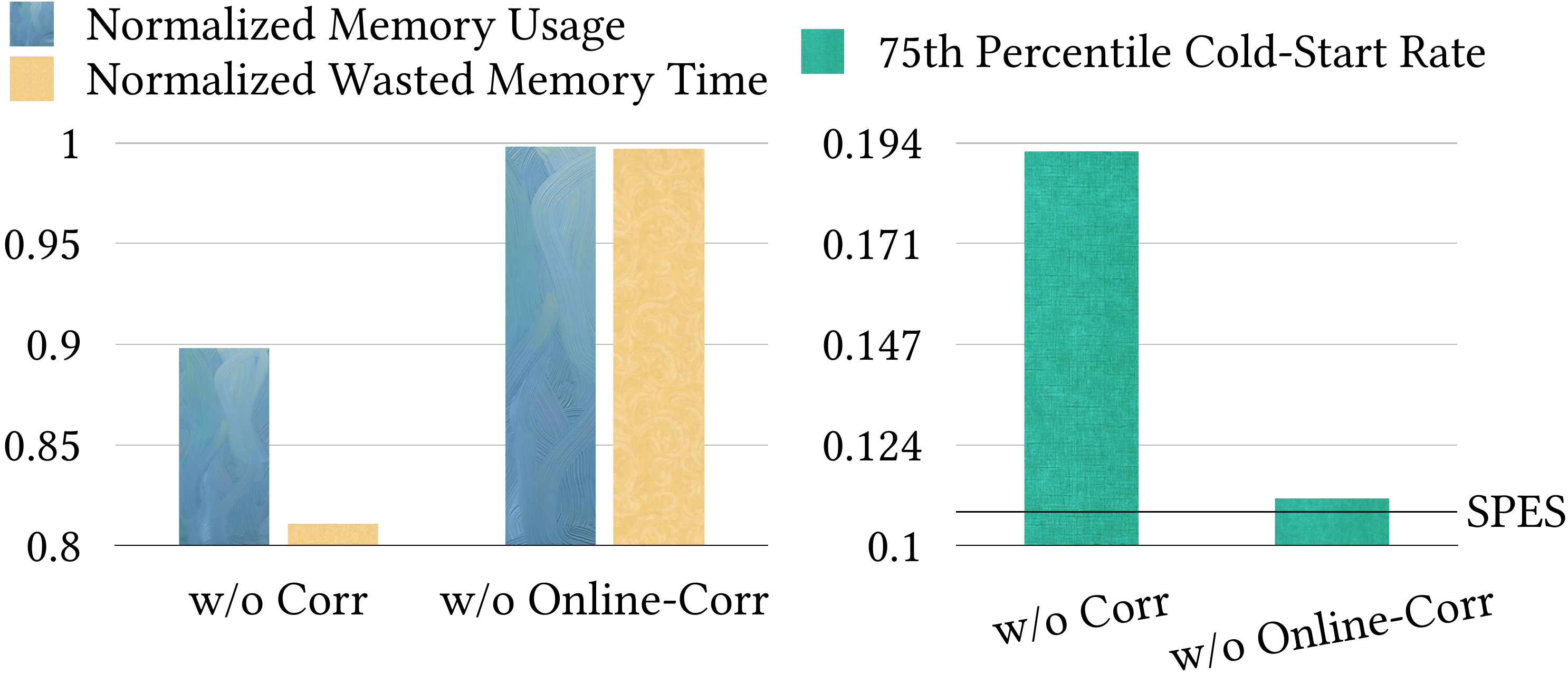}}
    \caption{The correlation strategy contributes to the cold-start and memory-waste reduction.}
    \vspace{-0.12in}
    \label{fig:impact_corr}
\end{figure}

\textbf{Impact of inter-function correlations.}\label{sec:exp:ablation:corr} 
This section gauges our design in processing ill-informed functions by developing inter-function correlations.
As introduced in $\S$\ref{sec:method:difficult}, we propose a simple yet effective $T$\textit{-lagged co-occurrence rate} metric to connect ill-informed functions with categorized ones. Those with closely related known functions are ``correlated''.
This strategy is applied during both training and simulation.
Figure~\ref{fig:impact_corr} presents the impact of this strategy.
\textit{w/o Corr} means re-categorizing ``correlated'' functions into ``pulsed'', ``possible'' or ``unknown'' during training, but we still deal with unseen functions during the simulation.
\textit{w/o Online-Corr} denotes removing the simulation-applied strategy, \textit{i.e.} regarding unseen functions as ``unknown'' but still retaining the ``correlated'' functions obtained during training.
It is shown that the latter strategy slightly reduces the Q3-CSR, whereas the former makes a significant contribution.
We attribute the results to the influence of the function number. 
4.71\% of the functions belong to ``correlated'' whereas only 1.89\% are unseen.

\begin{figure}[htb]

    \centering
    \vspace{-0.1in}
        {\includegraphics[width=0.6\linewidth]{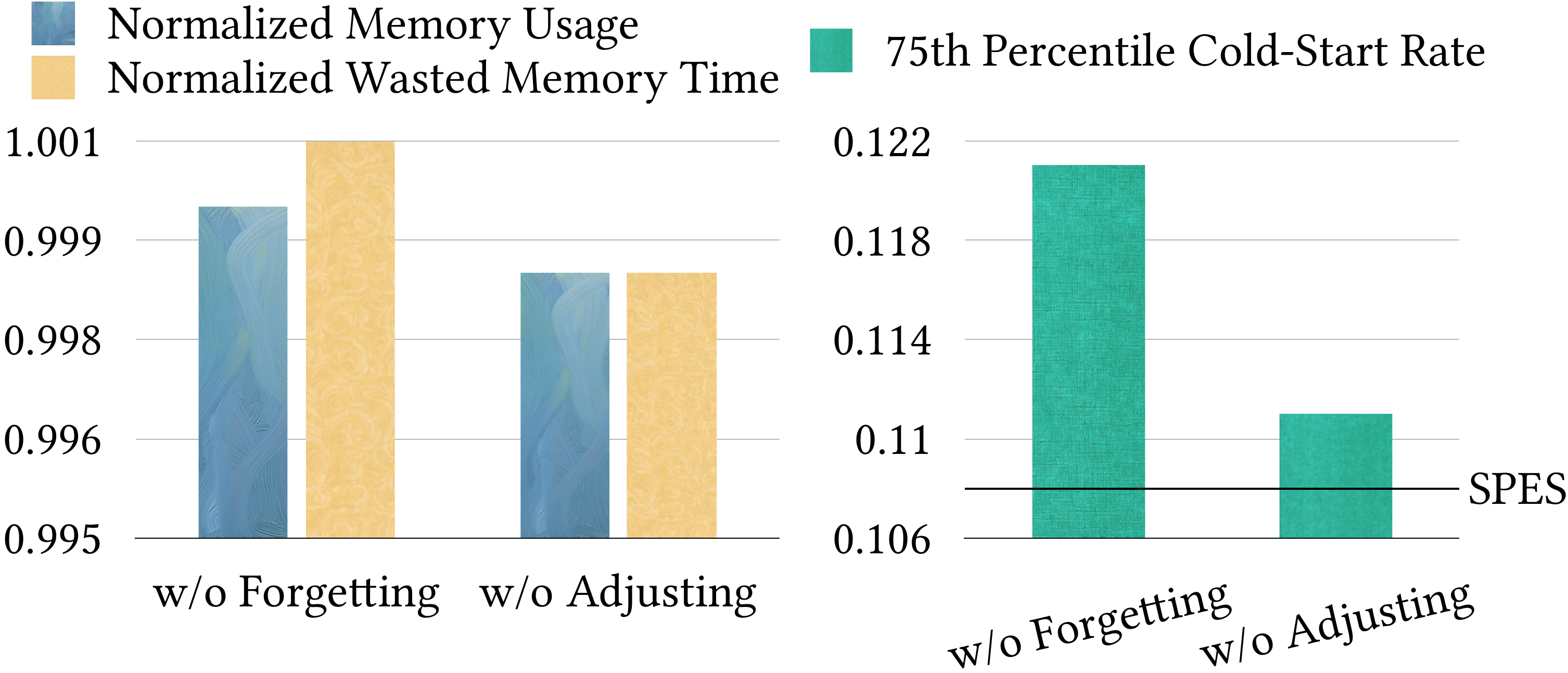}}
    \caption{\tool' adaptivity benefits cold-start reduction.}
    \vspace{-0.1in}
    \label{fig:impact_shift}
\end{figure}

\textbf{Impact of designs regarding concept shifts.}\label{sec:exp:ablation:shift}
This section studies the impact of two designs regarding concept shifts:
1) \textit{forgetting}: forcing unknown functions to a defined type by ignoring older invocations while focusing more on recent data ($\S$\ref{sec:method:difficult});
2) \textit{adjusting}: adaptively adjusting the predictive values during the simulation ($\S$\ref{sec:method:adaptive}).
Figure~\ref{fig:impact_shift} depicts the effectiveness after omitting these two adaptive designs, respectively. 
Removing the second shifting strategy has a slight impact. 
Similarly, this is attributed to the fact that the forgetting strategy involves more function categorization efforts.
We do not pre-load unknown functions, so the forgetting strategy categorizing 340 unknown functions has a larger impact. In contrast, the adjusting strategy only categorizes 174 unknown functions into the ``newly-possible'' type and updates the predictive values of 499 ``(appro-)regular/dense'' functions, resulting in less impact.
Nevertheless, both designs contribute to the effectiveness of \tool. These designs will play a greater role with more data and a longer simulation period (usually indicating larger shifts).

\subsection{Evaluating \newtool With Memory Limitation Under Multi-Worker Scenario}
We evaluate \newtool by answering the following research questions:
\label{sec:exp:eval:spes-ext}

\begin{itemize}
    \item \textbf{RQ5}: With memory limitation and multiple workers, how effectively does \newtool decrease cold starts?

    \item \textbf{RQ6}: With memory limitation and multiple workers, how effectively does \newtool balance the server load? 

    \item \textbf{RQ7}: With memory limitation and multiple workers, how much memory waste and computation overhead does \newtool indeed incur?

    \item \textbf{RQ8}: How does the balancing policy contribute to \newtool?

    \item \textbf{RQ9}: How does the number of workers influence the results?
\end{itemize}

\subsubsection{Baselines}
Given that \newtool integrates three novel components (provision policy, eviction decision, and load balancing) and no existing solution combines all three aspects, we adopt a component-wise comparison approach. 

\noindent\textbf{Provision Policy.} 
To compare with \newtool, we use the five state-of-the-art baselines mentioned in section \ref{sec:exp:baselines} as the provision policy of the baseline scheduler.

\noindent\textbf{Eviction Decision.}
For schedulers lacking explicit multi-worker eviction mechanisms (excluding FaaSCache which employs priority-based eviction), we implement first-in-first-out (FIFO) eviction: functions are evicted based on loading time (oldest first), and currently invoked functions are protected from eviction.
On the other hand, \newtool extends traditional FIFO eviction by applying priority-based eviction to PendingSet and PrewarmSet (Section~\ref{sec:method:evict}) and employing FIFO for worker pool eviction to maximize idle memory utilization.

\noindent\textbf{Load Balancing.}
Following~\cite{FaaSRank}, we compare \newtool with four common baselines:

\begin{itemize}
    \item Hashing scheduler. This is the OpenWhisk default scheduler. In our experiment, we calculate a hash value for each function and map it to the corresponding worker.
    
    \item Round-robin scheduling. With this method, workers take turns to load the coming functions.
    
    \item Least-loaded scheduling. In this scheduler, we always distribute the incoming invocation to the server with the least load.
    
    \item Most-loaded scheduling. This is the scheduling strategy of AWS Lambda. We utilize the same server greedily to load the coming functions. If it reaches its capacity, we turn to the next server.

\end{itemize}


This selection process yields 20 distinct baseline combinations. To maintain clarity in our presentation and discussion, we adopt the following representative strategy: For each provision policy, we identify the optimal combination of components (provision policy + eviction decision + load balancing) and use this configuration as the canonical representation of that policy class. 
For instance, if the FaaSCache/Hash combination demonstrates superior cold start mitigation compared to alternative FaaSCache pairings (Round Robin, Least Loaded, Most Loaded), we only exclusively report and analyze the FaaSCache/Hash results as the representative case for all FaaSCache-based configurations. 

\subsubsection{Metrics and parameters}
To comprehensively evaluate system performance, we augment the baseline metrics from Section~\ref{sec:exp:baselines} with an extra \textit{Normalized Worker Variance} (NWV) metric for load balancing assessment.
NWV is computed as $NWV = \frac{\sigma^2_{m_r}}{\mu_{m_r}}$, where $\sigma^2_{m_r}$ represents the variance of workers' remaining memory and $\mu_{mr}$ denotes the mean remaining memory at minute-level intervals. This normalization eliminates bias from absolute memory availability, focusing instead on relative distribution efficiency.

We set the server memory configuration following the two critical constraints.
The upper bound is that the memory must be constrained to properly evaluate provisioning policy effectiveness under resource competition.
The lower bound is that the memory must be adequate to handle peak historical demand without triggering invocation queuing.
This is because we decouple the invocation allocation problem from the queuing problem for greater clarity, according to Occam's Razor principle.
Therefore, we set the total memory of servers as $1.1 \times \lceil \max(M_{hist}) \rceil$ where $M_{hist}$ represents historical memory demand. The 10\% buffer ensures reliable operation while maintaining evaluation stringency.

The number of workers in our experiment is set to six. We will discuss the impact of the server count in Section~\ref{sec:exp:workernum}.

\subsubsection{RQ5: Effectiveness in Cold-Start Reduction}

\begin{figure}[htb]
    \centering
    \vspace{-0.12in}
        {\includegraphics[width=0.4\linewidth]{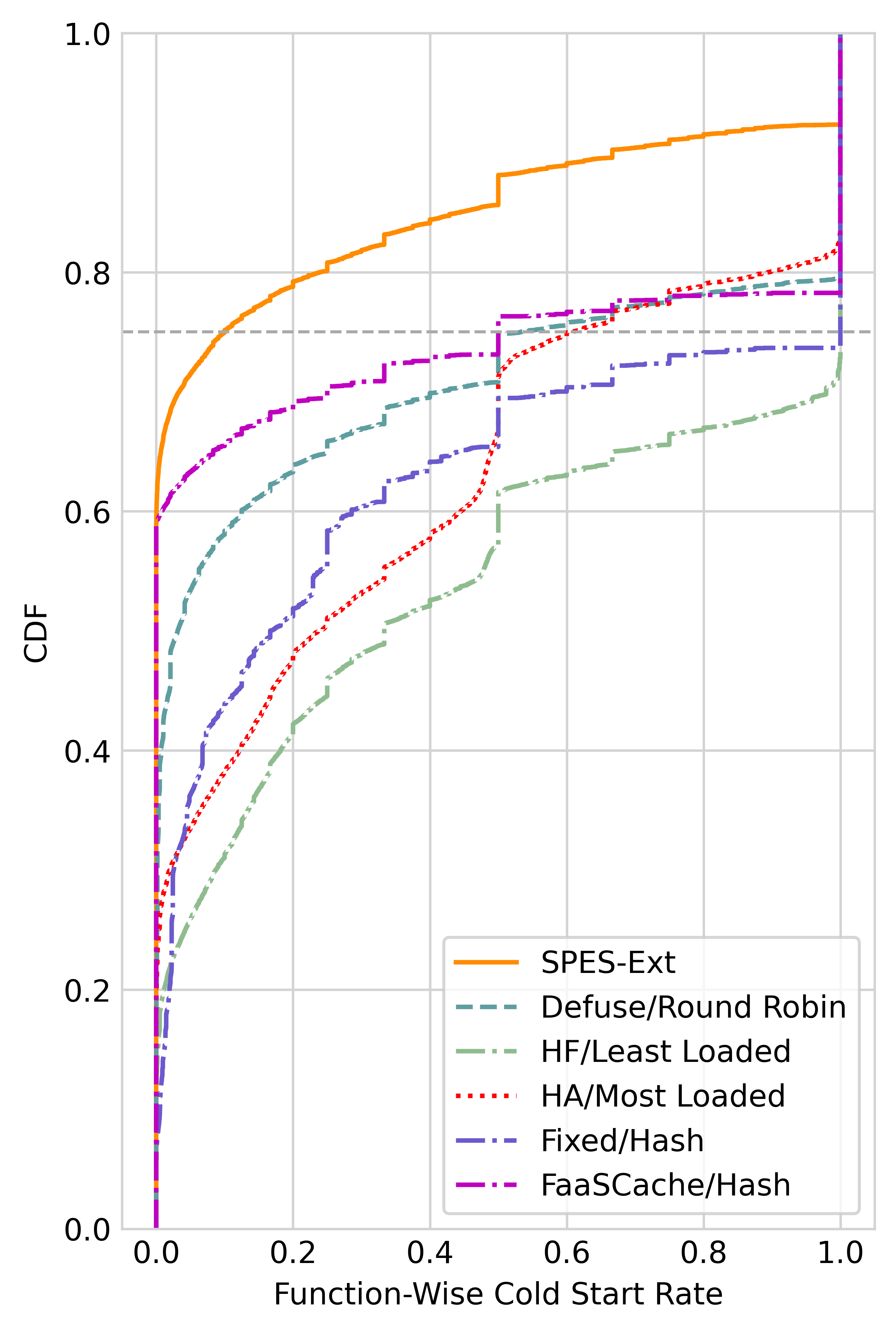}}
    \caption{Cold start behaviors of \newtool and baselines.}
    \vspace{-0.1in}
    \label{fig:q1_coldstart}
\end{figure}

Figure \ref{fig:q1_coldstart} presents the comparative analysis of cold start rate (CSR) distributions between \newtool and baseline schedulers using cumulative distribution functions (CDFs). The leftward shift of \newtool's CDF curve demonstrates its consistent superiority in cold start mitigation.

To be more specific, regarding infrequently invoked functions, on a function-wise level, \newtool reduces the 75th percentile cold-start rate (Q3-CSR, for simplicity) from 50\% to 10\% compared with FaasCache, the best-performing baseline, achieving 80\% improvement. Compared with other baselines, our policy reduces the Q3-CSR by 81.25\%-90\%. Furthermore, our policy reduces the 90th percentile cold-start rate by 9.76\% from 100\% of all other baselines. For warm-start functions, 57.94\% of the functions don't experience cold starts under our policy, which indicates that most functions can be invoked with warm startup. Meanwhile, for the baselines, only FaasCache has a 1.54\% higher proportion than \newtool, while for others,the proportion of warm start functions varies from 5.78\% to 16.80\%. These demonstrate that under such a scenario, our policy can utilize the limited memory to significantly improve cold start latency for both frequently and infrequently invoked functions.

Additionally, Figure \ref{fig:q1_coldstart} displays that our policy greatly reduces the percentage of ``always cold'' functions, i.e. functions with CSR equal to 100\%. Under \newtool policy, only 7.65\% functions experience a cold start every time they are invoked. Among the baselines, HA has the fewest always-cold functions, which is closest to that of \newtool, whereas HF, the function-grained method, increases always-cold functions dramatically. This can be attributed to infrequently invoked functions. About 3.82\% functions are invoked less than twice during training, and 6.14\% are only once during the simulation, so always-cold functions seem inevitable with limited records. HA mitigates this issue by grouping and loading functions together, yet results in more memory consumption and waste than HF. However, each application contains fewer than 3 functions on average, so it doesn't cost much more. \newtool instead connects unseen and unpredictable functions with known functions using trigger, application, and user information, and gradually updates the associated functions based on actual invocations. Hence, \newtool performs even better than HA with less memory.

\begin{figure}[htb]
    \centering
    \vspace{0.1in}
        {\includegraphics[width=0.4\linewidth]{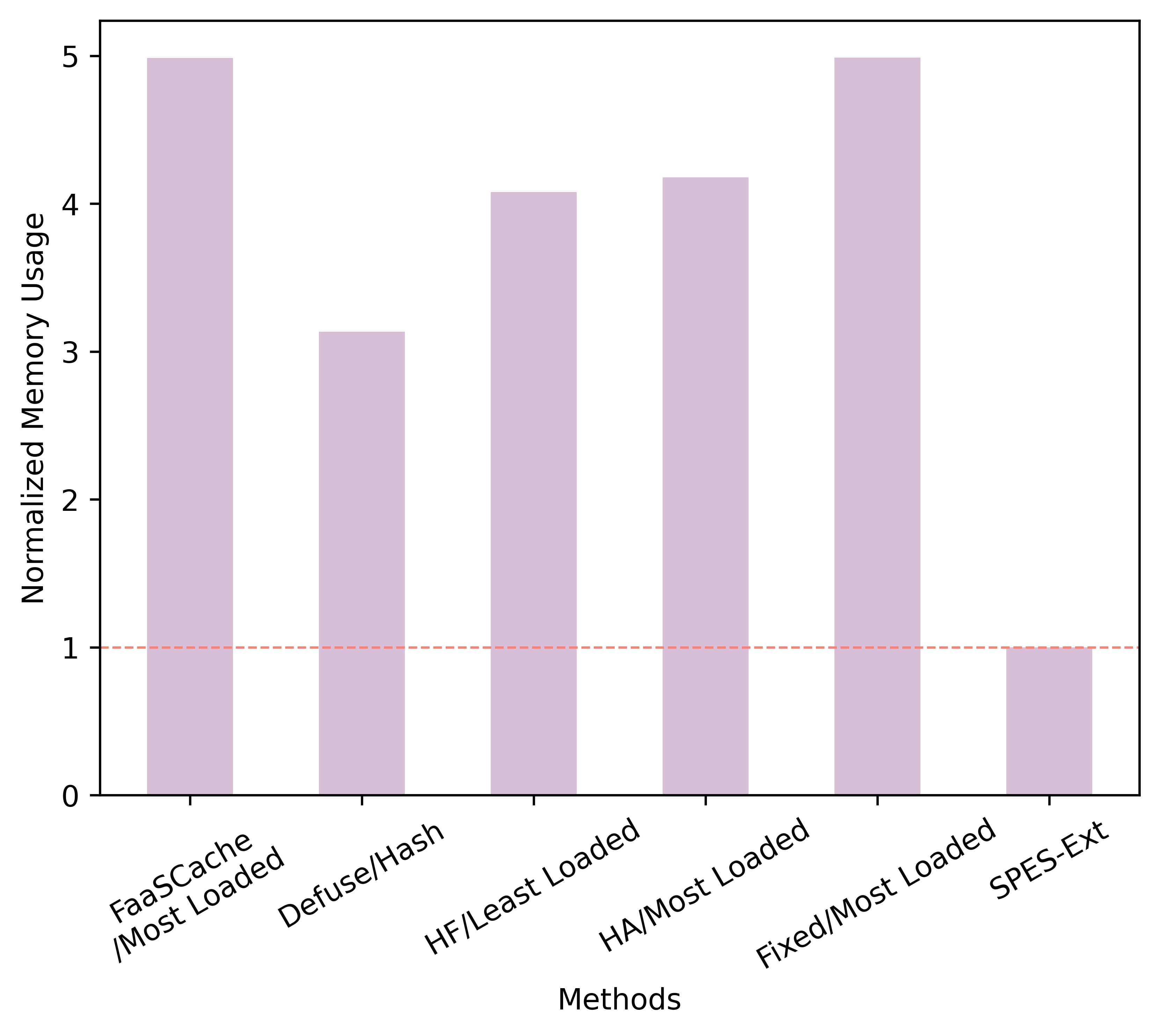}}
    \caption{Normalized memory usage of \newtool and baselines.}
    \vspace{-0.1in}
    \label{fig:q1_usage}
\end{figure}

Although the upper limit of the memory is fixed for these methods, the actual memory usage varies. Figure \ref{fig:q1_usage} shows the memory usage normalized by the one of \newtool. The memory usage of \newtool is 68.08\% less than Defuse, which is the most space-efficient. Compared with other baselines, \newtool saves 75.49\%-79.95\% memory. FaaSCache, the best baseline in cold start alleviation, consumes 4 times more memory than \newtool.

\subsubsection{RQ6: Load Balancing Effect}

Since there are many provision and load-balancing policy combinations, we compare and show the best combinations in each type of provision policy, which means the combination with the lowest mean variance. For example, HF-LeastLoaded has a lower mean variance than HF-Hash, HF-RoundRobin, HF-MostLoaded, so we choose HF-LeastLoaded as the representative of all HF combinations when measuring load balancing effect.

\begin{figure}[htbp]
\centering
\subfigure[Normalized Worker Load Variance for All Methods]{
\begin{minipage}{0.4\textwidth}
\centering 
\includegraphics[width=0.95\linewidth]{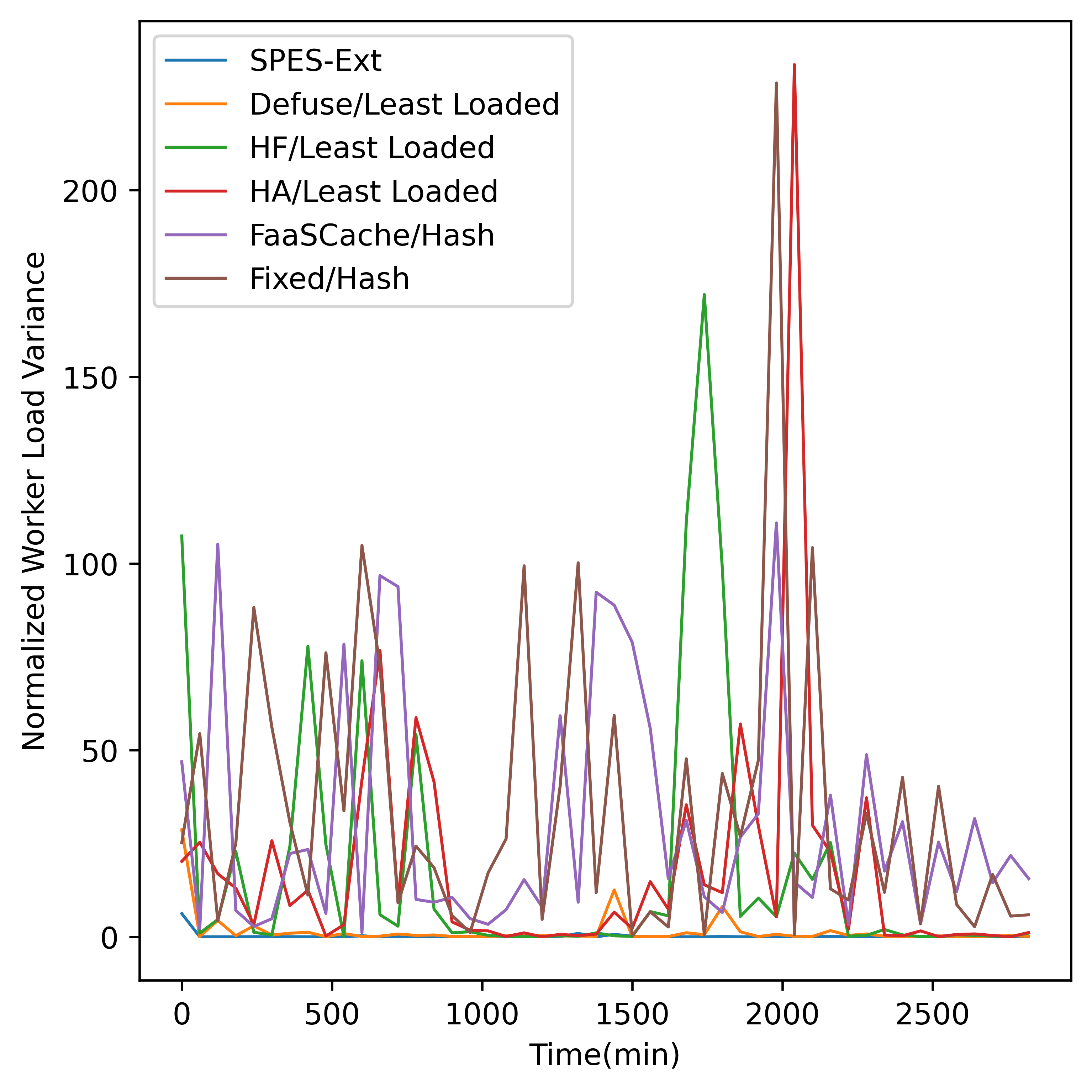}
\label{fig:q2_loadall}
\end{minipage}
}
\subfigure[Normalized Worker Load Variance for \newtool]{
\begin{minipage}{0.4\textwidth}
\centering 
\includegraphics[width=0.95\linewidth]{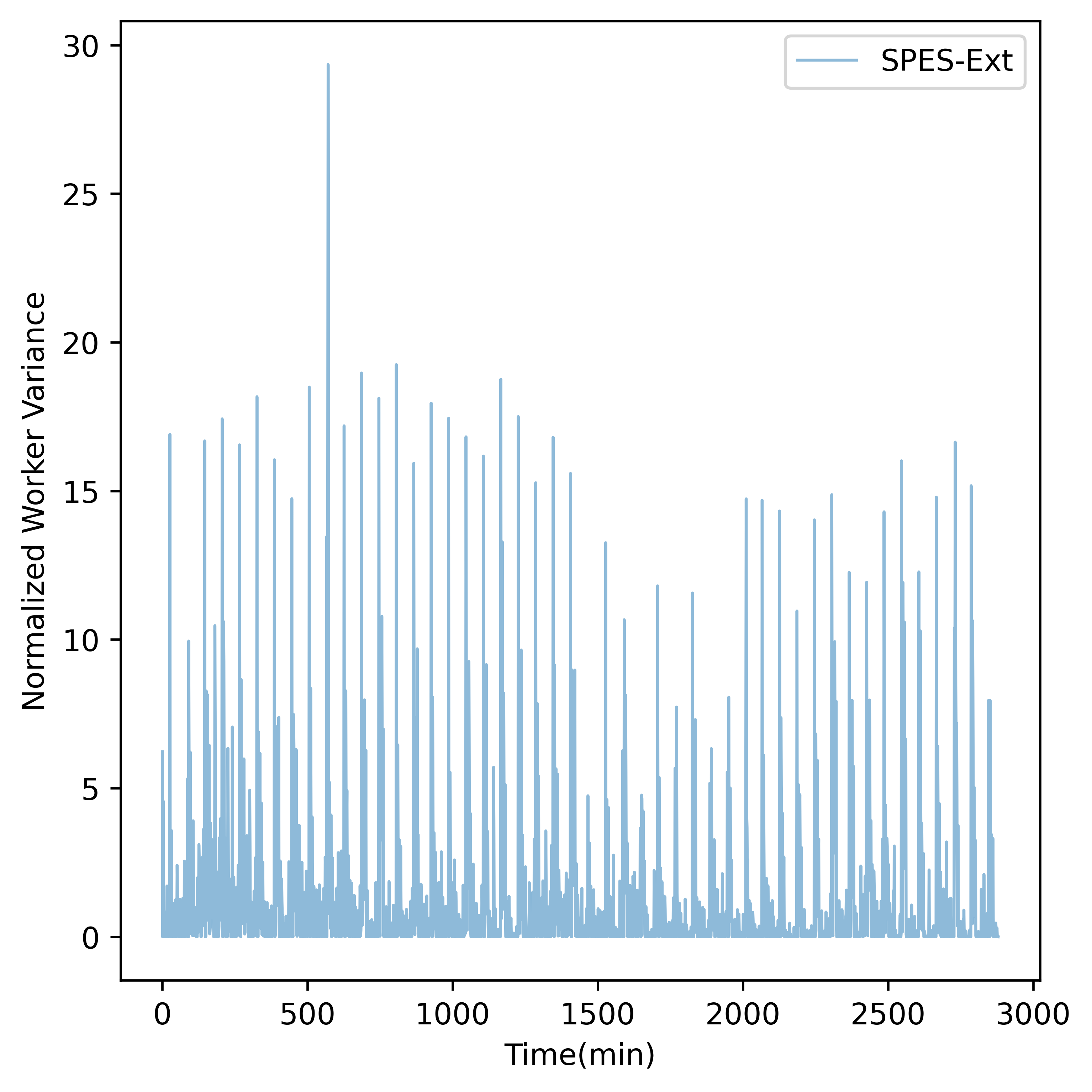}
\label{fig:q2_loadspes}
\end{minipage}
}
\caption{\newtool dramatically lowers the load variance of the workers.}
\vspace{-0.1in}
\label{fig:q2_load}
\end{figure}

As Figure \ref{fig:q2_loadall}  and \ref{fig:q2_loadspes} present, the normalized variance of \newtool remains under 30 in the 2880 minutes,  which lowers the baselines mean variances by 52.49\% - 96.14\%. Particularly, FaaSCache-MostLoaded has the highest mean variance and HA-LeastLoaded has the lowest. Among the baselines, three of five baselines have the best performance when combined with LeastLoaded, indicating that loading into the worker with the largest space first can help balance the load. The load balancing policy of \newtool also utilizes this, and it sorts the loading set first and prioritizes the loading of the function with the largest memory, which tries to coordinate the balance from the size of the functions. This further improves the load-balancing effect.


\subsubsection{RQ7: Wasted Memory Time and Overhead}

\begin{figure}[htbp]
\centering
\subfigure[Normalized wasted memory time (WMT, lower is better)]{
\begin{minipage}{0.4\textwidth}
\centering 
\includegraphics[width=0.95\linewidth]{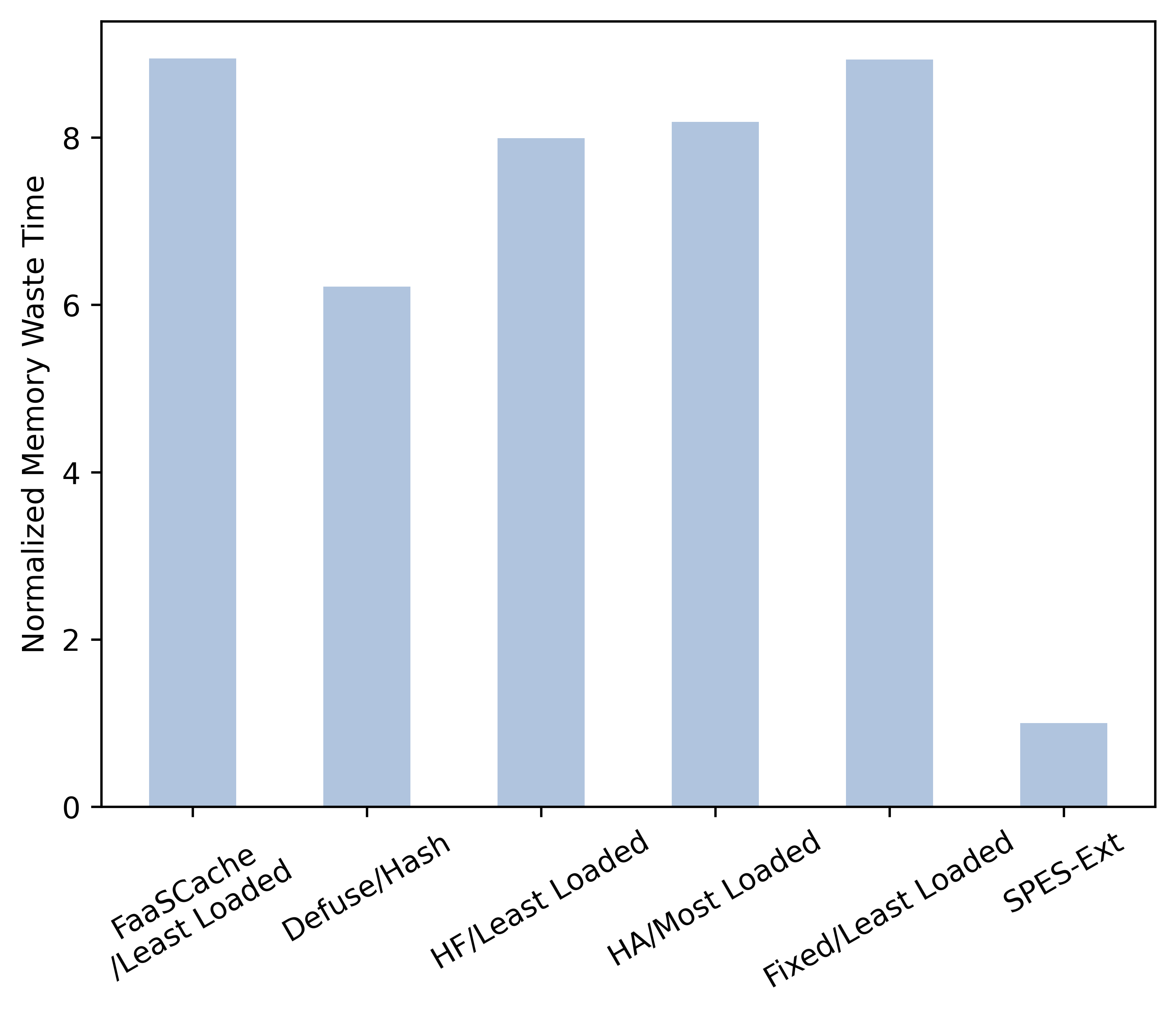}
\label{fig:q3_allwmt}
\end{minipage}
}
\subfigure[Effective memory consumption ratio (EMCR, higher is better)]{
\begin{minipage}{0.4\textwidth}
\centering 
\includegraphics[width=0.95\linewidth]{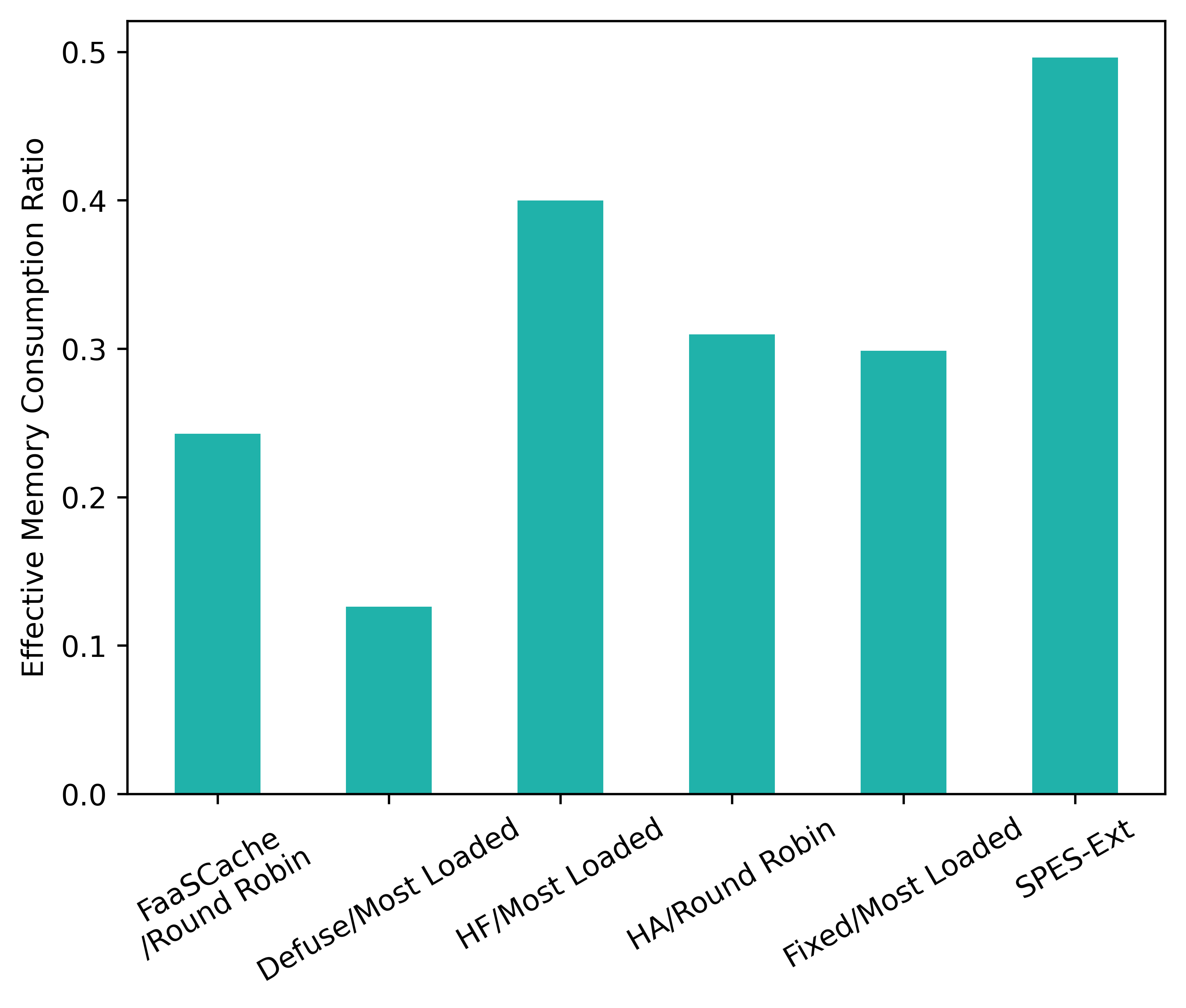}
\label{fig:q3_allemcr}
\end{minipage}
}
\caption{\newtool significantly reduces memory waste and achieves more effective memory consumption.}
\vspace{-0.1in}
\label{fig:q3_mem}
\end{figure}

Figure \ref{fig:q3_allwmt} illustrates the wasted memory time of all methods under a memory-constrained scenario. Our method \newtool significantly decreases the wasted memory time (WMT) by 83.92\%-88.82\% compared with other baselines. Among the baselines, FaaSCache wastes most memory time while it is most effective on cold-start reduction, indicating that it trades memory resources for efficiency.

Figure \ref{fig:q3_allemcr} presents that \newtool has a high EMCR of 49.61\%, which is 24.10\%-293.19\% higher than the baselines. Particularly, Hybrid-Function(HF) has the highest EMCR but the worst cold start, which means that it doesn't load either useful functions or useless functions and inefficient loading leads to a high EMCR but also a high cold start rate.


\subsubsection{RQ8: The contribution of balancing policy to \newtool}

We perform a component-wise analysis by isolating the load balancing policy while maintaining \newtool's other components (provisioning and eviction policies). This ablation study replaces only the balancing module with alternative strategies from baseline schedulers.

\begin{figure}[htbp]
\centering
\subfigure[CDF of Function-Wise Cold Start Rate]{
\begin{minipage}{0.4\textwidth}
\centering 
\includegraphics[width=0.95\linewidth]{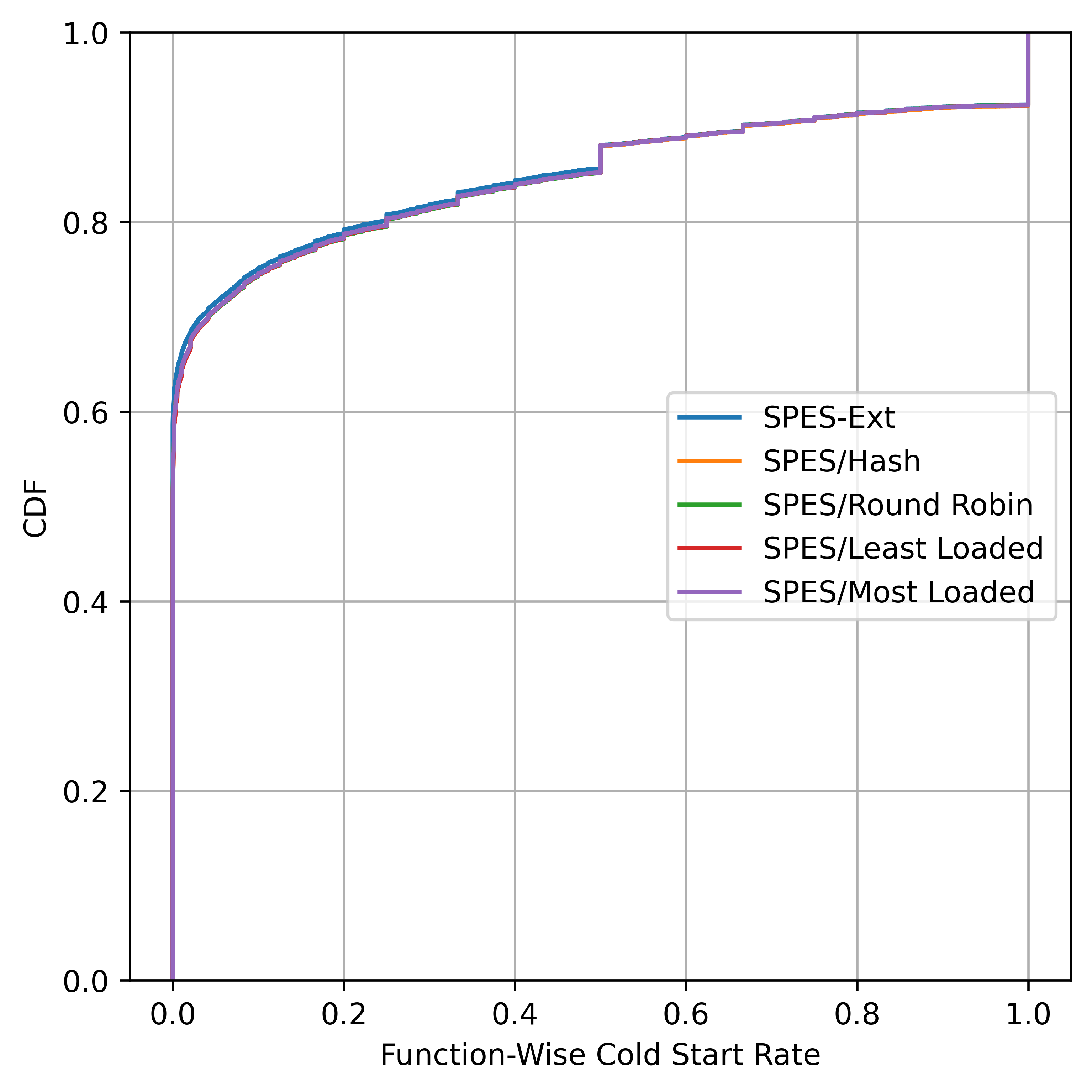}
\label{fig:q4_coldstart}
\end{minipage}
}
\hfil
\subfigure[Normalized wasted memory time (WMT, lower is better)]{
\begin{minipage}{0.4\textwidth}
\centering 
\includegraphics[width=0.95\linewidth]{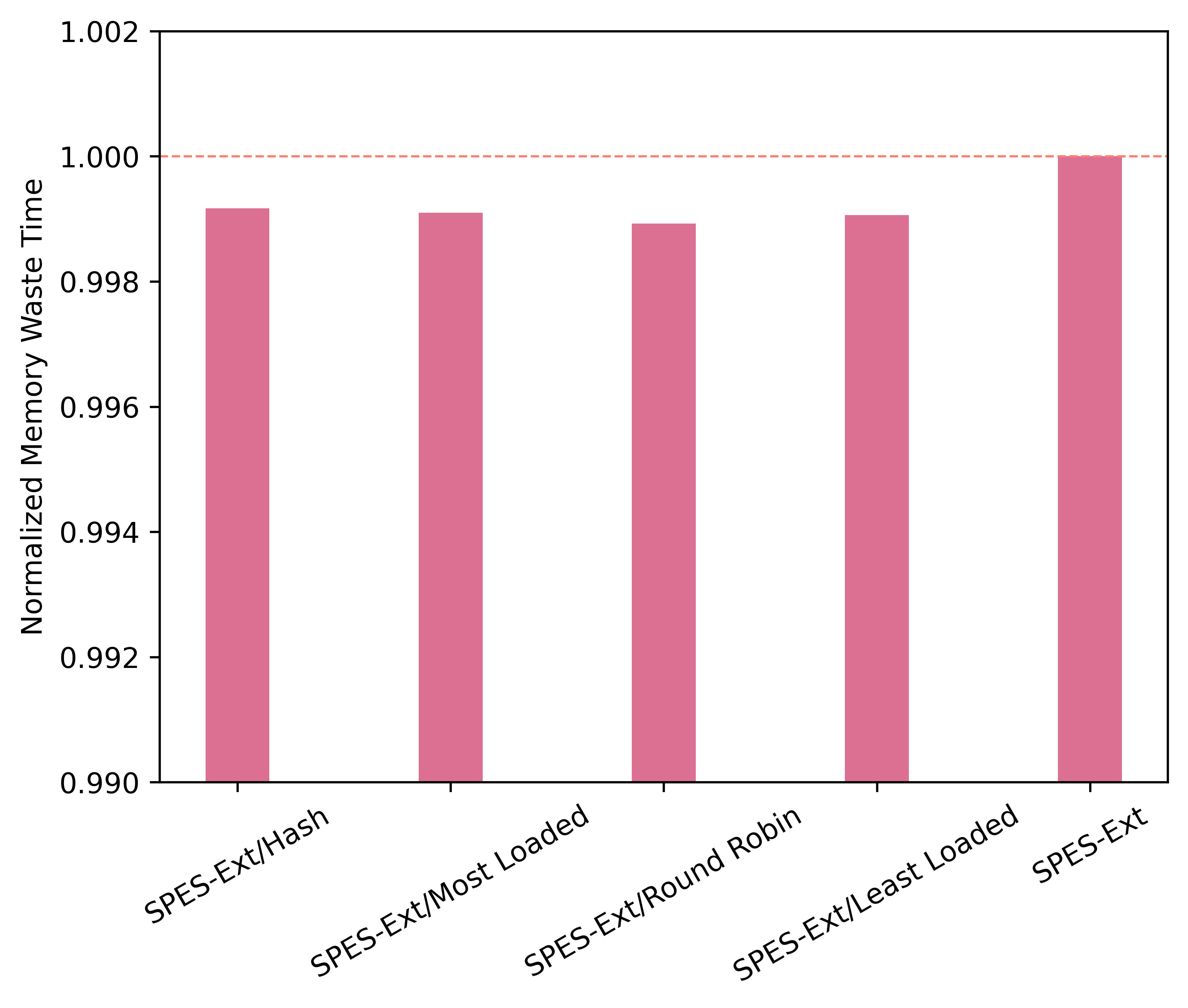}
\label{fig:q4_wmt}
\end{minipage}
}

\subfigure[Normalized Worker Load Variance]{
\begin{minipage}{0.4\textwidth}
\centering 
\includegraphics[width=0.95\linewidth]{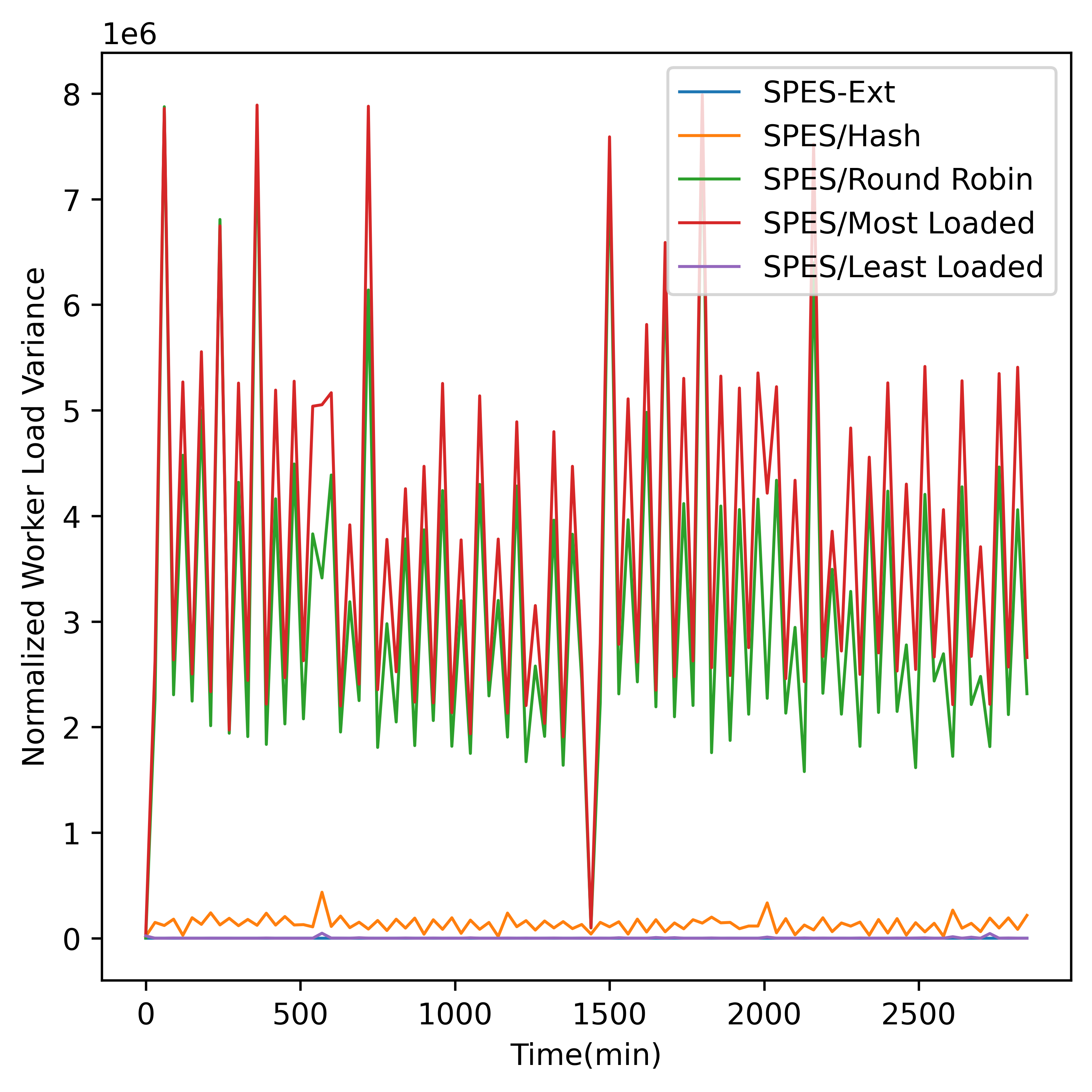}
\label{fig:q4_load}
\end{minipage}
}

\caption{The load-balancing policy \newtool has a slight impact on cold start and memory waste time, but greatly lowers the load variance.}
\vspace{-0.1in}
\label{fig:q4_cmp}
\end{figure}

We analyze the results from three aspects.
For cold start alleviation, Figure \ref{fig:q4_coldstart} reveals that \newtool has a marginal but consistent 9.9\% reduction in Q3-CSR across all baseline variants. 
It also achieves an equivalent performance at the 90th percentile and for always-cold rates.
In terms of memory efficiency, as quantified in Figure \ref{fig:q4_wmt}, \newtool consumes no more than 0.11\% WMT than baseline balancers. 
For load distribution analysis, Figure \ref{fig:q4_load} demonstrates that extreme variance is observed in baseline strategies.
SPES/Most-Loaded has the largest variance and SPES/Least-Loaded has the least variance among the baselines, which are respectively $1.1e6\times$ and $4e3\times$ larger than \newtool on average.

The experimental results confirm that our load balancing policy provides significant load variance reduction (up to 6 orders of magnitude) and consistent cold start improvements without introducing substantial memory waste.

\subsubsection{RQ9: The influence of the number of workers}\label{sec:exp:workernum}

\begin{figure}[htbp]
\centering
\subfigure[CDF of Function-Wise Cold Start Rate]{
\begin{minipage}{0.4\textwidth}
\centering 
\includegraphics[width=0.95\linewidth]{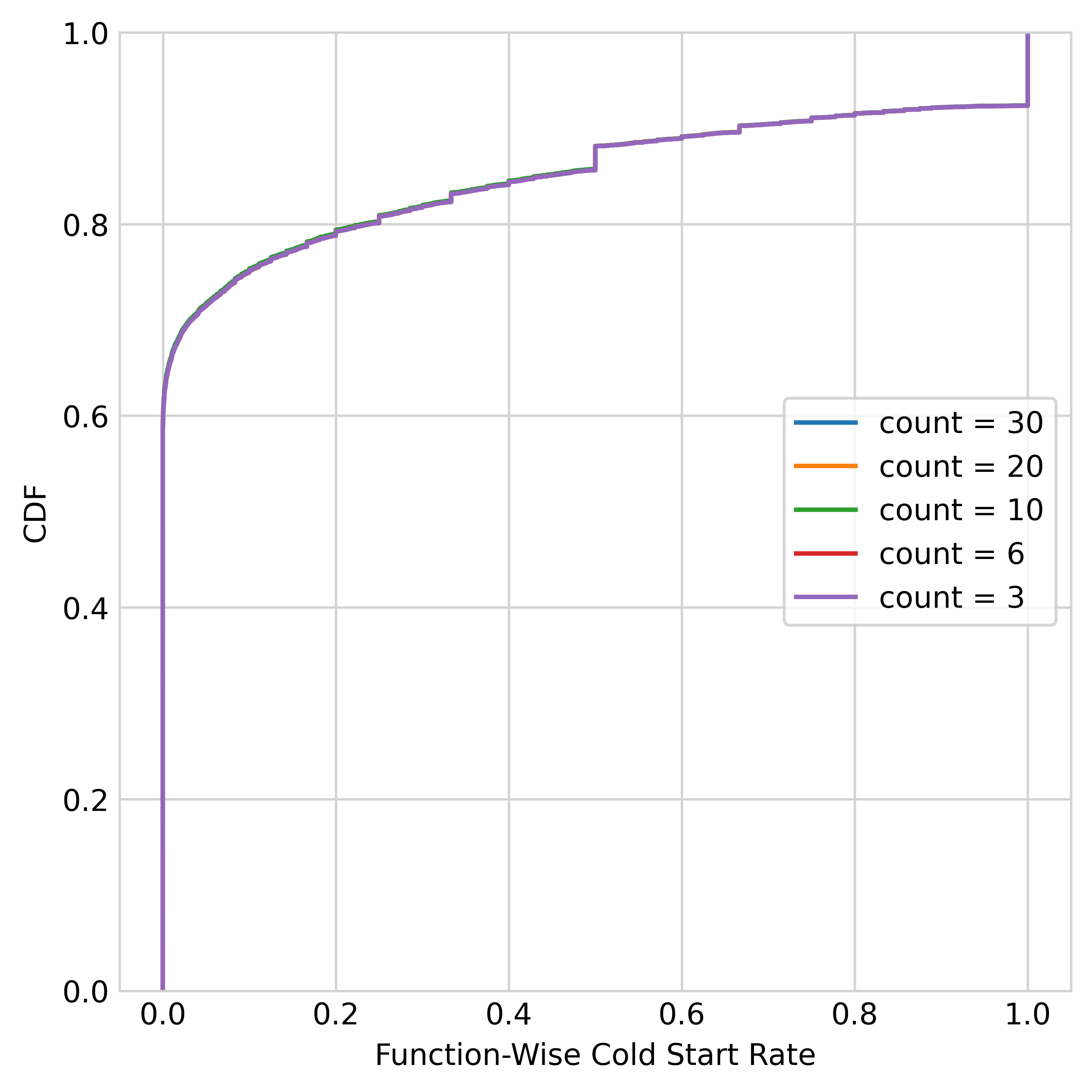}
\label{fig:q5_coldstart}
\end{minipage}
}
\hfil
\subfigure[Normalized wasted memory time (WMT, lower is better)]{
\begin{minipage}{0.4\textwidth}
\centering 
\includegraphics[width=0.95\linewidth]{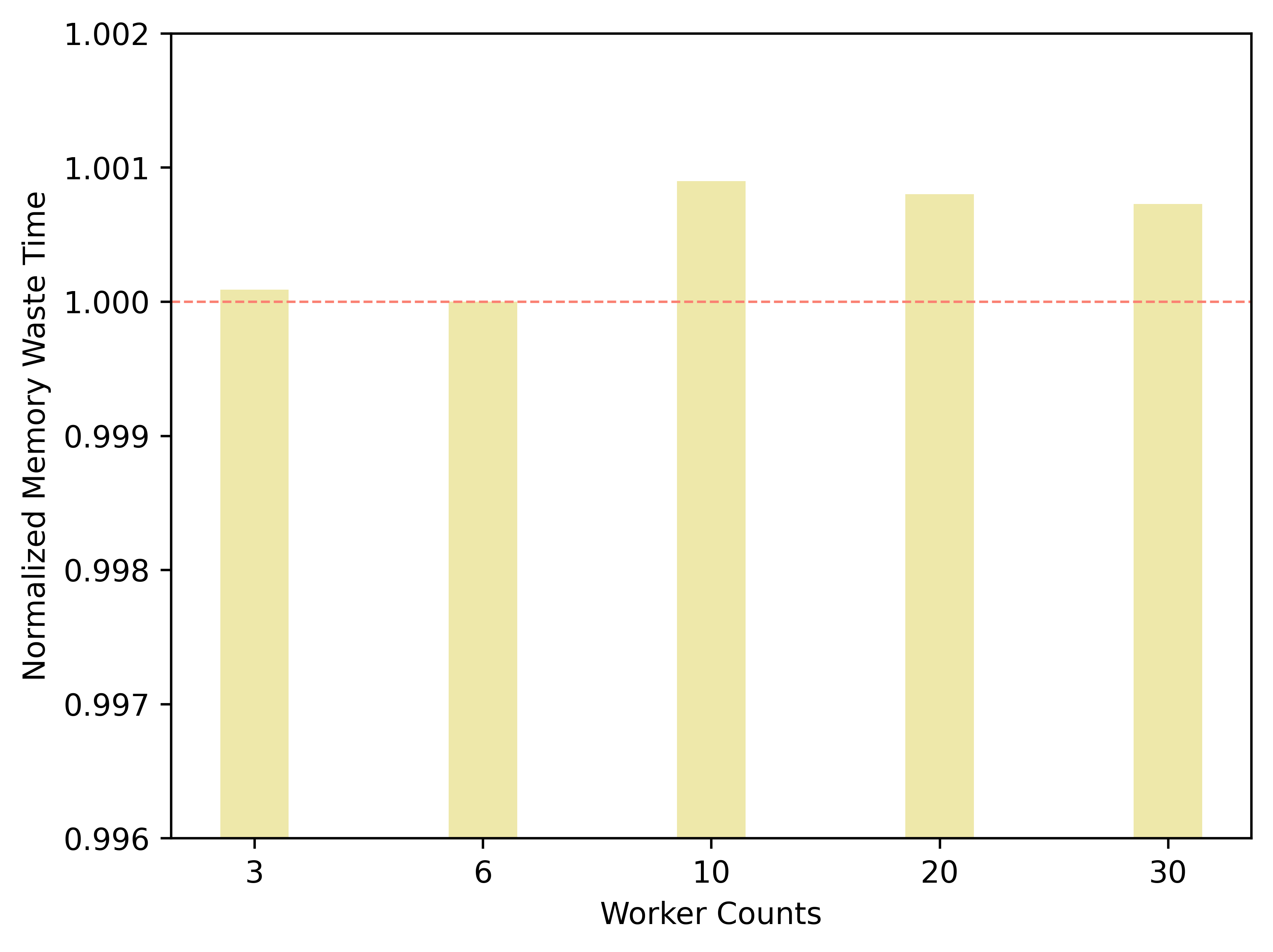}
\label{fig:q5_wmt}
\end{minipage}
}

\subfigure[Normalized Worker Load Variance]{
\begin{minipage}{0.4\textwidth}
\centering 
\includegraphics[width=0.95\linewidth]{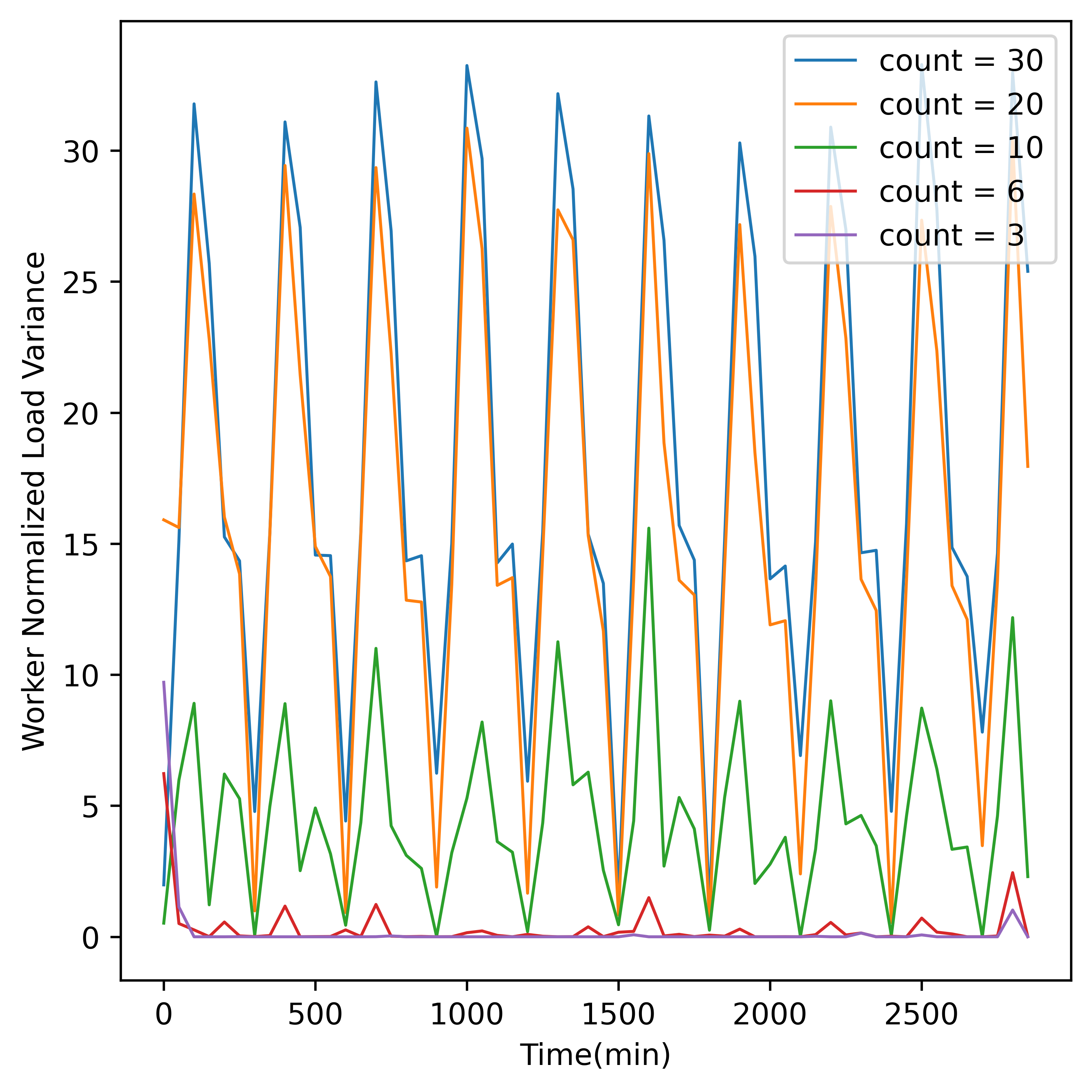}
\label{fig:q5_load}
\end{minipage}
}

\caption{The growth of the workers number has limited influence on cold start performance and memory consumption, but increases the load variance.}
\vspace{-0.1in}
\label{fig:q5_cmp}
\end{figure}

We compare the cold start performance of \newtool with the same total memory but a different number of workers. Figure \ref{fig:q5_coldstart} demonstrates that the number of workers has minimal impact on the function cold start latency, which is reasonable because both the total memory and the provision policy remain the same. Figure \ref{fig:q5_wmt} shows that they has similar memory waste, where the setting with 10 servers has most memory waste and setting with 6 servers has the least, while the latter is only 0.07\% more.

Regarding the load balancing effect, as shown in Figure \ref{fig:q5_load}, the normalized variance of workers is proportional to the number of workers. The mean value of normalized variance varies from 0.20 to 19.17 as the number of workers increases from three to thirty. An intuitive reason is that it's more difficult to balance the worker load when the number of workers is larger. Meanwhile, since the total space of all workers is fixed, to ensure that the worker can hold functions with a large memory that may not appear in the history, the number of workers should not be too large in our experiment. Hence, considering the analysis above, we choose the worker count as six. 
\section{Discussion}\label{sec:review}
\subsection{Limitations}
\subsubsection{Idealized Simulation Principles}
Our simulation adopts two key assumptions to simplify the process and mitigate dataset limitations. 
Firstly, we assume each function execution completes within one minute. For functions with longer execution, adjusting the counting unit as necessary to accommodate the vast majority (e.g., 95\%) of functions within this timeframe can ensure the assumption's relevance. 
For functions exceeding this one timeframe without new invocations—e.g., for a two-minute execution without activity in the second minute—we maintain memory allocation and increment the WMT despite no memory being actually wasted. 
Such deviation can particularly underestimate resource efficiency, especially for methods with frequent evictions, including ours and HF. 
Thus, we posit that such marginal deviations do not affect the overall integrity of our experimental findings.
Secondly, we assume all functions share the same cold-start latency. Our focus on function-specific metrics means that variations in cold-start latency across functions do not significantly affect our primary research insights, though a more detailed dataset could enhance our analysis of the aggregate cold-start latency and offer a deeper comparison of methodologies.

\subsubsection{Over-simplicity of memory consumption}
We mainly make three assumptions about the memory consumption.

First, we regard all function instances of the same applications as equally consuming the average memory limited by the data without function-wise memory consumption.
They actually require different memory allocations, yet the assumption is still reasonable as these functions are usually responsible for simple tasks with confined memory size~\cite{Defuse, ChallengeStudy}.

In addition, we simply assume that the workers can simultaneously load all the invoked functions in each minute, and thus it does not lead to function queueing problem. Otherwise, we should consider not only the cold start latency but also the queueing delay, which involves tradeoffs among loading different functions.

Furthermore, we make the assumption that when a function has multiple concurrent invocations, they have the same priority and can take turns to execute in the same container within one minute. If multiple containers are required for concurrent invocations, we may underestimate the memory consumption of frequently invoked functions and need to decide the provisioning of nearly identical containers \cite{FaasCache}.


\subsubsection{Ignorance of the worker dynamics}
We simply assume that the count and capacity of workers are fixed throughout the experiment, so we don’t involve worker autoscaling. If autoscaling is considered like \cite{suresh2019fnsched}, the number of invokers and invoker memory should autoscale in response to changes in workload demand. We plan to design a more elastic scheduler that covers the autoscaling module and refines our current design to optimize under different scales.

\subsection{Threats to Validity}\label{sec:threat}
\subsubsection{Internal}
The evolution of serverless functions is very fast. With only 14-day data, we can observe distinct concept shifts in function invocations. We currently do not conduct proactive re-categorization or rule generation. The performance of \newtool may degrade in the very long term.
To mitigate this threat, we design adaptive strategies to handle relatively small shifts, which can easily be extended to long-term shifts.

\subsubsection{External}
The effectiveness of \newtool on other datasets is yet unknown. 
To alleviate this concern, we choose Azure Function, the most widely used dataset from real-world industry, whose representativeness has been confirmed by many studies~\cite{ServerlessWild, Defuse, FaaST, FaaSRank, Harvested}.
Moreover, it is the only publicly available industrial dataset.
We will strengthen cooperation with the industry to evaluate \newtool on more datasets and prompt the implementation of \newtool on a real-world FaaS system.
We also release our code to facilitate reproducibility and more experiments in the community.

\section{Related Work}

\textbf{Cold Starts Alleviation.} Numerous efforts~\cite{Hermod,IceBreaker,Xanadu,ServerlessWild,Defuse,FaaSRank,Catalyzer,FaasCache, LCS} have focused on minimizing cold starts and memory overhead for FaaS platforms. Related studies~\cite{currentTrents,FaaSLight} classify current methods into two categories: speeding up function warming (involving renovations in the system layer) and reducing occurrences of runtime cold starts (at the application layer).
Methods of these two categories are orthogonal, so they can be combined for further cold start optimization.

This first category presents new systems to optimize infrastructure and resource management, employing techniques like load balancing among worker nodes~\cite{Hermod, IceBreaker}, snapshotting~\cite{Prebreaking, vHive}, sandbox (e.g., container) scheduling and management~\cite{Ensure, Xanadu, Catalyzer, SAND, Pagurus, Resuing, PCPM}.
One of the most popular research directions is sandbox sharing, i.e., sandboxes of the previous execution are reused for new invocations since reusing the idle sandbox incurs less delay than allocating a new one~\cite{SAND}.
A recent study, Pagurus~\cite{Pagurus}, proposed a container management scheme that allows one function's idle warm container to be forked by another function to alleviate cold starts, showing promising results.
However, implementing these methods requires significant engineering efforts, system expertise, and ongoing maintenance due to the underlying platform and sandbox modifications, which pose challenges in diverse infrastructures and platforms.

The second category mitigates cold-start occurrences~\cite{ServerlessWild, Defuse, FaasCache, FaaSRank, Ensure, LCS} through a non-intrusive way.
A primary step of function scheduling is to decide when and whether to pre-load/evict a function instance to reduce cold starts.
\cite{ServerlessWild} utilizes a small histogram to monitor function inter-invocation times, benefiting workloads with clear invocation patterns by optimizing keep-alive and pre-warming.
However, it is fully data-driven and ignores the underlying invocation patterns, leaving much room for domain-knowledge-involved invocation prediction.
Defuse~\cite{Defuse} employs dependency mining from function invocation histories to optimize the keep-alive time and pre-warming. Yet, it relies on the statistical histogram and turns to a fixed keep-alive policy for more than 32\% of the functions, delivering inadequate perfection of effectiveness.
LCS~\cite{LCS} selected the least recently warm container to reduce cold starts by keeping the containers alive for a longer period.
FaaSCache~\cite{FaasCache} innovatively establishes an equivalence between keeping functions alive and keeping objects in a cache, thereby implementing function provision based on Greedy-Dual-Size Frequency object caching. However, its fundamental idea is to use up the given resources as much as possible until they are insufficient, then evict containers to minimize cold starts. Thus, its way of optimizing resources is rigid without any predictions, failing to smoothly optimize cold starts and resource usage simultaneously. This leads to a more significant waste of memory resources.

There is room for improvement in these function provisioning methods, as most of them either lack fine-grained invocation prediction or rely on costly learning models without fully exploiting diverse invocation patterns.
Instead, \newtool adopts lightweight horses-for-courses rule-based strategies to enhance existing approaches via accurate invocation prediction. 
\newtool also proposes adaptive designs and builds up inter-function correlations, delivering better memory-economic cold-start reduction.
Other efforts target a downstream task to determine a suitable node for scheduling an individual function request. 
FaaSRank~\cite{FaaSRank} and ENSURE~\cite{Ensure} attempt to pack load on the adequate number of invokers, allowing the additional unneeded invokers to idle, so as to reduce the completion time and cold starts of functions.
Our method can be combined with these methods for finer-grained scheduling.

\textbf{Load Balancing Related Scheduling.} There are works that focus on data-aware or resource-aware load-balancing.
Package-aware load balancing~\cite{aumala2019beyond} considers the data locality of software packages by enhancing consistent hashing to avoid worker overload and ensure overall performance. 
Meanwhile,~\cite{abdi2023palette}~\cite{aumala2019beyond} concentrate on the locality of the data accessed by function invocations. ~\cite{fuerst2022locality} also refines consistent hashing to develop a locality-aware load-balancing policy. It mainly focuses on the locality of in-memory functions regarding how to choose the ``home'' server of each function, i.e. assigning the invocations of the same function to a certain subset of servers to guarantee locality. Palette~\cite{abdi2023palette} is a load balancer, especially for data-intensive applications, that introduces colors as locality hints and route invocations with the same color to the same set of instances. 
~\cite{suresh2019fnsched} takes the diversity in resource consumption and lifetime of functions into account, and classifies the serverless applications into two broad categories based on their runtime and resource usage. It also involves the multi-invoker scheduling and autoscales the number of hosts when needed.
CH-AL~\cite{lu2025ch} enhances consistent hashing by co-locating functions with shared container image layers and forwarding requests based on real-time node load estimation.
Hiku~\cite{akbari2025hiku} proposes pull-based scheduling where idle workers proactively request tasks, achieving function locality and load balance without consistent hashing.
Metronome~\cite{chen2025metronome} employs an online prediction model to forecast function execution times for locality-aware delay scheduling, distinguishing data locality from infrastructure locality.
\newtool is a priority-aware approach that categorizes functions by their invocation patterns so that we can better characterize the workload and prioritize different functions.

Learning-based approach like~\cite{FaaSRank} utilizes deep reinforcement learning to learn function scheduling decisions and select the invokers. However, RL decisions on the pay-as-you-go FaaS platforms may not be a perfect choice since they may consume much more resources and increase the expenditures of users, which may counteract the resources they save. \newtool is rule-based and costs fewer resources with higher efficiency.

\section{Conclusion}\label{sec:conclusion}
This paper proposes a general rule-based and priority-aware scheduler under the memory-limited and multi-worker scenario.
Our insight is that typical invocation patterns resulting from the underlying serverless architecture and triggers with commonness can lead to predictive invocation behaviors. 
The prediction information implies certain function characteristics, making it feasible to prioritize the functions and thus construct scheduling rules.
Hence, we propose \newtool, which provisions functions, makes eviction decisions, and balances workers' load based on categorization and invocation prediction using tailored strategies.
Our rule-based and priority-aware solution effectively mitigates the cold start and load-balancing problem by addressing the challenges of efficiency, scalability, imbalance, and evolution.
Experiments demonstrate that \newtool significantly improves three sides of the trade-off: economizing expensive resources, reducing cold starts to mitigate latency, and balancing workers' load.
Lastly, we make our code publicly available to facilitate future research and deployment of serverless computing.
\section*{Data Availability}
The artifact for \newtool is available at~\textcolor{myblue} {\url{https://github.com/HelloWorldGitHubUser/SPES-Ext}}.


\bibliography{sn-bibliography}

@inproceedings{LCS,
author = {Sethi, Biswajeet and Addya, Sourav Kanti and Ghosh, Soumya K.},
title = {LCS: Alleviating Total Cold Start Latency in Serverless Applications with LRU Warm Container Approach},
year = {2023},
publisher = {Association for Computing Machinery},
address = {New York, NY, USA},
url = {https://doi.org/10.1145/3571306.3571404},
booktitle = {Proceedings of the 24th International Conference on Distributed Computing and Networking},
pages = {197–206},
numpages = {10},
}

@inproceedings{PCPM,
  author       = {Anup Mohan and
                  Harshad S. Sane and
                  Kshitij Doshi and
                  Saikrishna Edupuganti and
                  Naren Nayak and
                  Vadim Sukhomlinov},
  title        = {Agile Cold Starts for Scalable Serverless},
  booktitle    = {11th {USENIX} Workshop on Hot Topics in Cloud Computing, HotCloud
                  2019, Renton, WA, USA, July 8, 2019},
  publisher    = {{USENIX} Association},
  year         = {2019},
  url          = {https://www.usenix.org/conference/hotcloud19/presentation/mohan},
  address={''},
}

@inproceedings{Prebreaking,
  author       = {Paulo Silva and
                  Daniel Fireman and
                  Thiago Emmanuel Pereira},
  title        = {Prebaking Functions to Warm the Serverless Cold Start},
  booktitle    = {Middleware '20: 21st International Middleware Conference, Delft, The
                  Netherlands, December 7-11, 2020},
  pages        = {1--13},
  publisher    = {{ACM}},
  year         = {2020},
  url          = {https://doi.org/10.1145/3423211.3425682},
  address={''},
}

@inproceedings{Resuing,
  author       = {Kun Suo and
                  Yong Shi and
                  Xiaohua Xu and
                  Dazhao Cheng and
                  Wei Chen},
  title        = {Tackling Cold Start in Serverless Computing with Container Runtime
                  Reusing},
  booktitle    = {Proceedings of the 2020 Workshop on Network Application Integration/CoDesign,
                  NAI@SIGCOMM 2020, Virtual Event, USA, August 14, 2020},
  pages        = {54--55},
  publisher    = {{ACM}},
  year         = {2020},
  url          = {https://doi.org/10.1145/3405672.3409493},
  address={''},
}

@inproceedings{SAND,
  author       = {Istemi Ekin Akkus and
                  Ruichuan Chen and
                  Ivica Rimac and
                  Manuel Stein and
                  Klaus Satzke and
                  Andre Beck and
                  Paarijaat Aditya and
                  Volker Hilt},
  title        = {{SAND:} Towards High-Performance Serverless Computing},
  booktitle    = {2018 {USENIX} Annual Technical Conference, {USENIX} {ATC} 2018, Boston,
                  MA, USA, July 11-13, 2018},
  pages        = {923--935},
  publisher    = {{USENIX} Association},
  year         = {2018},
  url          = {https://www.usenix.org/conference/atc18/presentation/akkus},
  address={''},
}

@inproceedings{Pagurus,
  author       = {Zijun Li and
                  Linsong Guo and
                  Quan Chen and
                  Jiagan Cheng and
                  Chuhao Xu and
                  Deze Zeng and
                  Zhuo Song and
                  Tao Ma and
                  Yong Yang and
                  Chao Li and
                  Minyi Guo},
  title        = {Help Rather Than Recycle: Alleviating Cold Startup in Serverless Computing
                  Through Inter-Function Container Sharing},
  booktitle    = {2022 {USENIX} Annual Technical Conference, {ATC} Carlsbad, CA, USA, July 11-13, 2022},
  pages        = {69--84},
  publisher    = {{USENIX} Association},
  year         = {2022},
  url          = {https://www.usenix.org/conference/atc22/presentation/li-zijun-help},
  address={''},
}

@inproceedings{ServerlessBFT,
  author       = {Suyash Gupta and
                  Sajjad Rahnama and
                  Erik Linsenmayer and
                  Faisal Nawab and
                  Mohammad Sadoghi},
  title        = {Reliable Transactions in Serverless-Edge Architecture},
  booktitle    = {39th {IEEE} International Conference on Data Engineering, {ICDE} 2023,
                  Anaheim, CA, USA, April 3-7, 2023},
  pages        = {301--314},
  publisher    = {{IEEE}},
  year         = {2023},
  url          = {https://doi.org/10.1109/ICDE55515.2023.00030},
  address={''},
}

@inproceedings{Valve,
  author       = {Pubali Datta and
                  Prabuddha Kumar and
                  Tristan Morris and
                  Michael Grace and
                  Amir Rahmati and
                  Adam Bates},
  title        = {Valve: Securing Function Workflows on Serverless Computing Platforms},
  booktitle    = {{WWW} '20: The Web Conference 2020, Taipei, Taiwan, April 20-24, 2020},
  pages        = {939--950},
  publisher    = {{ACM} / {IW3C2}},
  year         = {2020},
  url          = {https://doi.org/10.1145/3366423.3380173},
  address={''},
}

@inproceedings{Freyr,
  author       = {Hanfei Yu and
                  Hao Wang and
                  Jian Li and
                  Xu Yuan and
                  Seung{-}Jong Park},
  title        = {Accelerating Serverless Computing by Harvesting Idle Resources},
  booktitle    = {{WWW} '22: The {ACM} Web Conference 2022, Virtual Event, Lyon, France,
                  April 25 - 29, 2022},
  pages        = {1741--1751},
  publisher    = {{ACM}},
  year         = {2022},
  url          = {https://doi.org/10.1145/3485447.3511979},
  address={''},
}

@inproceedings{Cirrus,
  author       = {Jo{\~{a}}o Carreira and
                  Pedro Fonseca and
                  Alexey Tumanov and
                  Andrew Zhang and
                  Randy H. Katz},
  title        = {Cirrus: a Serverless Framework for End-to-end {ML} Workflows},
  booktitle    = {Proceedings of the {ACM} Symposium on Cloud Computing, SoCC 2019,
                  Santa Cruz, CA, USA, November 20-23, 2019},
  pages        = {13--24},
  publisher    = {{ACM}},
  year         = {2019},
  url          = {https://doi.org/10.1145/3357223.3362711},
  address={''},
}

@inproceedings{WebServerless,
  author       = {Khanista Namee and
                  Rittiphon Phoarun and
                  Ghadeer Mohsen Albadrani and
                  Jantima Polpinij and
                  Sarayoot Tanessakulwattana and
                  Pongpol Sphanphong},
  title        = {A Form and {API} Data Management Platform for Progressive Web Application
                  and Serverless Application Architecture},
  booktitle    = {{CIIS} 2019: The 2nd International Conference on Computational Intelligence
                  and Intelligent Systems, Bangkok, Thailand, November, 2019},
  pages        = {144--149},
  publisher    = {{ACM}},
  year         = {2019},
  url          = {https://doi.org/10.1145/3372422.3372452},
  address={''},
}

@inproceedings{Firestore,
  author       = {Ram Kesavan and
                  David Gay and
                  Daniel Thevessen and
                  Jimit Shah and
                  C. Mohan},
  title        = {Firestore: The NoSQL Serverless Database for the Application Developer},
  booktitle    = {39th {IEEE} International Conference on Data Engineering, {ICDE} 2023,
                  Anaheim, CA, USA, April 3-7, 2023},
  pages        = {3376--3388},
  publisher    = {{IEEE}},
  year         = {2023},
  url          = {https://doi.org/10.1109/ICDE55515.2023.00259},
  address={''},
}

@article{Netherite,
  author       = {Sebastian Burckhardt and
                  Badrish Chandramouli and
                  Chris Gillum and
                  David Justo and
                  Konstantinos Kallas and
                  Connor McMahon and
                  Christopher Meiklejohn and
                  Xiangfeng Zhu},
  title        = {Netherite: Efficient Execution of Serverless Workflows},
  journal      = {Proc. {VLDB} Endow.},
  volume       = {15},
  number       = {8},
  pages        = {1591--1604},
  year         = {2022},
  url          = {https://www.vldb.org/pvldb/vol15/p1591-burckhardt.pdf},
address={''},
}

@article{RiseServerless,
  author       = {Paul C. Castro and
                  Vatche Ishakian and
                  Vinod Muthusamy and
                  Aleksander Slominski},
  title        = {The rise of serverless computing},
  journal      = {Commun. {ACM}},
  volume       = {62},
  number       = {12},
  pages        = {44--54},
  year         = {2019},
  url          = {https://doi.org/10.1145/3368454},
address={''},
}

@inproceedings{Defuse,
  author       = {Jiacheng Shen and
                  Tianyi Yang and
                  Yuxin Su and
                  Yangfan Zhou and
                  Michael R. Lyu},
  title        = {Defuse: {A} Dependency-Guided Function Scheduler to Mitigate Cold
                  Starts on FaaS Platforms},
  booktitle    = {41st {IEEE} International Conference on Distributed Computing Systems,
                  {ICDCS} 2021, Washington DC, USA, July 7-10, 2021},
  pages        = {194--204},
  publisher    = {{IEEE}},
  year         = {2021},
  url          = {https://doi.org/10.1109/ICDCS51616.2021.00027},
address={''},
}

@inproceedings{ServerlessWild,
  author       = {Mohammad Shahrad and
                  Rodrigo Fonseca and
                  I{\~{n}}igo Goiri and
                  Gohar Chaudhry and
                  Paul Batum and
                  Jason Cooke and
                  Eduardo Laureano and
                  Colby Tresness and
                  Mark Russinovich and
                  Ricardo Bianchini},
  title        = {Serverless in the Wild: Characterizing and Optimizing the Serverless
                  Workload at a Large Cloud Provider},
  booktitle    = {2020 {USENIX} Annual Technical Conference, {USENIX} {ATC} 2020, July
                  15-17, 2020},
  pages        = {205--218},
  publisher    = {{USENIX} Association},
  year         = {2020},
  url          = {https://www.usenix.org/conference/atc20/presentation/shahrad},
address={''},
}

@inproceedings{vHive,
  author       = {Dmitrii Ustiugov and
                  Plamen Petrov and
                  Marios Kogias and
                  Edouard Bugnion and
                  Boris Grot},
  title        = {Benchmarking, analysis, and optimization of serverless function snapshots},
  booktitle    = {{ASPLOS} '21: 26th {ACM} International Conference on Architectural
                  Support for Programming Languages and Operating Systems, Virtual Event,
                  USA, April 19-23, 2021},
  pages        = {559--572},
  publisher    = {{ACM}},
  year         = {2021},
  url          = {https://doi.org/10.1145/3445814.3446714},
address={''},
}

@online{AzureDataset,
title = {Azure Functions Trace 2019},
author = {Microsoft Azure and Microsoft Research},
url={https://azurecloudpublicdataset2.blob.core.windows.net/azurepublicdatasetv2/azurefunctions_dataset2019/},
year = {2019},
organization={Microsoft Research},
}

@online{AzureTriggers,
title = {Azure Functions triggers and bindings concepts},
author = {Microsoft},
url={https://learn.microsoft.com/en-us/azure/azure-functions/functions-triggers-bindings},
year = {2023}
}

@online{AzureScenario,
title = {Azure Functions scenarios},
author = {Microsoft},
url={https://learn.microsoft.com/en-us/azure/azure-functions/functions-scenarios?pivots=programming-language-csharp},
year = {2023}
}

@inproceedings{Hermod,
  author       = {Kostis Kaffes and
                  Neeraja J. Yadwadkar and
                  Christos Kozyrakis},
  title        = {Hermod: principled and practical scheduling for serverless functions},
  booktitle    = {Proceedings of the 13th Symposium on Cloud Computing, SoCC 2022, San
                  Francisco, California, November 7-11, 2022},
  pages        = {289--305},
  publisher    = {{ACM}},
  year         = {2022},
  url          = {https://doi.org/10.1145/3542929.3563468},
address={''},
}

@inproceedings{IceBreaker,
  author       = {Rohan Basu Roy and
                  Tirthak Patel and
                  Devesh Tiwari},
  title        = {IceBreaker: warming serverless functions better with heterogeneity},
  booktitle    = {{ASPLOS} '22: 27th {ACM} International Conference on Architectural
                  Support for Programming Languages and Operating Systems, Lausanne,
                  Switzerland, 28 February 2022 - 4 March 2022},
  pages        = {753--767},
  publisher    = {{ACM}},
  year         = {2022},
  url          = {https://doi.org/10.1145/3503222.3507750},
address={''},
}

@inproceedings{Xanadu,
  author       = {Nilanjan Daw and
                  Umesh Bellur and
                  Purushottam Kulkarni},
  title        = {Xanadu: Mitigating cascading cold starts in serverless function chain
                  deployments},
  booktitle    = {Middleware '20: 21st International Middleware Conference, Delft, The
                  Netherlands, December 7-11, 2020},
  pages        = {356--370},
  publisher    = {{ACM}},
  year         = {2020},
  url          = {https://doi.org/10.1145/3423211.3425690},
address={''},
}

@online{Reproduce,
title = {reproducing-serverless-in-the-wild},
year = {2022}, 
author = {Andreas Kuster and Jiyuan Shen and Magina Dai and Jiale Guo},
url = {https://github.com/andreaskuster/reproducing-serverless-in-the-wild},
}

@inproceedings{ChallengeStudy,
  author       = {Jinfeng Wen and
                  Zhenpeng Chen and
                  Yi Liu and
                  Yiling Lou and
                  Yun Ma and
                  Gang Huang and
                  Xin Jin and
                  Xuanzhe Liu},
  title        = {An empirical study on challenges of application development in serverless
                  computing},
  booktitle    = {{ESEC/FSE} '21: 29th {ACM} Joint European Software Engineering Conference
                  and Symposium on the Foundations of Software Engineering, Athens,
                  Greece, August 23-28, 2021},
  pages        = {416--428},
  publisher    = {{ACM}},
  year         = {2021},
  url          = {https://doi.org/10.1145/3468264.3468558},
address={''},
}

@inproceedings{FaaST,
author = {Romero, Francisco and Chaudhry, Gohar Irfan and Goiri, \'{I}\~{n}igo and Gopa, Pragna and Batum, Paul and Yadwadkar, Neeraja J. and Fonseca, Rodrigo and Kozyrakis, Christos and Bianchini, Ricardo},
title = {Faa\$T: A Transparent Auto-Scaling Cache for Serverless Applications},
year = {2021},
publisher = {Association for Computing Machinery},
address = {New York, NY, USA},
url = {https://doi.org/10.1145/3472883.3486974},
booktitle = {Proceedings of the ACM Symposium on Cloud Computing},
pages = {122–137},
numpages = {16},
series = {SoCC '21}
}

@inproceedings{FaaSRank,
  author       = {Hanfei Yu and
                  Athirai A. Irissappane and
                  Hao Wang and
                  Wes J. Lloyd},
  title        = {FaaSRank: Learning to Schedule Functions in Serverless Platforms},
  booktitle    = {{IEEE} International Conference on Autonomic Computing and Self-Organizing
                  Systems, {ACSOS} 2021, Washington, DC, USA, September 27 - Oct. 1,
                  2021},
  pages        = {31--40},
  publisher    = {{IEEE}},
  year         = {2021},
  url          = {https://doi.org/10.1109/ACSOS52086.2021.00023},
address={''},
}

@inproceedings{AQUATOPE,
  author       = {Zhuangzhuang Zhou and
                  Yanqi Zhang and
                  Christina Delimitrou},
  title        = {{AQUATOPE:} QoS-and-Uncertainty-Aware Resource Management for Multi-stage
                  Serverless Workflows},
  booktitle    = {Proceedings of the 28th {ACM} International Conference on Architectural
                  Support for Programming Languages and Operating Systems, Volume 1,
                  {ASPLOS} 2023, Vancouver, BC, Canada, March 25-29, 2023},
  pages        = {1--14},
  publisher    = {{ACM}},
  year         = {2023},
  url          = {https://doi.org/10.1145/3567955.3567960},
address={''},
}

@inproceedings{Catalyzer,
  author       = {Dong Du and
                  Tianyi Yu and
                  Yubin Xia and
                  Binyu Zang and
                  Guanglu Yan and
                  Chenggang Qin and
                  Qixuan Wu and
                  Haibo Chen},
  title        = {Catalyzer: Sub-millisecond Startup for Serverless Computing with Initialization-less
                  Booting},
  booktitle    = {{ASPLOS} '20: Architectural Support for Programming Languages and
                  Operating Systems, Lausanne, Switzerland, March 16-20, 2020},
  pages        = {467--481},
  publisher    = {{ACM}},
  year         = {2020},
  url          = {https://doi.org/10.1145/3373376.3378512},
address={''},
}

@inproceedings{FaaSFlow,
  author       = {Zijun Li and
                  Yushi Liu and
                  Linsong Guo and
                  Quan Chen and
                  Jiagan Cheng and
                  Wenli Zheng and
                  Minyi Guo},
  title        = {FaaSFlow: enable efficient workflow execution for function-as-a-service},
  booktitle    = {{ASPLOS} '22: 27th {ACM} International Conference on Architectural
                  Support for Programming Languages and Operating Systems, Lausanne,
                  Switzerland, 28 February 2022 - 4 March 2022},
  pages        = {782--796},
  publisher    = {{ACM}},
  year         = {2022},
  url          = {https://doi.org/10.1145/3503222.3507717},
  address={''},
}

@inproceedings{FaasCache,
  author       = {Alexander Fuerst and
                  Prateek Sharma},
  title        = {FaasCache: keeping serverless computing alive with greedy-dual caching},
  booktitle    = {{ASPLOS} '21: 26th {ACM} International Conference on Architectural
                  Support for Programming Languages and Operating Systems, Virtual Event,
                  USA, April 19-23, 2021},
  pages        = {386--400},
  publisher    = {{ACM}},
  year         = {2021},
  url          = {https://doi.org/10.1145/3445814.3446757},
  address={''},
}

@inproceedings{currentTrents,
  author       = {Parichehr Vahidinia and
                  Bahar J. Farahani and
                  Fereidoon Shams Aliee},
  title        = {Cold Start in Serverless Computing: Current Trends and Mitigation
                  Strategies},
  booktitle    = {2020 International Conference on Omni-layer Intelligent Systems, {COINS} 2020, Barcelona, Spain, Aug 31 - Sept 2, 2020},
  pages        = {1--7},
  publisher    = {{IEEE}},
  year         = {2020},
  url          = {https://doi.org/10.1109/COINS49042.2020.9191377},
  address={''},
}

@inproceedings{SOCK,
  author       = {Edward Oakes and
                  Leon Yang and
                  Dennis Zhou and
                  Kevin Houck and
                  Tyler Harter and
                  Andrea C. Arpaci{-}Dusseau and
                  Remzi H. Arpaci{-}Dusseau},
  title        = {{SOCK:} Rapid Task Provisioning with Serverless-Optimized Containers},
  booktitle    = {2018 {USENIX} Annual Technical Conference, {USENIX} {ATC} 2018, Boston,
                  MA, USA, July 11-13, 2018},
  pages        = {57--70},
  publisher    = {{USENIX} Association},
  year         = {2018},
  url          = {https://www.usenix.org/conference/atc18/presentation/oakes},
  address={''},
}

@inproceedings{FaaSNet,
  author       = {Ao Wang and
                  Shuai Chang and
                  Huangshi Tian and
                  Hongqi Wang and
                  Haoran Yang and
                  Huiba Li and
                  Rui Du and
                  Yue Cheng},
  title        = {FaaSNet: Scalable and Fast Provisioning of Custom Serverless Container
                  Runtimes at Alibaba Cloud Function Compute},
  booktitle    = {2021 {USENIX} Annual Technical Conference, {USENIX} {ATC} 2021, July
                  14-16, 2021},
  pages        = {443--457},
  publisher    = {{USENIX} Association},
  year         = {2021},
  url          = {https://www.usenix.org/conference/atc21/presentation/wang-ao},
address={''},
}

@article{KStest,
  title={Note on the kolmogorov statistic in the discrete case},
  author={Gottfried E. Noether},
  journal={Metrika},
  year={1963},
  volume={7},
  pages={115-116},
  url={https://api.semanticscholar.org/CorpusID:120687545}
}

@article{Formalfoundations,
  author       = {Abhinav Jangda and
                  Donald Pinckney and
                  Yuriy Brun and
                  Arjun Guha},
  title        = {Formal foundations of serverless computing},
  journal      = {Proc. {ACM} Program. Lang.},
  volume       = {3},
  number       = {{OOPSLA}},
  pages        = {149:1--149:26},
  year         = {2019},
  url          = {https://doi.org/10.1145/3360575},
}

@article{Economics,
  author       = {Adam Eivy},
  title        = {Be Wary of the Economics of "Serverless" Cloud Computing},
  journal      = {{IEEE} Cloud Comput.},
  volume       = {4},
  number       = {2},
  pages        = {6--12},
  year         = {2017},
  url          = {https://doi.org/10.1109/MCC.2017.32},
}

@inproceedings{Firecracker,
  author       = {Alexandru Agache and
                  Marc Brooker and
                  Alexandra Iordache and
                  Anthony Liguori and
                  Rolf Neugebauer and
                  Phil Piwonka and
                  Diana{-}Maria Popa},
  title        = {Firecracker: Lightweight Virtualization for Serverless Applications},
  booktitle    = {17th {USENIX} Symposium on Networked Systems Design and Implementation,
                  {NSDI} 2020, Santa Clara, CA, USA, February 25-27, 2020},
  pages        = {419--434},
  publisher    = {{USENIX} Association},
  year         = {2020},
  url          = {https://www.usenix.org/conference/nsdi20/presentation/agache},
  address={''},
}

@article{StaticAnalysis,
  author       = {Roman Haas and
                  Rainer Niedermayr and
                  Tobias Roehm and
                  Sven Apel},
  title        = {Is Static Analysis Able to Identify Unnecessary Source Code?},
  journal      = {{ACM} Trans. Softw. Eng. Methodol.},
  volume       = {29},
  number       = {1},
  pages        = {6:1--6:23},
  year         = {2020},
  url          = {https://doi.org/10.1145/3368267},
}

@article{FaaSLight,
author = {Liu, Xuanzhe and Wen, Jinfeng and Chen, Zhenpeng and Li, Ding and Chen, Junkai and Liu, Yi and Wang, Haoyu and Jin, Xin},
title = {FaaSLight: General Application-Level Cold-Start Latency Optimization for Function-as-a-Service in Serverless Computing},
year = {2023},
publisher = {Association for Computing Machinery},
address = {New York, NY, USA},
volume = {32},
number = {5},
issn = {1049-331X},
url = {https://doi.org/10.1145/3585007},
journal = {ACM Trans. Softw. Eng. Methodol.},
month = {jul},
articleno = {119},
numpages = {29},
}

@inproceedings{ServerlessData,
  author       = {Josep Samp{\'{e}} and
                  Gil Vernik and
                  Marc S{\'{a}}nchez Artigas and
                  Pedro Garc{\'{\i}}a L{\'{o}}pez},
  title        = {Serverless Data Analytics in the {IBM} Cloud},
  booktitle    = {Proceedings of the 19th International Middleware Conference, Middleware
                  Industrial Track 2018, Rennes, France, December 10-14, 2018},
  pages        = {1--8},
  publisher    = {{ACM}},
  year         = {2018},
  url          = {https://doi.org/10.1145/3284028.3284029},
  address={''},
}

@inproceedings{ServerlessDataScience,
  author       = {Yuncheng Wu and
                  Tien Tuan Anh Dinh and
                  Guoyu Hu and
                  Meihui Zhang and
                  Yeow Meng Chee and
                  Beng Chin Ooi},
  title        = {Serverless Data Science - Are We There Yet? {A} Case Study of Model
                  Serving},
  booktitle    = {{SIGMOD} '22: International Conference on Management of Data, Philadelphia,
                  PA, USA, June 12 - 17, 2022},
  pages        = {1866--1875},
  publisher    = {{ACM}},
  year         = {2022},
  url          = {https://doi.org/10.1145/3514221.3517905},
  address={''},
}

@inproceedings{TowardsML,
  author       = {Jiawei Jiang and
                  Shaoduo Gan and
                  Yue Liu and
                  Fanlin Wang and
                  Gustavo Alonso and
                  Ana Klimovic and
                  Ankit Singla and
                  Wentao Wu and
                  Ce Zhang},
  title        = {Towards Demystifying Serverless Machine Learning Training},
  booktitle    = {{SIGMOD} '21: International Conference on Management of Data, Virtual
                  Event, China, June 20-25, 2021},
  pages        = {857--871},
  publisher    = {{ACM}},
  year         = {2021},
  url          = {https://doi.org/10.1145/3448016.3459240},
  address={''},
}

@inproceedings{Atoll,
  author       = {Arjun Singhvi and
                  Arjun Balasubramanian and
                  Kevin Houck and
                  Mohammed Danish Shaikh and
                  Shivaram Venkataraman and
                  Aditya Akella},
  title        = {Atoll: {A} Scalable Low-Latency Serverless Platform},
  booktitle    = {SoCC '21: {ACM} Symposium on Cloud Computing, Seattle, WA, USA, November
                  1 - 4, 2021},
  pages        = {138--152},
  publisher    = {{ACM}},
  year         = {2021},
  url          = {https://doi.org/10.1145/3472883.3486981},
  address={''},
}

@inproceedings{Peeking,
  author       = {Liang Wang and
                  Mengyuan Li and
                  Yinqian Zhang and
                  Thomas Ristenpart and
                  Michael M. Swift},
  title        = {Peeking Behind the Curtains of Serverless Platforms},
  booktitle    = {2018 {USENIX} Annual Technical Conference, {USENIX} {ATC} 2018, Boston,
                  MA, USA, July 11-13, 2018},
  pages        = {133--146},
  publisher    = {{USENIX} Association},
  year         = {2018},
  url          = {https://www.usenix.org/conference/atc18/presentation/wang-liang},
  address={''},
}

@inproceedings{AJAX,
  author       = {Christoffer Quist Adamsen and
                  Anders M{\o}ller and
                  Saba Alimadadi and
                  Frank Tip},
  title        = {Practical {AJAX} race detection for JavaScript web applications},
  booktitle    = {Proceedings of the 2018 {ACM} Joint Meeting on European Software Engineering
                  Conference and Symposium on the Foundations of Software Engineering,
                  {ESEC/SIGSOFT} {FSE} 2018, Lake Buena Vista, FL, USA, November 04-09,
                  2018},
  pages        = {38--48},
  publisher    = {{ACM}},
  year         = {2018},
  url          = {https://doi.org/10.1145/3236024.3236038},
  address={''},
}

@inproceedings{Kraken,
  author       = {Vivek M. Bhasi and
                  Jashwant Raj Gunasekaran and
                  Prashanth Thinakaran and
                  Cyan Subhra Mishra and
                  Mahmut Taylan Kandemir and
                  Chita R. Das},
  title        = {Kraken: Adaptive Container Provisioning for Deploying Dynamic DAGs
                  in Serverless Platforms},
  booktitle    = {SoCC '21: {ACM} Symposium on Cloud Computing, Seattle, WA, USA, November
                  1 - 4, 2021},
  pages        = {153--167},
  publisher    = {{ACM}},
  year         = {2021},
  url          = {https://doi.org/10.1145/3472883.3486992},
  address={''},
}

@inproceedings{LaptopLambda,
  author       = {Sadjad Fouladi and
                  Francisco Romero and
                  Dan Iter and
                  Qian Li and
                  Shuvo Chatterjee and
                  Christos Kozyrakis and
                  Matei Zaharia and
                  Keith Winstein},
  title        = {From Laptop to Lambda: Outsourcing Everyday Jobs to Thousands of Transient
                  Functional Containers},
  booktitle    = {2019 {USENIX} Annual Technical Conference, {ATC} Renton,
                  WA, USA, July 10-12, 2019},
  pages        = {475--488},
  publisher    = {{USENIX} Association},
  year         = {2019},
  url          = {https://www.usenix.org/conference/atc19/presentation/fouladi},
  address={''},
}

@inproceedings{INFless,
  author       = {Yanan Yang and
                  Laiping Zhao and
                  Yiming Li and
                  Huanyu Zhang and
                  Jie Li and
                  Mingyang Zhao and
                  Xingzhen Chen and
                  Keqiu Li},
  title        = {INFless: a native serverless system for low-latency, high-throughput
                  inference},
  booktitle    = {{ASPLOS} '22: 27th {ACM} International Conference on Architectural
                  Support for Programming Languages and Operating Systems, Lausanne,
                  Switzerland, 28 February 2022 - 4 March 2022},
  pages        = {768--781},
  publisher    = {{ACM}},
  year         = {2022},
  url          = {https://doi.org/10.1145/3503222.3507709},
  address={''},
}

@inproceedings{SCache,
  author       = {Chen Chen and
                  Lars Nagel and
                  Lin Cui and
                  Fung Po Tso},
  title        = {S-Cache: Function Caching for Serverless Edge Computing},
  booktitle    = {Proceedings of the 6th International Workshop on Edge Systems, Analytics
                  and Networking, EdgeSys 2023, Rome, Italy, 8 May 2023},
  pages        = {1--6},
  publisher    = {{ACM}},
  year         = {2023},
  url          = {https://doi.org/10.1145/3578354.3592865},
  address={''},
}

@inproceedings{Harvested,
  author       = {Yanqi Zhang and
                  I{\~{n}}igo Goiri and
                  Gohar Irfan Chaudhry and
                  Rodrigo Fonseca and
                  Sameh Elnikety and
                  Christina Delimitrou and
                  Ricardo Bianchini},
  title        = {Faster and Cheaper Serverless Computing on Harvested Resources},
  booktitle    = {{SOSP} '21: {ACM} {SIGOPS} 28th Symposium on Operating Systems Principles,
                  Virtual Event / Koblenz, Germany, October 26-29, 2021},
  pages        = {724--739},
  publisher    = {{ACM}},
  year         = {2021},
  url          = {https://doi.org/10.1145/3477132.3483580},
  address={''},
}

@inproceedings{Ensure,
  author       = {Amoghavarsha Suresh and
                  Gagan Somashekar and
                  Anandh Varadarajan and
                  Veerendra Ramesh Kakarla and
                  Hima Upadhyay and
                  Anshul Gandhi},
  title        = {{ENSURE:} Efficient Scheduling and Autonomous Resource Management
                  in Serverless Environments},
  booktitle    = {{IEEE} International Conference on Autonomic Computing and Self-Organizing
                  Systems, {ACSOS} 2020, Washington, DC, USA, August 17-21, 2020},
  pages        = {1--10},
  publisher    = {{IEEE}},
  year         = {2020},
  url          = {https://doi.org/10.1109/ACSOS49614.2020.00020},
  address={''},
}

@inproceedings{FaaSnap,
  author       = {Lixiang Ao and
                  George Porter and
                  Geoffrey M. Voelker},
  title        = {FaaSnap: FaaS made fast using snapshot-based VMs},
  booktitle    = {EuroSys '22: Seventeenth European Conference on Computer Systems,
                  Rennes, France, April 5 - 8, 2022},
  pages        = {730--746},
  publisher    = {{ACM}},
  year         = {2022},
  url          = {https://doi.org/10.1145/3492321.3524270},
  address={''},
}

@article{lee2024spes,
  title={SPES: Towards Optimizing Performance-Resource Trade-Off for Serverless Functions},
  author={Lee, Cheryl and Zhu, Zhouruixin and Yang, Tianyi and Huo, Yintong and Su, Yuxin and He, Pinjia and Lyu, Michael R},
  journal={arXiv preprint arXiv:2403.17574},
  year={2024}
}

@inproceedings{suresh2019fnsched,
  title={Fnsched: An efficient scheduler for serverless functions},
  author={Suresh, Amoghvarsha and Gandhi, Anshul},
  booktitle={Proceedings of the 5th international workshop on serverless computing},
  pages={19--24},
  year={2019}
}

@inproceedings{aumala2019beyond,
  title={Beyond load balancing: Package-aware scheduling for serverless platforms},
  author={Aumala, Gabriel and Boza, Edwin and Ortiz-Avil{\'e}s, Luis and Totoy, Gustavo and Abad, Cristina},
  booktitle={2019 19th IEEE/ACM International Symposium on Cluster, Cloud and Grid Computing (CCGRID)},
  pages={282--291},
  year={2019},
  organization={IEEE}
}

@inproceedings{abdi2023palette,
  title={Palette load balancing: Locality hints for serverless functions},
  author={Abdi, Mania and Ginzburg, Samuel and Lin, Xiayue Charles and Faleiro, Jose and Chaudhry, Gohar Irfan and Goiri, Inigo and Bianchini, Ricardo and Berger, Daniel S and Fonseca, Rodrigo},
  booktitle={Proceedings of the Eighteenth European Conference on Computer Systems},
  pages={365--380},
  year={2023}
}

@inproceedings{fuerst2022locality,
  title={Locality-aware load-balancing for serverless clusters},
  author={Fuerst, Alexander and Sharma, Prateek},
  booktitle={Proceedings of the 31st International Symposium on High-Performance Parallel and Distributed Computing},
  pages={227--239},
  year={2022}
}

@inproceedings{akbari2025hiku,
  title={Hiku: Pull-based scheduling for serverless computing},
  author={Akbari, Saman and Hauswirth, Manfred},
  booktitle={2025 IEEE 25th International Symposium on Cluster, Cloud and Internet Computing (CCGrid)},
  pages={450--461},
  year={2025},
  organization={IEEE}
}

@inproceedings{lu2025ch,
  title={CH-AL: A Consistent Hashing-Based Adaptive Load Balancing Strategy for Serverless Computing},
  author={Lu, Xinxi and Han, Ziyao and Liu, Yang},
  booktitle={2025 IEEE International Conference on Joint Cloud Computing (JCC)},
  pages={1--7},
  year={2025},
  organization={IEEE}
}

@article{chen2025metronome,
  title={Metronome: Differentiated Delay Scheduling for Serverless Functions},
  author={Chen, Zhuangbin and Zheng, Juzheng and Zheng, Zibin},
  journal={arXiv preprint arXiv:2512.05703},
  year={2025}
}

\end{document}